\documentclass[a4paper,11pt]{article}
\pdfoutput=1 

\usepackage{jheppub} 
\usepackage{hyperref}
\hypersetup{
    bookmarks=true,         
    unicode=false,          
    pdftoolbar=true,        
    pdfmenubar=true,        
    pdffitwindow=false,     
    pdfstartview={FitH},    
    pdftitle={My title},    
    pdfauthor={Author},     
    pdfsubject={Subject},   
    pdfcreator={Creator},   
    pdfproducer={Producer}, 
    pdfkeywords={keyword1} {key2} {key3}, 
    pdfnewwindow=true,      
    colorlinks=true,       
    linkcolor=red,          
    citecolor=purple,        
    filecolor=magenta,      
    urlcolor=blue,           
    linktocpage=true
}
\usepackage{graphics}
\usepackage{rotating}
\usepackage{subfigure}
\usepackage{pstricks}
\usepackage{color}
\usepackage{amsfonts}
\usepackage{mathrsfs,amsmath}
\usepackage{epsfig}
\usepackage{amsmath,amssymb,amsthm,graphicx,latexsym}
\usepackage{ulem}

\newcommand{\be}{\begin{equation}}
\newcommand{\ee}{\end{equation}}
\newcommand{\bea}{\begin{eqnarray}}
\newcommand{\eea}{\end{eqnarray}}
\newcommand{\bml}{\begin{subequations}}
\newcommand{\eml}{\end{subequations}}
\newcommand{\bfig}{\begin{figure}}
\newcommand{\efig}{\end{figure}}

\newcommand{\slashed}{\hspace{-1.1ex}/}

\newcommand{\bmat}{\begin{pmatrix}}
\newcommand{\emat}{\end{pmatrix}}
\begin{document}
	$~~~~~~~~~~~~~~~~~~~~~~~~~~~~~~~~~~~~~~~~~~~~~~~~~~~~~~~~~~~~~~~~~~~~~~~~~~~~~~~~~~~~$\textcolor{red}{\Large\bf TIFR/TH/17-01}
	\title{\textsc{\fontsize{25}{17}\selectfont \sffamily \bfseries \textcolor{purple}{Entangled de Sitter from Stringy  \\ Axionic Bell pair I: An analysis using Bunch-Davies vacuum}}}
	
	\author[a,b]{Sayantan Choudhury,
		\footnote{\textcolor{purple}{\bf Presently working as a Post-Doctoral fellow at IUCAA, Pune, \\$~~~~~$Alternative
				E-mail: sayanphysicsisi@gmail.com}. ${}^{}$}}
	\
	
\author[c,d,e]{Sudhakar Panda
}
\affiliation[a]{
	Inter-University Centre for Astronomy and Astrophysics, Post Bag 4,
	Ganeshkhind, Pune 411007, India.
}
	\affiliation[b]{Department of Theoretical Physics, Tata Institute of Fundamental Research, Colaba, Mumbai - 400005, India.
	}
	\affiliation[c]{Institute of Physics, Sachivalaya Marg, Bhubaneswar, Odisha - 751005, India.
		}
		\affiliation[d]{	
		National Institute of Science Education and Research,
		Jatni, Bhubaneswar, Odisha - 752050, India.}
		\affiliation[e]{Homi Bhabha National Institute, Training School Complex,
		Anushakti Nagar, Mumbai-400085, India.
		}
	\emailAdd{sayantan@iucaa.in, panda@iopb.res.in}

	\abstract{In this work, we study the quantum entanglement and compute  entanglement entropy in de Sitter space for a bipartite quantum field theory driven by axion originating from {\bf Type IIB} string compactification on a Calabi-Yau three fold (${\bf CY^3}$) and in presence of ${\bf NS5}$ brane. For this compuation, we consider a spherical surface ${\bf S}^2$, which divide the spatial slice of de Sitter (${\bf dS_4}$) into exterior and interior sub regions. We also consider the initial choice of vaccum to be Bunch-Davies state. First we derive the solution of the wave function of axion in a hyperbolic open chart by constructing a suitable basis for Bunch-Davies vacuum state using Bogoliubov transformation. We then, derive the expression for density matrix by tracing over the exterior region. This allows us to compute entanglement entropy and R$\acute{e}$nyi entropy in $3+1$ dimension.  Further we quantify the UV finite contribution of entanglement entropy which contain the physics of long range quantum correlations of our expanding universe. Finally, our analysis compliments the necessary condition for generating non vanishing entanglement entropy in primordial cosmology due to the axion. 
	}

	\maketitle
	\flushbottom
	\section{\textcolor{blue}{Introduction}}
	
	The information encoded in entanglement entropy is very powerful to distinguish quantum states with long range correlation as appearing in the context of condensed matter physics \cite{Amico:2007ag,Horodecki:2009zz,Laflorencie:2015eck}. It plays a  significant role in the field of quantum information \cite{Plenio:2007zz,Cerf:1996nb,Cerf:1995sa,Calabrese:2004eu,Horodecki:2004ez} and cosmoloogy \cite{MartinMartinez:2012sg,Nambu:2008my,Campo:2005qn,Nambu:2011ae,VerSteeg:2007xs,Mazur:2008wa,Maldacena:2012xp,Maldacena:2015bha,Choudhury:2016cso,Choudhury:2016pfr,Kanno:2014lma,Kanno:2017dci,Kanno:2016gas,Kanno:2014bma,Kanno:2014ifa,Fischler:2013fba,Fischler:2014tka}. It is also useful in quantum field theory to identify the specific nature of the long range quantum correlations that we generally get using the standard vacuum states like the Chernikov-Tagirov, Bunch-Davies, Hartle-Hawking or the most general $\alpha$ vacuum \cite{Chernikov:1968zm,Bunch:1978zm,Hartle:1983ai}. Till date quantum entanglement has been one of the most mysterious and at the same time fascinating features of  foundational aspects of quantum mechanics. This is because, it is believed that performing a local measurement may instantaneously affect the outcome of the measurement beyond the physical light cone under consideration. This is interpreted as the violation of causality and commonly known as the Einstein-Podolsky-Rosen (EPR) paradox \cite{Bell:1964kc}. However, it is important to note that neither quantum information gets transferred in such a physical measurement, nor it changes the causality constraints. Nevertheless, quantum entanglement may have applications to explain the origin of various physical phenomena \cite{Bombelli:1986rw,Srednicki:1993im,Casini:2009sr}. The prime example is the well known Schwinger effect in de Sitter \cite{Frob:2014zka,Fischler:2014ama} space describing pair production of particles in presence of uniform electric field \cite{Kanno:2014lma}. In this specific context, the pair production occurs spontaneously with a certain finite coordinate separation in field space. Such physical states of the particle pairs are required to be correlated quantum mechanically. 
	
			    \begin{figure*}[htb]
			    \centering
			    {
			        \includegraphics[width=12.5cm,height=6.5cm] {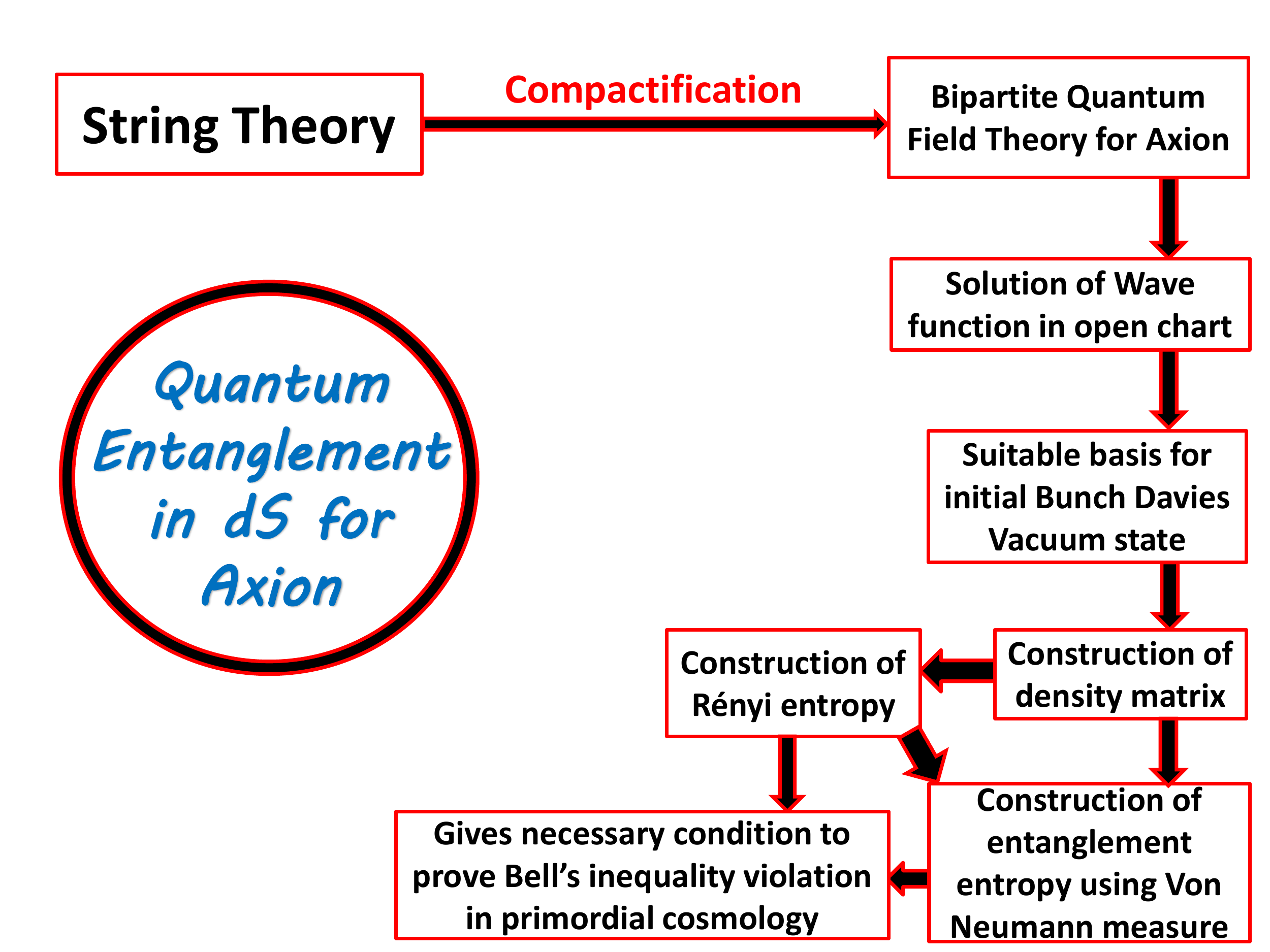}
			    }
			    \caption[Optional caption for list of figures]{Schematic diagram for the computation algorithm of entanglement entropy in de Sitter space due to axion.} 
			    \label{fzaa}
			    \end{figure*}
	 
	In the present work we consider an axion field theory obtained from compactification of {\bf Type IIB} string theory on a Calabi-Yau three fold ({\bf CY$^3$}) but in presence of {\bf NS5} brane. So that the model consist of effective potential which breaks the shift symmetry associated with the axion field in a controled manner. This model has been analyzed as a candidate for driving inflation \cite{McAllister:2008hb,Silverstein:2008sg,McAllister:2014mpa} as well as a quintessence model for the late time acceleration of our universe \cite{Panda:2010uq}. Furthermore, in a recent analysis it was shown that this model captures an interesting phenomena of violation of Bell's inequality in primordial cosmology \cite{Maldacena:2015bha,Choudhury:2016cso,Choudhury:2016pfr} by formation of EPR pair. In the present analysis we will strengthen the formation of EPR pair by demonstrating that the entanglement entropy of axionic EPR pair is non zero, which confirms the existence of superhorizon long range correlation in primordial cosmology. This connection also will be helpful in future to provide an algorithm to contrast various models of inflation \cite{Choudhury:2015hvr,Maharana:1997cz,Baumann:2009ds,Agarwal:2011wm}. Additionally, we we compute R$\acute{e}$nyi entropy in $3+1$ D de Sitter space and its connection with quantum entanglement in presence of axionic model.

	In fig.~(\ref{fzaa}), we have schematically shown the logic involved in computing the entanglement entropy in de Sitter space. The plan of the rest of the paper is as follows. In \underline{\textcolor{purple}{\bf Section \ref{ka1}}} we briefly review the strategy for computing the entanglement entropy in de Sitter space. In \underline{\textcolor{purple}{\bf Section \ref{ka33}}} we follow the above strategy for the afore mentioned stringy axionic model. In \underline{\textcolor{purple}{\bf Section \ref{ka44}}} we present our conclusion and future prospects.  Finally, in \underline{\textcolor{purple}{\bf Appendix}} we give the detailed derivation of entanglement entropy and R$\acute{e}$nyi entropy in the present context.

	\section{\textcolor{blue}{ Computational strategy: Brief review}}
	\label{ka1}
	Here we briefly review the method to derive entanglement entropy in 3+1 dimensional de Sitter space following ref.~\cite{Maldacena:2012xp} and references therein. We consider a closed surface $\Sigma$ in a hypersurface where time is frozen. Consequently, the space-like hypersurface is divided into an interior and exterior region which are identified as \textcolor{red}{\bf RI} (which is \textcolor{red}{\bf L} in open chart) and \textcolor{red}{\bf RII} (which is \textcolor{red}{\bf R} in open chart) region in the representative schematic diagram shown in fig.~(\ref{fzaca}). If we represent the total Hilbert space of the system under consideration by ${\bf \cal H}$, then in the present context one can use the following approximate bipartite decomposition \cite{Callan:1994py}, given by~\footnote{In general, for a general quantum field theoretic system one cannot use the direct product decomposition ansatz for the total Hilbert space i.e. ${\bf \cal H}\neq {\bf \cal H}_{\bf INT}\otimes {\bf \cal H}_{\bf EXT},$ 
	which implies that in general quantum field theory is not bipartite. But if we introduce a short distance {\bf UV cut-off} in the local QFT $\epsilon_{\bf UV}$, which sometimes identified to be the lattice {\bf UV cut-off} then only one can treat the local quantum field theoretic system as a bipartite system.},
	${\bf \cal H}={\bf \cal H}_{\bf INT}\otimes {\bf \cal H}_{\bf EXT}.$
	This implies that ${\bf \cal H}$ can be written as a direct product of two Hilbert spaces ${\bf \cal H}_{\bf INT}$ and ${\bf \cal H}_{\bf EXT}$ which is consistent with the requirement of local QFT. In this context, the building blocks of ${\bf \cal H}_{\bf INT}$ and ${\bf \cal H}_{\bf EXT}$ are the localized modes in the \textcolor{red}{\bf RI} and \textcolor{red}{\bf RII} region respectively. This allows us to construct the density matrix for the internal \textcolor{red}{\bf RI} region by tracing over all degrees of freedom in the external \textcolor{red}{\bf RII} region and is given by:
	\bea \rho={\bf Tr}_{\bf R}|{\bf BD}\rangle \langle {\bf BD}|.\eea
	Here the vacuum state $|{\bf BD}\rangle$ is the Bunch-Davies vacuum. Using the well known Von Neumann entropy formula, the entanglement entropy in de Sitter space can be expressed in terms of the density matrix as:
	\bea S&=& -{\bf Tr}\left[\rho\ln \rho\right],\eea
	which quantifies how much a given quantum state is
	quantum mechanically entangled.
	In this paper we compute this expression explicitly and establish its relationship with Bell's inequality violation in cosmology \cite{Maldacena:2015bha,Choudhury:2016cso,Choudhury:2016pfr}. 
    
    Here we consider $\Sigma$ to be a closed surface ${\bf S}^2$ which
        			has radius $R_{{\bf S}^2}$ with the restriction, $R_{{\bf S}^2}>>R_{\bf dS}=H^{-1}$, where $H$ is the Hubble parameter and $R_{\bf dS}$ is the de Sitter radius. This surface ${\bf S}^2$ separates the spatial slice into the interior (\textcolor{red}{\bf L}) and
        			exterior (\textcolor{red}{\bf R}) region which is shown in fig.~(\ref{fzaca}).  The reduced density matrix, which is a key ingredient  for computing entanglement entropy, is obtained by tracing over the exterior (\textcolor{red}{\bf R}) region. 	Also it is important to note that the total entanglement entropy can be expressed as a sum of {\bf UV} divergent and {\bf UV} finite contribution as:
    \bea S&=&S_{\bf UV-divergent}+S_{\bf UV-finite}.\eea
     When the ${\bf S}^2$ is taken at the boundary of $3+1$ D de Sitter space then ${\bf  SO(1,3)}$ symmetry~\footnote{In arbitrary D dimensions entanglement entropy in de Sitter space is invariant under ${\bf  SO(1,D)}$ isometry group.} plays significant role to define the entanglement entropy for a given principal quantum number. We note that entanglement entropy in de Sitter space is completely different from that of the dynamical gravity \cite{Ryu:2006bv,Ryu:2006ef,Nishioka:2009un,Rangamani:2016dms,Hubeny:2007xt,Dong:2013qoa,Camps:2013zua,Banerjee:2014oaa,Bhattacharyya:2014yga,Pal:2015mda}, which we are not considering. 
     
    		 
    
    In $3+1$ dimensions, the {\bf UV-divergent} contribution is appearing due to the local quantum field theoretic effects and it takes the following simplified form \cite{Srednicki:1993im,Bombelli:1986rw,Maldacena:2012xp,Kanno:2014lma}:
    \bea \label{et} S_{\bf UV-divergent}&=&{\bf c_1}\frac{{\cal A}_{\bf ENT}}{\epsilon^2_{\bf UV}}+\left[{\bf c_2}+\left({\bf c_3}m^2+{\bf c_4}H^2\right){\cal A}_{\bf ENT}\right]\ln\left(\epsilon_{\bf UV}H\right),\eea
    where $\epsilon_{\bf UV}$ is the short distance lattice {\bf UV cut-off} of the bipartite local quantum field theory under consideration, ${\cal A}_{\bf ENT}$ represents the proper area of the entangling region and ${\bf c_i}\forall i=1,2,3,4$ are the numerical coefficients which carries different physical significance in the present computation. For an example, in Eqn~(\ref{et}), the first term represent area dependent contribution to the {\bf UV divergent} part of the entropy, which is originated from entanglement between particles in the vicinity of the entangling surface under consideration. Further taking flat space limit, where the Hubble parameter $H\rightarrow 0$, the {\bf UV-divergent} part of the entropy of the bipartite local quantum field theoretic system for scalar field (axion) can be expressed as~\footnote{  Here we in the flat space limit $H\rightarrow 0$ we get:
            \bea \lim_{H\rightarrow 0}\ln\left(\epsilon_{\bf UV}H\right)&=&2\ln\left(\epsilon_{\bf UV}\right)+\Delta,\eea
            where $\Delta$ characterizes the following finite contribution in the flat space limit, which is given by:
            \bea \Delta&=&\lim_{H\rightarrow 0}\sum^{\infty}_{n=1}\frac{(-1)^{n+1}}{n!}\left(\frac{H}{\epsilon_{\bf UV}}-1\right)^{n}=-\sum^{\infty}_{n=1}\frac{1}{n!}=1-e.\eea}:
    \bea \lim_{H\rightarrow 0}S_{\bf UV-divergent}&=&{\bf c_1}\frac{{\cal A}_{\bf ENT}}{\epsilon^2_{\bf UV}}+\left[{\bf c_2}+{\bf c_3}m^2{\cal A}_{\bf ENT}\right]\ln\left(\epsilon^2_{\bf UV}e^{1-e}\right),\eea
    In this context the coefficients ${\bf c_2}$ and ${\bf c_3}$ contribute in the flat limiting result of the entropy of the system under consideration. Also it is important to note that, the coefficients ${\bf c_2}$ and ${\bf c_3}$ are short distance lattice {\bf UV cut-off} independent and therefore universal in nature. On the other hand, the last term in Eqn~(\ref{et}), explicitly incorporates the effect of bulk extrinsic and intrinsic curvature contribution in de Sitter background. 
    
     In our computation of entanglement entropy in de Sitter space we only restrict our discussion within the {\bf UV-finite} contributions which contain the physics of long range correlations of the quantum mechanical state. Most important fact is that, the total entanglement entropy in de Sitter space is invariant under ${\bf SO(1,4)}$ isometry group and consequently both {\bf UV-divergent} and {\bf UV-finite} contributions remain invariant under the same group. However, for the computation we only use the connected subgroup of ${\bf SO(1,4)}$, which is ${\bf SO(1,3)}$ in this context. For more details see ref.~\cite{Maldacena:2012xp}. Additionally, in the present context in the subhorizon scale, where $kc_S \eta>>-1$~\footnote{Here $k$ is the momentum scale of the corresponding Fourier modes in De Sitter space, $c_S$ is the speed of sound and $\eta$ is the conformal time.}, plays crucial role as the long range entanglement is valid within this scale. In the superhorizon scale, where $kc_S \eta<<-1$ or more precisely at late time scale i.e. at the limiting case $\eta\rightarrow 0$, long range contributions in the de Sitter entanglement entropy freezes. However, in this limit some additional contribution appears which only contributes to short distance scale and the associated entanglement can be expressed in a local form in (quasi) de Sitter space. Consequently, at late time scale, $\eta\rightarrow 0$ of the {\bf UV-finite} contributions in the entanglement entropy can be written as \cite{Maldacena:2012xp,Kanno:2014lma}:
     \bea \label{etw} S_{\bf UV-finite}&=&{\bf c_5}{\cal A}_{\bf ENT}H^2-\frac{\bf c_6}{2}\ln\left({\cal A}_{\bf ENT}H^2\right)+{\bf finite~contributions}.\eea
     Here the {\bf UV finite} part gets contributions from two different sources. 
      First part being proportional to the area of entangling surface of ${\bf S}^2$ and the proportionality factor is dependent on a overall coefficient and Hubble parameter $H$. Second part is proportional to the logarithm of the product of area of ${\bf S}^2$ and Hubble parameter $H$ and the proportionality factor is identified to be the interesting part of the entanglement entropy which we compute explicitly from the prescribed setup.
        
      Additionally it is important to note the overall characteristic coefficient as appearing in the first term in {\bf UV finite} contribution is not completely independent. Once we know the coefficient of the logarithmic term, one can determine the other coefficient in terms of this contribution.
     
     Further taking the flat space limit, $H\rightarrow 0$, the {\bf UV-finite} contributions to the entanglement entropy can be expressed as~\footnote{Here we use the flat space limit $H\rightarrow 0$ we get:
                   \bea \lim_{H\rightarrow 0}\ln\left({\cal A}_{\bf ENT}H^2\right)&=&3\ln\left({\cal A}_{\bf ENT}\right)+2\Theta,\eea
                   where $\Theta$ characterizes the following contribution in the flat space limit, which is given by:
                   \bea \Theta&=&\lim_{H\rightarrow 0}\sum^{\infty}_{n=1}\frac{(-1)^{n+1}}{n}\left(\frac{H}{{\cal A}_{\bf ENT}}-1\right)^{n}=-\sum^{\infty}_{n=1}\frac{1}{n},\eea
                   which is in general divergent but can be regularized by putting the previously introduced lattice {\bf UV} cut-off $\epsilon_{\bf UV}$ as:
                    \bea \Theta~~~\underrightarrow{\bf Lattice~regulator~\epsilon_{\bf UV}}~~~\Theta_{\bf REG}&=&\lim_{H\rightarrow 0}\sum^{\epsilon_{\bf UV}}_{n=1}\frac{(-1)^{n+1}}{n}\left(\frac{H}{{\cal A}_{\bf ENT}}-1\right)^{n}=-\sum^{\epsilon_{\bf UV}}_{n=1}\frac{1}{n}=-{\cal H}_{\epsilon_{\bf UV}},~~~~~\eea
                    where ${\cal H}_{\epsilon_{\bf UV}}$ is the harmonic number of order $\epsilon_{\bf UV}$.}:
     \bea \label{etwss}  \lim_{H\rightarrow 0}S_{\bf UV-finite}&=&-\frac{3{\bf c_6}}{2}\ln\left({\cal A}_{\bf ENT}\right)+{\bf c_6}{\cal H}_{\epsilon_{\bf UV}}+{\bf finite~contributions}.\eea
     Further, we introduce a new parameter ${\cal A}_{\bf C}$ representing the entangling area in comoving coordinates, which is defined as, 
          ${\cal A}_{\bf C}={\cal A}_{\bf ENT}H^2\eta^2.$
          In terms of newly introduced parameter the {\bf UV-finite} part of the entanglement entropy can be written as:
    \bea \label{etwaaa} S_{\bf UV-finite}&=&{\bf c_5}\frac{{\cal A}_{\bf C}}{\eta^2}+{\bf c_6}\ln\eta-\frac{\bf c_6}{2}\ln\left({\cal A}_{\bf C}\right) +{\bf finite~contributions},\eea
     where 
      the co-efficient of the logarithmic contribution, ${\bf c_6}$, characterizes the long range correlations. In our computation our prime objective is to compute the expression for ${\bf c_6}$ in de Sitter space. On general ground, ${\bf c_6}$ can be expressed as a sum of two possible
      contributions of the extrinsic curvature of the entangling surface ($\Sigma\equiv {\bf S^2}$) in comoving coordinate system as \cite{Solodukhin:2008dh}:
      \bea {\bf c_6}&=&\left[{\bf g_1}\int_{\Sigma\equiv {\bf S^2}}K_{AB}K_{AB}+{\bf g_2}\int_{\Sigma\equiv {\bf S^2}}\left(K_{AA}\right)^2\right]\equiv S_{\bf intr},\eea 
      where ${\bf g_1}$ and ${\bf g_2}$ are two undetermined co-efficients and $K_{AB}$ is the extrinsic curvature of entangling surface. Here we note that, the co-efficient ${\bf c_6}$ is identified as the interesting part of the ${\bf UV-finite}$ contribution to the entanglement entropy, as represented by $S_{\bf intr}$, which we compute in this paper. 
            
      In the next section,we compute the entanglement entropy for axionic Bell pairs using the results of this section.

    \section{\textcolor{blue}{Entanglement for axionic Bell pair}}
    \label{ka33}

    \subsection{Geometrical construction and underlying symmetries}
    \label{ka33a}
    \begin{figure*}[htb]
    	\centering
    	{
    		\includegraphics[width=9.5cm,height=6.5cm] {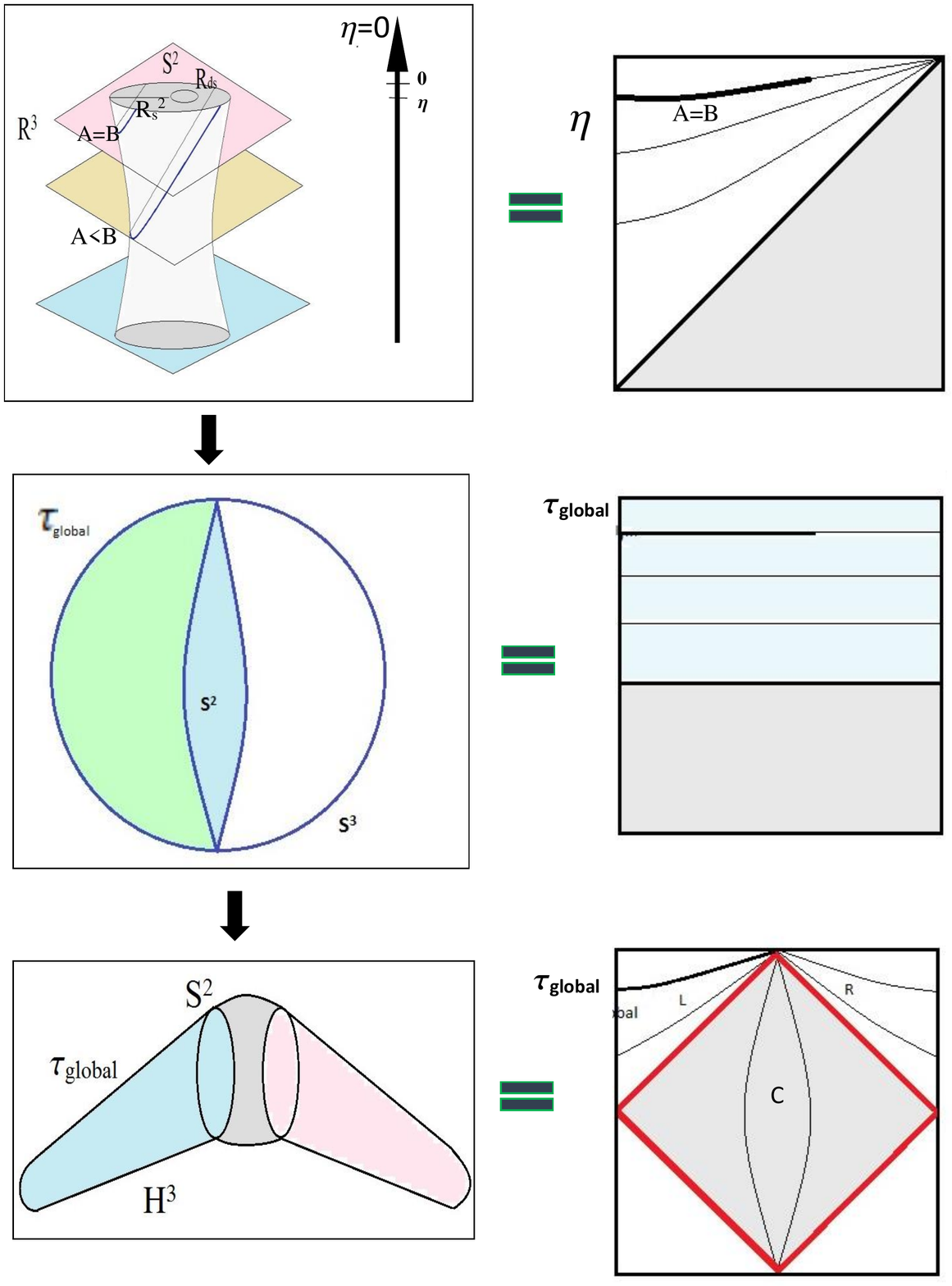}
    	}
    	\caption[Optional caption for list of figures]{Schematic diagram for the geometrical construction of the bipartite system. First we consider ${\bf S^2}$ with radius $R>>R_{\bf dS}=H^{-1}$, which at late global time $\tau_{global}$ can mapped to half of a ${\bf S^3}$
    		with boundary ${\bf S^2}$ at the equator. For simplicity we can also use hyperbolic slices where the interior of the ${\bf S^3}$ mimics the role of the ``left'' (L) slices. Corresponding Penrose
    		diagrams are drawn for completeness \cite{Maldacena:2012xp}.} 
    	\label{fzaca}
    \end{figure*}
   Let us first discuss about the underlying symmetries of the present setup in de Sitter space. For this let us consider, a surface ${\rm \bf S}^2$ defined by the relation,
    $\sum^{3}_{i=1}x^2_{i}=R^2_{{\rm \bf S}^2},$
where $R_{{\rm \bf S}^2}$ is the radius of the spherical surface ${\rm \bf S}^2$ and $x_{i}\forall i=1,2,3$ are the flat coordinates. To consider a spherical surface sufficiently large compared to the size of the horizon one need to consider here $R_{{\rm \bf S}^2}>>\eta$, where $\eta$ is the conformal time defined as, 
$\eta=\int dt/a(t).$
Here $a(t)$ is the scale factor which is appearing in the spatially flat $k=0$ FLRW metric in $3+1$ dimensions, written in the conformally flat form as:
\bea ds^2&=&\left(-dt^2+a^2(t)\sum^{3}_{i=1}dx^2_{i}\right)=a^2(\eta)\left(-d\eta^2+\sum^{3}_{i=1}dx^2_{i}\right).\eea
We note that for $R_{{\rm \bf S}^2}>>\eta$ one can suppress the conformal time coordinate and we achieve this by fixing the value of $R_{{\rm \bf S}^2}$ in the $\eta\rightarrow 0$ limit. This surface in this context is invariant by an ${\rm \bf SO(1,3)}$ subgroup, which is the ${\rm \bf SO(1,4)}$ quasi de Sitter isometry group in left region. Technically this is commonly known as left invariant in which we perform our rest of the computation presented in this paper.
Also for our purpose we use de Sitter global coordinate system in which the equal time slices are three sphere ${\bf S}^3$. In this context the entangling surface is two sphere ${\bf S}^2$, which is the equator of three sphere ${\bf S}^3$. Further using quasi de Sitter isometry group as $\eta=0$ one can consider a map which transform ${\bf S}^2$ to the equator of 
three sphere ${\bf S}^3$ on the boundary of $3+1$ D quasi de Sitter space. However, this gives rise to divergent contributions, which can be regularized by transforming back to the ${\bf S}^2$ to a surface at global time $\tau_{\rm global}$.
  	  
    In the present situation we also use hyperbolic slices which are geometrically characterized by the equation, 
    $\sum^{5}_{j=1}Y^2_{j}= H^{-2},$
    where for Euclidean signature we can write down the following parametric equations:
    \bea
           \label{r1}
  \displaystyle  Y_{j}&=&\displaystyle H^{-1}\times\left\{\begin{array}{ll}
 \displaystyle  \cos\tau_{\rm E}~\sin\rho_{\rm E}~\hat{\bf n}_{j}~~~~~~~~~~~~ &
                                                           \mbox{\small {\textcolor{red}{\bf  for $j=1,2,3$}}}  
                                                          \\ 
          \displaystyle \sin\tau_{\rm E} & \mbox{\small { \textcolor{red}{\bf for $j=4$}}}\\ 
                    \displaystyle \cos\tau_{\rm E}~\cos\rho_{\rm E} & \mbox{\small { \textcolor{red}{\bf for $j=5$}}}.~~~~~~~~
                                                                    \end{array}
                                                          \right.
                                                          \eea 
     Here $\hat{\bf n}_{j}\forall j=1,2,3$
    are the $j$-th component of the unit vector defined in ${\bf R}^{3}$.
    
    In this hyperbolic slices the metric with Euclidean signature can be expressed as:
    \bea ds^2_{\rm E}&=& H^{-2}\left[d\tau^2_{\rm E}+\cos^2\tau_{\rm E}\left(d\rho^2_{\rm E}+\sin^2\rho_{\rm E}~d\Omega^2_{\bf 2}\right)\right],
    \eea
    where $d\Omega^2_{\bf 2}$ is defined in the two sphere ${\bf S}^2$. Further we perform the following analytic continuation in the fifth coordinate:
    \bea Y_{5}&=&H^{-1}\cos\tau_{\rm E}~\cos\rho_{\rm E}\rightarrow X_{0}=iY_{5}=iH^{-1}\cos\tau_{\rm E}~\cos\rho_{\rm E}\eea
    and redefine the rest of the coordinates as, 
    $X_{k}=Y_{k}~{ \forall k=1,2,3,4}.$
    Consequently, in the Lorentzian signature one can write the following geometrical equation:
    \bea \sum^{4}_{l=0}X^2_{l}=\left(-X^2_0+\sum^{4}_{k=1}X^2_k\right)=H^{-2}.\eea
    On the other hand, in the Lorentzian signature one can consider three different region, which are characterized as:
    \bea
               \label{r2}
      \displaystyle \textcolor{red}{\bf R}&:&\displaystyle\left\{\begin{array}{ll}
     \displaystyle \tau_{\rm E}=\frac{\pi}{2}-it_{\bf R}~~~~~~~~~~~~ &
                                                               \mbox{\small {\textcolor{red}{\bf for $t_{\bf R}\geq 0$}}}  
                                                              \\ 
              \displaystyle \rho_{\rm E}=-ir_{\bf R} & \mbox{\small { \textcolor{red}{\bf for $r_{\bf R}\geq 0$}}}.~~~~~~~~
                                                                        \end{array}
                                                              \right.\\
\label{r3a}
      \displaystyle \textcolor{red}{\bf C}&:&\displaystyle\left\{\begin{array}{ll}
     \displaystyle \tau_{\rm E}=t_{\bf C}~~~~~~~~~~~~ &
                                                               \mbox{\small {\textcolor{red}{\bf for $-\frac{\pi}{2}\leq t_{\bf C}\leq \frac{\pi}{2}$}}}  
                                                              \\ 
              \displaystyle \rho_{\rm E}=\frac{\pi}{2}-ir_{\bf C} & \mbox{\small { \textcolor{red}{\bf for $-\infty<r_{\bf c}< \infty$}}}.~~~~~~~~
                                                                        \end{array}
                                                              \right. \\
\label{r4a}
      \displaystyle \textcolor{red}{\bf L}&:&\displaystyle\left\{\begin{array}{ll}
     \displaystyle \tau_{\rm E}=-\frac{\pi}{2}+it_{\bf L}~~~~~~~~~~~~ &
                                                               \mbox{\small { \textcolor{red}{\bf for $t_{\bf L}\geq 0$}}}  
                                                              \\ 
              \displaystyle \rho_{\rm E}=-ir_{\bf L} & \mbox{\small { \textcolor{red}{\bf for $r_{\bf L}\geq 0$}}}.~~~~~~~~
                                                                        \end{array}
                                                              \right.                                         \eea
 Further, in hyperbolic slices the metric with Lorentzian signature can be expressed for the three regions as:                                                             
  \bea
                 \label{r2z}
        \displaystyle \textcolor{red}{\bf R}&:&\displaystyle\left\{\begin{array}{ll}
       \displaystyle ds^2_{\bf R}=H^{-2}\left[-dt^2_{\bf R}+\sinh^2t_{\bf R}\left(dr^2_{\bf R}+\sinh^2r_{\bf R}~d\Omega^2_{\bf 2}\right)\right], 
                                                                          \end{array}
                                                                \right.\\
  \label{r3}
        \displaystyle \textcolor{red}{\bf C}&:&\displaystyle\left\{\begin{array}{ll}
       \displaystyle  ds^2_{\bf C}=H^{-2}\left[dt^2_{\bf C}+\cos^2t_{\bf C}\left(-dr^2_{\bf C}+\cosh^2r_{\bf C}~d\Omega^2_{\bf 2}\right)\right], \end{array}
                                                                \right. \\
  \label{r4}
        \displaystyle \textcolor{red}{\bf L}&:&\displaystyle\left\{\begin{array}{ll}
       \displaystyle  ds^2_{\bf L}=H^{-2}\left[-dt^2_{\bf L}+\sinh^2t_{\bf L}\left(dr^2_{\bf L}+\sinh^2r_{\bf L}~d\Omega^2_{\bf 2}\right)\right].  \end{array}
                                                                \right.                                         \eea 
        In fig.~(\ref{fzaca}) schematic diagram for the geometrical construction and underlying symmetries of the bipartite system.                                                                                                      
    \subsection{Wave function of axion in an open chart}
    \label{ka33b}
    
    In this section our prime objective is to compute the expression for entanglement entropy in
        de Sitter space driven by an axion field. This axion field originates from the R-R sector of {\bf Type IIB} string theory compactified on a Calabi-Yau three fold (${\bf CY^3}$) in presence of ${\bf NS~5}$ brane. For details of the compactification and the resulting effective action, in $3+1$ dimensions, for the axion field, see refs.~\cite{McAllister:2008hb,Silverstein:2008sg,McAllister:2014mpa,Panda:2010uq,Svrcek:2006yi}. Using this effective action we discuss that such stringy axions can be treated as necessary ingredient to study quantum entanglement 
        in the context of primordial cosmology~\footnote{From the view of quantum information theory violation of Bell's inequality is more appropriately connected to non-locality. Violation of Bell’s inequality (CHSH) in bipartite quantum systems necessarily requires quantum entanglement, but the reverse statement is not always true \cite{Bartkiewicz:1306.6504,Verstraete:0112012}. In a bipartite quantum system described by a specific class
        	of state, for instance say for the Werner states with non vanishing entanglement entropy can be easily connected to the violation in Bell's inequality \cite{Horodecki:2009zz}. However, without complete understanding
      of the quantum state being considered in such analysis, it is not possible
        	in general to determine exactly the extent of
        	Bell's inequality violation. }.

        
        
\begin{figure*}[htb]
    \centering
        \includegraphics[width=12.2cm,height=7cm] {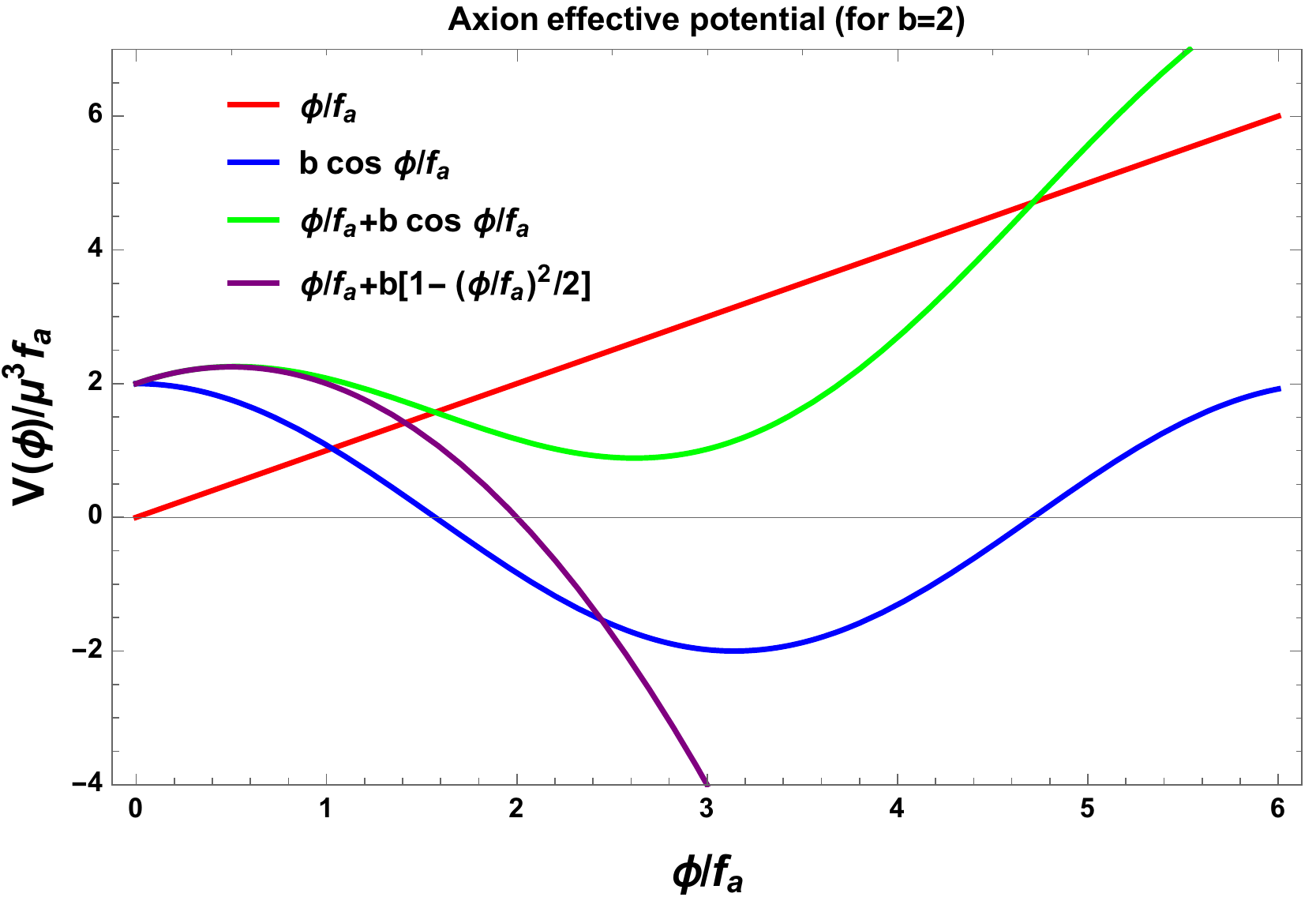}
        \label{fig1}
    \caption[Optional caption for list of figures]{Schematic behaviour of the axion effective potential obtained from String Theory.} 
    \label{aafza}
    \end{figure*}

        Let us start with the aforementioned effective action for axion field:
    \bea\label{ax}  S_{axion}&=& \int d^{4}x \sqrt{-g}\left[-\frac{1}{2}(\partial \phi)^2 +V(\phi)\right],\eea
    where $\phi$ is the axion field and the corresponding effective potential can be expressed as:
    \bea\label{axion} V(\phi)&=&\mu^3\phi+\Lambda^4_{G}\cos\left(\frac{\phi}{f_{a}}\right)=\mu^3\left[\phi+bf_{a}\cos\left(\frac{\phi}{f_{a}}\right)\right],\eea
    where $\mu^3$ is the mass scale, $f_a$ is the axion decay constant and we have defined a new parameter $b$ as, 
    $b= \Lambda^4_{G}/\mu^3 f_{a}$. In this case scale $\Lambda_{G}$ is given by,  $\Lambda_{G}=\sqrt{m_{SUSY} L^3/ \sqrt{\alpha^{'}}g_{s}}~e^{-cS_{inst}},$
               where $S_{inst}$ is the instanton action, $c\sim{\cal O}(1)$ is a constant factor, $m_{SUSY}$ is the supersymmetry breaking scale, $\alpha^{'}$ is the Regge slope parameter, $g_s$ is the string coupling constant and $L^6$ is the volume factor in string units. 
               Using this action we compute the wave function of axion in open chart. In fig.~(\ref{aafza}) we have shown the schematic behaviour of the axion potential. 
   For the computation of the wave function here we consider the following two possibilities:
    \begin{enumerate}
    \item \underline{\textcolor{red}{\bf Case~I:}}\\
    In the first approximation of the computation we restrict ourself up to the linear term of the effective potential:
        \bea\label{axionaa} V(\phi)&\approx&\mu^3\phi.\eea
         Here the linear term of the effective potential can be treated as a source term in the equation of motion where the axion has no mass.
    
    \item \underline{\textcolor{red}{\bf Case~II:}}\\
    On the other hand, in the small field limit $\phi<<f_a$, where analytic solution of the equation of motion is possible, one can approximate the non-perturbative contribution as, $\cos\left(\frac{\phi}{f_{a}}\right)\approx 1-\frac{1}{2}\left(\frac{\phi}{f_{a}}\right)^2$.
        Consequently in small field limit $\phi<<f_a$ one can approximate the total effective potential for axion as:
        \bea\label{axion2} V(\phi)&\approx&\Lambda^4_{C}+\mu^3\phi-\frac{\Lambda^4_{C}}{2}\left(\frac{\phi}{f_{a}}\right)^2=\mu^3\left[bf_{a}+\phi\right]-\frac{m^2_{axion}}{2}\phi^2,\eea
        Here we define the effective mass of the axion as:
                         \bea 
                         m^2_{axion}&=&\frac{\mu^3 b}{f_{a}}=\frac{\Lambda^4_{G}}{f^2_{a}},\eea
                         where the axion decay consatnt in general follow a time dependent profile and for our purpose we use the following choice,
                         $f_a=\sqrt{100-\frac{80}{1+\left(\ln\frac{\eta}{\eta_c}\right)^2}}~H$,
                         which was used in refs.~\cite{Maldacena:2015bha,Choudhury:2016cso,Choudhury:2016pfr} to demonstrate Bell's inequality violation in primordial cosmology. It is important to mention here that, the one point function from axionic fluctuation in primordial cosmology is exactly proportional to the parameter $m_{axion}/f_a$ and such contributions are apeparing in \textcolor{red}{\bf Case II}, which we further use for the computation of entanglement entropy to prove the existance of EPR Bell pairs in cosmology.
        \end{enumerate}

    Further using Eqn~(\ref{ax}) one can write down the following equation of motion for the axion for the above mentioned two cases:
    \bea \underline{\textcolor{red}{\bf For ~Case~I}:}~~~~~~~~~~~~~~~~~~~~~~~~~~~~~~~\Box\phi&=&\mu^3,\\
    \underline{\textcolor{red}{\bf For ~Case~II}:}~~~~~~~~~~~~~~~\left(\Box+m^2_{axion}\right)\phi&=&\mu^3=\frac{m^2_{axion}f_a}{b},\eea
    where $\Box$ is the $3+1$ D D'Alembertian operator which can be expressed in open chart as \cite{Sasaki:1994yt}:
    \bea \Box&=&\left[\frac{1}{a^3(t)}\partial_{t}\left(a^3(t)\partial_{t}\right)-\frac{1}{H^2a^2(t)}\hat{\bf L}^2_{\bf H^3}\right].\eea
   Also the exact time dependence in the scale factor $a(t)$ for an open chart of the de Sitter background is given by:
 \bea
                   \label{r2zzaa}
          \displaystyle a(t)&=&\displaystyle\frac{1}{H}\sinh t.\eea
   Additionally, the Laplacian operator $\hat{\bf L}^2_{\bf H^3}$ in hyperbolic slice ${\bf H^3}$ can be written as \cite{Sasaki:1994yt}:
          \bea \hat{\bf L}^2_{\bf H^3}&=&\frac{1}{\sinh^2r}\left[\partial_{r}\left(\sinh^2r~\partial_{r}\right)+\frac{1}{\sin\theta}\partial_{\theta}\left(\sin\theta~\partial_{\theta}\right)+\frac{1}{\sin^2\theta}\partial^2_{\Phi}\right],\eea     
                              which satisfies the following eigenvalue equation:
          \bea \hat{\bf L}^2_{\bf H^3}{\rm\cal Y}_{plm}(r,\theta,\Phi)&=&-(1+p^2){\rm\cal Y}_{plm}(r,\theta,\Phi).     \eea
          Here ${\cal Y}_{plm}(r,\theta,\Phi)$ are the eigenfunctions of $\hat{\bf L}^2_{\bf H^3}$ which can be written using method of separation of variable here one can decompose ${\cal Y}_{plm}(r,\theta,\Phi)$ into a radial and angular part as:
                  \bea  {\cal Y}_{plm}(r,\theta,\Phi)&=&\frac{\Gamma\left(ip+l+1\right)}{\Gamma\left(ip+1\right)}~\frac{p}{\sqrt{\sinh r}}~{\cal P}^{-\left(l+\frac{1}{2}\right)}_{\left(ip-\frac{1}{2}\right)}\left(\cosh r\right)Y_{lm}(\theta,\Phi),\eea          
          where $Y_{lm}(\theta,\Phi)$ is the spherical harmonic function and ${\cal P}^{-\left(l+\frac{1}{2}\right)}_{\left(ip-\frac{1}{2}\right)}\left(\cosh r\right)$ represents associated Legendre polynomial.
        
          The full solution of the equations of motions for the $\textcolor{red}{\bf  ~Case~I}$ and $\textcolor{red}{\bf  ~Case~II}$ can be expressed in the following forms:
          \bea\label{dfull} \phi(t,r,\theta,\Phi)&=&\sum_{\Lambda}\left[a_{\Lambda}{\cal U}_{\Lambda}(t,r,\theta,\Phi)+a^{\dagger}_{\Lambda}{\cal U}^{*}_{\Lambda}(t,r,\theta,\Phi)\right],\eea  
          where ${\cal U}_{\Lambda}(t,r,\theta,\Phi)$ forms a complete basis of mode function labeled by index $\Lambda$ which is fixed by the three quantum numbers $p$, $l$ and $m$ and an index $\sigma=\pm$ for $\textcolor{red}{\bf R}$ and $\textcolor{red}{\bf L}$ hyperboloid in the present context. 
          
          It is important to note that equation of motions for the $\textcolor{red}{\bf  ~Case~I}$ and $\textcolor{red}{\bf  ~Case~II}$ are inhomogeneous differential equations with constant source and for this the total solutions of these equations can be expressed as:                                                                             \bea {\cal U}_{\Lambda}(t,r,\theta,\Phi)={\cal U}_{\sigma plm}(t,r,\theta,\Phi)&=&\frac{1}{a(t)}\chi_{p,\sigma}(t){\cal Y}_{plm}(r,\theta,\Phi)=\frac{H}{\sinh t}\chi_{p,\sigma}(t){\cal Y}_{plm}(r,\theta,\Phi),~~~~~~~~\eea 
          where the time dependent part of the wave function $\chi_{p,\sigma}(t)$ forms a complete set of positive frequency function which can be written as a sum of complementary part ($\chi^{(c)}_{p,\sigma}(t)$) and particular integral part ($\chi^{(p)}_{p,\sigma}(t)$):
          \bea \chi_{p,\sigma}(t)&=&\chi^{(c)}_{p,\sigma}(t)+\chi^{(p)}_{p,\sigma}(t).\eea
          Here our prime objective is to explicitly compute both of the contributions and use them further in the computation of entanglement entropy.
          
          Further, one can write down the following equation of motion for the complementary part of the time dependent contribution in the total solution as:
              \bea {\bf \cal D}_{t}\chi^{(c)}_{p,\sigma}(t)&=&0,\eea
              where ${\bf \cal D}_{t}$ is the differential operator in the present context, which can be expressed for the above mentioned two cases as:
              \bea
                                 \label{cc}
                        \displaystyle {\bf \cal D}_{t}&=&\footnotesize\displaystyle\left\{\begin{array}{ll}
                       \displaystyle \left[\partial^2_t +3\coth t~ \partial_t+\frac{(1+p^2)}{\sinh^2t}\right]~~~~~~~~~~~~~~~~~ &
                                                                                 \mbox{\small {\textcolor{red}{\bf for Case I}}}  
                                                                                \\ 
                                \displaystyle \left[\partial^2_t +3\coth t~ \partial_t+\frac{(1+p^2)}{\sinh^2t}+\frac{m^2_{axion}}{H^2}\right] & \mbox{\small {\textcolor{red}{\bf for Case II}}}.~~~~~~~~
                                                                                          \end{array}
                                                                                \right.\eea
                                                                                The final solution for the
   complementary time dependent part for \textcolor{red}{\bf Case I}  and \textcolor{red}{\bf Case II} are given by:
 \bea
                                  \label{ccd}
                         \displaystyle \underline{\textcolor{red}{\bf Case~ I:}}\nonumber&&\\ \chi^{(c)}_{p,\sigma}(t)&=&\footnotesize\displaystyle\left\{\begin{array}{ll}
                        \displaystyle \frac{1}{2\sinh\pi p}\left[\frac{\left(e^{\pi p}+\sigma\right)}{\Gamma(2+ip)}{\cal P}^{ip}_{1}(\cosh t_{\bf R})-\frac{\left(e^{-\pi p}+\sigma\right)}{\Gamma(2-ip)}{\cal P}^{-ip}_{1}(\cosh t_{\bf R})\right]~~~~~~~~ &
                                                                                  \mbox{\small {\textcolor{red}{\bf for R}}}  
                                                                                 \\ 
                                 \displaystyle \frac{\sigma}{2\sinh\pi p}\left[\frac{\left(e^{\pi p}+\sigma\right)}{\Gamma(2+ip)}{\cal P}^{ip}_{1}(\cosh t_{\bf L})-\frac{\left(e^{-\pi p}+\sigma\right)}{\Gamma(2-ip)}{\cal P}^{-ip}_{1}(\cosh t_{\bf L})\right] & \mbox{\small {\textcolor{red}{\bf for L}}}.~~~~~
                                                                                           \end{array}                          \right.\\
\label{ccdz}
                         \displaystyle \underline{\textcolor{red}{\bf Case~ II:}}\nonumber&&\\ \chi^{(c)}_{p,\sigma}(t)&=&\footnotesize\displaystyle\left\{\begin{array}{ll}
                        \displaystyle \frac{1}{2\sinh\pi p}\left[\frac{\left(e^{\pi p}-i\sigma~e^{-i\pi\nu}\right)}{\Gamma\left(\nu+\frac{1}{2}+ip\right)}{\cal P}^{ip}_{\left(\nu-\frac{1}{2}\right)}(\cosh t_{\bf R})\displaystyle-\frac{\left(e^{-\pi p}-i\sigma~e^{-i\pi\nu}\right)}{\Gamma\left(\nu+\frac{1}{2}-ip\right)}{\cal P}^{-ip}_{\left(\nu-\frac{1}{2}\right)}(\cosh t_{\bf R})\right]~~ &
                                                                                  \mbox{\small {\textcolor{red}{\bf for R}}}  
                                                                                 \\ 
                                 \displaystyle \frac{\sigma}{2\sinh\pi p}\left[\frac{\left(e^{\pi p}-i\sigma~e^{-i\pi\nu}\right)}{\Gamma\left(\nu+\frac{1}{2}+ip\right)}{\cal P}^{ip}_{\left(\nu-\frac{1}{2}\right)}(\cosh t_{\bf L})\displaystyle-\frac{\left(e^{-\pi p}-i\sigma~e^{-i\pi\nu}\right)}{\Gamma\left(\nu+\frac{1}{2}-ip\right)}{\cal P}^{-ip}_{\left(\nu-\frac{1}{2}\right)}(\cosh t_{\bf L})\right] & \mbox{\small {\textcolor{red}{\bf for L}}},~~
                                                                                           \end{array}                          \right.                                                                                           \eea 
where in \textcolor{red}{\bf Case II}
we introduce a parameter $\nu$, which is defined for axion as:
\bea \nu&=&\sqrt{\frac{9}{4}-\frac{m^2_{axion}}{H^2}}=\sqrt{\frac{9}{4}-\frac{\mu^3 b}{f_a H^2}}=\sqrt{\frac{9}{4}-\frac{\Lambda^4_G}{f^2_a H^2}}.\eea
From the above solutions obtained for \textcolor{red}{\bf Case I}  and \textcolor{red}{\bf Case II} we observe the following characteristic features:
\begin{itemize}
\item In the above solutions $\sigma$ can take two values i.e. $\sigma=\pm 1$. For each values of $\sigma$ one can write down the final results in two regions- $\textcolor{red}{\bf R}$ and $\textcolor{red}{\bf L}$ regions of the hyperboloid (${\bf H^3}$). 
 
\item In the result for \textcolor{red}{\bf Case II}, in the limit $m_{axion}<<H$ the parameter $\nu$ is given by, $\nu=3/2$. For this value of $\nu$, the solutions for  
\textcolor{red}{\bf Case I} and \textcolor{red}{\bf Case II} coincide. For this reason one can treat \textcolor{red}{\bf Case I} as a special situation of \textcolor{red}{\bf Case II}.  
\item Similarly in the limit $m_{axion}=\sqrt{2}H$ we have $\nu=1/2$. This is an important class of solution where axion is conformally coupled. Using this class of solution one can get back the flat space limiting result of the entanglement entropy as the metric for de Sitter space is conformally flat.
\item In the case where $m_{axion}<\sqrt{2}H$ parameter $\nu$ is lying within the range, $1/2<\nu<3/2$ and this is also an important class of solution which is commonly known as the solution in the low mass region.
\item  Further in the limit $m_{axion}=H$ parameter $\nu$ is given by, $\nu=5/2$ and this is a restrictive class of solution where axion mass is exactly equal to the Hubble friction.
\item For $m_{axion}>>H$, we have, $ \nu=i\sqrt{\frac{m^2_{axion}}{H^2}
-\frac{9}{4}}\approx im_{axion}/H$. This is an important class of solution in the present context corresponding to high mass. Note that, for $\sqrt{2}H<m_{axion}<3H/2$ the parameter $\nu$ is lying within the window, $0<\nu<1/2$.
\item For $\textcolor{red}{\bf R}$ and $\textcolor{red}{\bf L}$ hyperboloid regions, the solutions have singularities at, $\nu\equiv\nu_{1}=-\frac{1}{2}- ip$ and 
$\nu\equiv\nu_{2}=-\frac{1}{2} +ip$.
This implies that to get non-singular solution the magnitude of the axion mass must satisfy the constraint, 
$\left|\frac{m_{axion}}{H}\right|\neq \sqrt{(p^2+4)(p^2+1)}$.
\item Additionally it is important to note that, the final solution for the
   complementary time dependent part is symmetric under the exchange of the sign of the quantum number $p$ i.e. $\chi^{(c)}_{p,\sigma}(t)=\chi^{(c)}_{-p,\sigma}(t)$.
\item  In this context the overall normalization constant of the time dependent complementary part of the solution is fixed by the following Klien-Gordon inner product
\cite{Sasaki:1994yt},
$\left(\left(\chi^{(c)}_{p,\sigma}(t),\chi^{(c)}_{p,\sigma^{'}}(t)\right)\right)={\cal N}_{p\sigma}\delta_{\sigma\sigma^{'}}$, 
where ${\cal N}_{p\sigma}$ is the overall normalization constant, which is given by:
\bea
                                 \label{c1c}
                        \displaystyle {\cal N}_{p\sigma}&=&\displaystyle\left\{\begin{array}{ll}
                       \displaystyle \frac{4}{\pi}\frac{\left[\cosh\pi p+\sigma\right]}{|\Gamma\left(2+ip\right)|^2}~~~~~~~~~~~~~~~~~ &
                                                                                 \mbox{\small {\textcolor{red}{\bf for Case I}}}  
                                                                                \\ 
                                \displaystyle \frac{4}{\pi}\frac{\left[\cosh\pi p-\sigma\cos\left(\nu-\frac{1}{2}\right)\right]}{|\Gamma\left(\nu+\frac{1}{2}+ip\right)|^2}~~~~~~~~~~~~~~~~~ & \mbox{\small {\textcolor{red}{\bf for Case II}}}.~~~~~~~~
                                                                                          \end{array}
                                                                                \right.\eea
\end{itemize}                                                                                      Further, one can write down the following equation of motion for the particular integral part of the time dependent contribution in the total solution as:
              \bea {\bf \cal D}_{t}\chi^{(p)}_{p,\sigma}(t)&=&{\cal J},\eea
              where ${\bf \cal D}_{t}$ is the differential operator as given in Eqn~(\ref{cc}) and ${\cal J}=\mu^3$.
              For \textcolor{red}{\bf Case II} depending on the time dependence of the axion effective mass as well as the time dependent axion decay constant, source function can be time dependent or may be constant. But for general consideration we assume that the source function is time dependent. 
              
                                                                                The final solution for the
   time dependent "particular integral" part is given by:
   \bea  \chi^{(p)}_{p,\sigma}(t)&=&\int dt^{'}~G_{\sigma}(t,t^{'})~{\cal J}(t^{'}),    \eea
   where $G_{\sigma}(t,t^{'})$ is the Green's function for axion field, given by:
   \bea\label{gf} G_{\sigma}(t,t^{'})&=&\sinh^2 t\sum^{\infty}_{n=0}\frac{1}{\left(p^2-p^2_{n}\right)}\chi^{(c)}_{p_{n},\sigma}(t)\chi^{(c)}_{p_{n},\sigma}(t^{'}).\eea 
   Here, $\chi^{(c)}_{p_{n},\sigma}(t)$ represent the solution for the complementary part with slight modification due to the replacement of $p$ by $p_{n}~\forall ~n=0,\cdots, \infty$ and is represented by:
 \bea
                                   \label{ccd1}
                          \displaystyle \underline{\textcolor{red}{\bf Case~ I:}}\nonumber&&\\ \chi^{(c)}_{p_{n},\sigma}(t)&=&\footnotesize\displaystyle\left\{\begin{array}{ll}
                         \displaystyle \frac{1}{2\sinh\pi p_{n}}\left[\frac{\left(e^{\pi p_{n}}+\sigma\right)}{\Gamma(2+ip_{n})}{\cal P}^{ip_{n}}_{1}(\cosh t_{\bf R})-\frac{\left(e^{-\pi p_{n}}+\sigma\right)}{\Gamma(2-ip_{n})}{\cal P}^{-ip_{n}}_{1}(\cosh t_{\bf R})\right]~~~~~~~~ &
                                                                                   \mbox{\small {\textcolor{red}{\bf for R}}}  
                                                                                  \\ 
                                  \displaystyle \frac{\sigma}{2\sinh\pi p_{n}}\left[\frac{\left(e^{\pi p_{n}}+\sigma\right)}{\Gamma(2+ip_{n})}{\cal P}^{ip_{n}}_{1}(\cosh t_{\bf L})-\frac{\left(e^{-\pi p_{n}}+\sigma\right)}{\Gamma(2-ip_{n})}{\cal P}^{-ip_{n}}_{1}(\cosh t_{\bf L})\right] & \mbox{\small {\textcolor{red}{\bf for L}}}.~~~~~
                                                                                            \end{array}                          \right.\\
 \label{ccdz1}
                          \displaystyle \underline{\textcolor{red}{\bf Case~ II:}}\nonumber&&\\ \chi^{(c)}_{p_{n},\sigma}(t)&=&\footnotesize\displaystyle\left\{\begin{array}{ll}
                         \displaystyle \frac{1}{2\sinh\pi p_{n}}\left[\frac{\left(e^{\pi p_{n}}-i\sigma~e^{-i\pi\nu}\right)}{\Gamma\left(\nu+\frac{1}{2}+ip_{n}\right)}{\cal P}^{ip_{n}}_{\left(\nu-\frac{1}{2}\right)}(\cosh t_{\bf R})-\frac{\left(e^{-\pi p_{n}}-i\sigma~e^{-i\pi\nu}\right)}{\Gamma\left(\nu+\frac{1}{2}-ip_{n}\right)}{\cal P}^{-ip_{n}}_{\left(\nu-\frac{1}{2}\right)}(\cosh t_{\bf R})\right]~~ &
                                                                                   \mbox{\small {\textcolor{red}{\bf for R}}}  
                                                                                  \\ 
                                  \displaystyle \frac{\sigma}{2\sinh\pi p_{n}}\left[\frac{\left(e^{\pi p_{n}}-i\sigma~e^{-i\pi\nu}\right)}{\Gamma\left(\nu+\frac{1}{2}+ip_{n}\right)}{\cal P}^{ip_{n}}_{\left(\nu-\frac{1}{2}\right)}(\cosh t_{\bf L})-\frac{\left(e^{-\pi p_{n}}-i\sigma~e^{-i\pi\nu}\right)}{\Gamma\left(\nu+\frac{1}{2}-ip_{n}\right)}{\cal P}^{-ip}_{\left(\nu-\frac{1}{2}\right)}(\cosh t_{\bf L})\right] & \mbox{\small {\textcolor{red}{\bf for L}}}.~
                                                                                            \end{array}                          \right. \eea     
Further, using the results obtained for the total solution of the EOM given in Eqn~(\ref{dfull}) we expand the field in terms of creation and annihilation operators in Bunch-Davies vacuum, which is given by:
\bea \label{ass2} \phi(r,t,\theta,\Phi)&=&\int^{\infty}_{0} dp \sum_{\sigma=\pm 1}\sum^{p-1}_{l=0}\sum^{+l}_{m=-l}\left[a_{\sigma plm}{\cal U}_{\sigma plm}(r,t,\theta,\Phi)+a^{\dagger}_{\sigma plm}{\cal U}^{*}_{\sigma plm}(r,t,\theta,\Phi)\right],~~~~~~ \eea
where the Bunch-Davies vacuum state is defined as:
\bea  a_{\sigma p l m}|{\bf BD}\rangle&=&0~~~~~~~~~~ \forall \sigma=(+1,-1);p=0,\cdots,\infty;l=0,\cdots,p-1,m=-l,\cdots,+l.~~~~~~~~~~~ \eea
In this context, the Bunch-Davies mode function ${\cal U}_{\sigma plm}(r,t,\theta,\phi)$ can be expressed as:
\bea \label{ass1} {\cal U}_{\sigma plm}(r,t,\theta,\Phi)&=&\frac{H}{\sinh t}\chi_{p,\sigma}(t){\cal Y}_{plm}(r,\theta,\Phi).\eea 
After substituting Eqn~(\ref{ass1}) in Eqn~(\ref{ass2}) we get the following expression for the wave function:
\bea \phi(r,t,\theta,\Phi)&=&\frac{H}{\sinh t}\int^{\infty}_{0} dp \sum^{p-1}_{l=0}\sum^{+l}_{m=-l}\sum_{\sigma=\pm 1}\left[a_{\sigma plm}\chi_{p,\sigma}(t)+a^{\dagger}_{\sigma plm}\chi^{*}_{p,\sigma}(t)\right]{\cal Y}_{plm}(r,\theta,\Phi).~~~~~~ \eea
Below we use the following sets of shorthand notations to denote various associated Legendre polynomials as appearing for the solutions of the wave functions:
\bea \underline{\textcolor{red}{\bf Complementary~solution}:}~~~~~
                                   \label{ccvvc}
                          \displaystyle {\cal P}^{q}&=&\displaystyle\left\{\begin{array}{ll}
                         \displaystyle {\cal P}^{ip}_{1}(\cosh t_{q})~~~~~~~~~~~~~~~~~ &
                                                                                   \mbox{\small {\textcolor{red}{\bf for Case I}}}  
                                                                                  \\ 
                                  \displaystyle {\cal P}^{ip}_{\left(\nu-\frac{1}{2}\right)}(\cosh t_{q}) & \mbox{\small {\textcolor{red}{\bf for Case II}}}.~~~~~~~~
                                                                                            \end{array}
                                                                                  \right.\\         
 \underline{\textcolor{red}{\bf Particular~solution}:}~~~~~
                                   \label{ccvvc1}
                          \displaystyle {\cal P}^{{q},n}&=&\displaystyle\left\{\begin{array}{ll}
                         \displaystyle {\cal P}^{ip_n}_{1}(\cosh t_{q})~~~~~~~~~~~~~~~~~ &
                                                                                   \mbox{\small {\textcolor{red}{\bf for Case I}}}  
                                                                                  \\ 
                                  \displaystyle {\cal P}^{ip_n}_{\left(\nu-\frac{1}{2}\right)}(\cosh t_{q}) & \mbox{\small {\textcolor{red}{\bf for Case II}}}.~~~~~~~~
                                                                                            \end{array}
                                                                                  \right.
                           \eea                                                                                                                    
 where the index $q={\bf R},{\bf L}$ for the right and left handed hyperbolic region in the open chart.
      
                                                   Thus, the total time dependent part of the solution can be simplified to the following form:
   \bea
                                   \label{cvvc}
                          \displaystyle \boxed{\chi_{p,\sigma}(t)=\sum_{q={\bf R},{\bf L}}\left\{\underbrace{\frac{1}{{\cal N}_{p}}\left[\alpha^{\sigma}_{q}~{\cal P}^{q}+\beta^{\sigma}_{q}~{\cal P}^{q*}\right]}_{\textcolor{red}{\bf Complementary~solution}}+\underbrace{\sum^{\infty}_{n=0}\frac{1}{{\cal N}_{p_n}\left(p^2-p^2_n\right)}\left[\bar{\alpha}^{\sigma}_{q,n}~\bar{\cal P}^{q,n}+\bar{\beta}^{\sigma}_{q,n}~\bar{\cal P}^{*q,n}\right]}_{\textcolor{red}{\bf Particular~solution}}\right\}},~~~~~~~~~\eea 
                          where we use the following redefined symbol:
                                            \bea\bar{\cal P}^{q,n}&=& \sinh^2t~  {\cal P}^{q,n}\int dt^{'}~\chi^{(c)}_{p_n,\sigma,q}(t^{'})~{\cal J}(t^{'}).\eea
             Here  $p_n\forall n=0,\cdots,\infty$, $\sigma=\pm 1$ and $q={\bf R},{\bf L}$.    
                                Also the normalization constant ${\cal N}_{p}$ for the complementary part is defined as, ${\cal N}_{p}=2\sinh \pi p ~\sqrt{{\cal N}_{p\sigma}}$, where ${\cal N}_{p\sigma}$ is defined in Eqn.~(\ref{c1c}). For the particular part if we replace $p$ by $p_n$ then the normalization constant ${\cal N}_{p_n}$ can be defined as, ${\cal N}_{p_n}=2\sinh \pi p_n ~\sqrt{{\cal N}_{p_n\sigma}}$.

 In Eqn~(\ref{cvvc}), expansion coefficient functions for complementary solution ($\alpha^{\sigma}_{q}$,$\beta^{\sigma}_{q}$) for $q={\bf R},{\bf L}$ are defined as:
   \bea 
   \label{d1}\alpha^{\sigma}_{\bf R}&=&\frac{1}{\sigma}\alpha^{\sigma}_{\bf L}= \displaystyle\left\{\begin{array}{ll}
                                                     \displaystyle \frac{\left(\sigma+e^{\pi p}\right)}{\Gamma\left(2+ip\right)}~~~~~~~~~~~~~~~~~ &
                                                                                                               \mbox{\small {\textcolor{red}{\bf for Case I}}}  
                                                                                                              \\ 
                                                              \displaystyle \frac{\left(e^{\pi p}-i\sigma e^{-i\pi\nu}\right)}{\Gamma\left(\nu+\frac{1}{2}+ip\right)}~~~~~~~ & \mbox{\small {\textcolor{red}{\bf for Case II}}}.~~~~~~~~
                                                                                                                        \end{array}
                                                                                                              \right.\\  
 \label{d2}\beta^{\sigma}_{\bf R}&=& \frac{1}{\sigma}\beta^{\sigma}_{\bf L}=\displaystyle\left\{\begin{array}{ll}
                                                     \displaystyle -\frac{\left(\sigma+e^{-\pi p}\right)}{\Gamma\left(2-ip\right)}~~~~~~~~~~~~~~~~~ &
                                                                                                               \mbox{\small {\textcolor{red}{\bf for Case I}}}  
                                                                                                              \\ 
                                                              \displaystyle -\frac{\left(e^{-\pi p}-i\sigma e^{-i\pi\nu}\right)}{\Gamma\left(\nu+\frac{1}{2}-ip\right)}~~~~~~~ & \mbox{\small {\textcolor{red}{\bf for Case II}}}.~~~~~~~~
                                                                                                                        \end{array}
                                                                                                              \right.\eea 
             
                                                                                               Exapnsion coefficients for particular solution ($\bar{\alpha}^{\sigma}_{q}$,$\bar{\beta}^{\sigma}_{q}$) can be obtained by replacing $p$ by $p_n$ in Eqn.~(\ref{d1}) aand Eqn.~(\ref{d2}).                                                                                                                               Before going to the further details, let us first analyze Eqn~(\ref{cvvc}):                                                                                         \begin{itemize}


\item  The full solution is valid for $p\neq p_{n}\forall n=0,\cdots,\infty$, which is the necessary condition to solve the inhomogeneous differential equation using Green's function technique.                                                                                                                                                                                           \item  If the source is absent, then the solution for the complementary part is perfectly consistent with ref.~\cite{Maldacena:2012xp}. 
                               \end{itemize}                                                    
 Further using Eqn~(\ref{cvvc}) we write the following matrix equation in a compact form as:
 \bea\label{xxxc} \boxed{{\bf \chi}^{I}=\frac{1}{{\cal N}_p}{\cal  M}^{I}_{J}{\cal P}^{J}+\sum^{\infty}_{n=0}\frac{1}{{\cal N}_{p,(n)}}\left({\cal  M}_{(n)}\right)^{I}_{J}{\cal P}^{J}_{(n)}} \eea 
 where for the complementary part of the solution we define the following matrices:
 \bea {\cal M}^{I}_{J}&=&\left(\begin{array}{ccc} \alpha^{\sigma}_{q} &~~~ \beta^{\sigma}_{q} \\ \beta^{\sigma^{*}}_{q} &~~~ \alpha^{\sigma^{*}}_{q}  \end{array}\right),~~~~~~
  \chi^{I}=\left(\begin{array}{ccc} \chi_{\sigma}(t) \\ \chi^{*}_{\sigma}(t),
   \end{array}\right),~~~~~~
   {\cal P}^{J}=\left(\begin{array}{ccc} {\cal P}^{q} \\ {\cal P}^{{q^*}},\\
      \end{array}\right).\eea
      Similarly for the partucluar solution we also define following matrices:
     \bea \left({\cal M}_{(n)}\right)^{I}_{J}&=&\left(\begin{array}{ccc} \bar{\alpha}^{\sigma}_{q,n} &~~~ \bar{\beta}^{\sigma}_{q,n} \\ \bar{\beta}^{\sigma^{*}}_{q,n} &~~~ \bar{\alpha}^{\sigma^{*}}_{q,n}  \end{array}\right),~~~~~~
       {\cal P}^{J}_{(n)}=\left(\begin{array}{ccc} {\cal P}^{q,n} \\ {\cal P}^{{q^*},n}\\
          \end{array}\right),\eea                 where $\sigma=\pm 1$, $q={\bf R},{\bf L}$ and $I,J=1,2,3,4 $.
          
           On the other hand the redefined normalization constant for the particular part of the solution ${\cal N}_{p,(n)}$ can be expressed as, 
          ${\cal N}_{p,(n)}=2\sinh \pi p_n ~\sqrt{{\cal N}_{p_n\sigma}} ~\left(p^2-p^2_n\right)$. Further using Eqn~(\ref{xxxc}) the Bunch-Davies mode function can be written as:  
          \bea   \frac{H}{\sinh t}a_{I}\chi^{I}=\frac{H}{\sinh t}a_{I}\left[\frac{1}{{\cal N}_p}{\cal  M}^{I}_{J}{\cal P}^{J}+\sum^{\infty}_{n=0}\frac{1}{{\cal N}_{p,(n)}}\left({\cal  M}_{(n)}\right)^{I}_{J}{\cal P}^{J}_{(n)} \right],~~~~~~
           \eea                                   where $a_{I}=(a_{\sigma},
           a^{\dagger}_{\sigma})$ represents a set of creation and annihilation operator.  
           
           On the other hand we define:
           \bea\label{oq1} b_{J} &=& a^{(c)}_{I}{\cal M}^{I}_{J},~~~ b_{J(n)} = a^{(p)}_{I(n)}\left({{\cal M}_{(n)}}\right)^{I}_{J},
           \eea
           where $a^{(c)}_{I}=(a^{(c)}_{\sigma},
                      a^{(c)\dagger}_{\sigma})$ and $a^{(p)}_{I(n)}=(a^{(p)}_{\sigma,n},
                                            a^{(p)\dagger}_{\sigma,n})$  are 
 the set of creation and annihilation operators which act on the complementary and particular part respectively. Thus, the operator contribution for the total solution is:
 \bea a_{I}&=& \left[a^{(c)}_{I}+\sum^{\infty}_{n=0}a^{(p)}_{I(n)}\right],\eea 
 where by inverting Eqn~(\ref{oq1}) we have expressed:
  \bea\label{oq1a} a^{(c)}_{I} &=& b_{J}\left({\cal M}^{-1}\right)^{I}_{J},~~~ a^{(p)}_{I(n)} = b_{J(n)}\left({\cal M}^{-1}_{(n)}\right)^{I}_{J}.
            \eea
  We define the following inverse matrices:
   \bea \left({\cal M}^{-1}\right)^{I}_{J}&=&\left(\begin{array}{ccc} \gamma_{\sigma q} &~~~ \delta_{\sigma q} \\ \delta^{*}_{\sigma q} &~~~ \gamma^{*}_{\sigma q}  \end{array}\right),
        ~~~~\left({\cal M}^{-1}_{(n)}\right)^{I}_{J}=\left(\begin{array}{ccc} \bar{\gamma}_{\sigma q,n} &~~~ \bar{\delta}_{\sigma q,n} \\ \bar{\delta}^{*}_{\sigma q,n} &~~~ \bar{\gamma}^{*}_{\sigma q,n}  \end{array}\right),\eea
                        where $\sigma=\pm 1$, $q={\bf R},{\bf L}$ and $I,J=1,2,3,4 $.   
     The entries of these inverse matrices are given by:
           \bea 
           \label{w1}\gamma_{j\sigma}= \displaystyle\footnotesize \left\{\begin{array}{ll}
                                                             \displaystyle \frac{\Gamma\left(2+ip\right)~e^{\pi p}}{4\sinh\pi p}\left(\begin{array}{ccc} \frac{1}{e^{\pi p}+1} &~~~ \frac{1}{e^{\pi p}-1} \\ \frac{1}{e^{\pi p}+1} &~~~ -\frac{1}{e^{\pi p}-1}  \end{array}\right)~~~~~~~~~~~~~~~~~ &
                                                                                                                       \mbox{\small {\textcolor{red}{\bf for Case I}}}  
                                                                                                                      \\ 
                                                                      \displaystyle \frac{\Gamma\left(\nu+\frac{1}{2}+ip\right)~e^{\pi p+i\pi\left(\nu+\frac{1}{2}\right)}}{4\sinh\pi p}\left(\begin{array}{ccc} \frac{1}{e^{\pi p+i\pi \left(\nu+\frac{1}{2}\right)}+1} &~~~ \frac{1}{e^{\pi p+i\pi \left(\nu+\frac{1}{2}\right)}-1} \\ \frac{1}{e^{\pi p+i\pi \left(\nu+\frac{1}{2}\right)}+1} &~~~ -\frac{1}{e^{\pi p+i\pi \left(\nu+\frac{1}{2}\right)}-1}  \end{array}\right)~~~~~~~ & \mbox{\small {\textcolor{red}{\bf for Case II}}}.~~
                                                                                                                                \end{array}
                                                                                                                      \right.\\
         \label{w2}\delta^{*}_{j\sigma}=\footnotesize\displaystyle\left\{\begin{array}{ll}
                                                                      \displaystyle \frac{\Gamma\left(2-ip\right)}{4\sinh\pi p}\left(\begin{array}{ccc} \frac{1}{e^{\pi p}+1} &~~~ -\frac{1}{e^{\pi p}-1} \\ \frac{1}{e^{\pi p}+1} &~~~ \frac{1}{e^{\pi p}-1}  \end{array}\right)~~~~~~~~~~~~~~~~~ &
                                                                                                                                \mbox{\small {\textcolor{red}{\bf for Case I}}}  
                                                                                                                               \\ 
                                                                               \displaystyle \frac{\Gamma\left(\nu+\frac{1}{2}-ip\right)~e^{i\pi\left(\nu+\frac{1}{2}\right)}}{4\sinh\pi p}\left(\begin{array}{ccc} \frac{1}{e^{\pi p}+e^{i\pi \left(\nu+\frac{1}{2}\right)}} &~~~ -\frac{1}{e^{\pi p}-e^{i\pi \left(\nu+\frac{1}{2}\right)}} \\  \frac{1}{e^{\pi p}+e^{i\pi \left(\nu+\frac{1}{2}\right)}} &~~~ \frac{1}{e^{\pi p}-e^{\pi p+i\pi \left(\nu+\frac{1}{2}\right)}}  \end{array}\right)~~~~~~~ & \mbox{\small {\textcolor{red}{\bf for Case II}}}.~~~~~~~~
                                                                                                                                         \end{array}
                                                                                                                               \right.                          \eea 
                                                                                                                                             Similarly $\bar{\gamma}_{j\sigma,n}$ and $\bar{\delta}^{*}_{j\sigma,n}$ can be obtained by replacing $p$ by $p_n$ in Eqn.~(\ref{w1}) and Eqn.~(\ref{w2}).
  Further we introduce set of rules which are useful to differentiate the operations of these operators in the complementary and particular solution. These set of rules are:
  \begin{itemize}
  \item \underline{\textcolor{red}{\bf Rule~I:}}\\
  The operator corresponding to the complementary solution is insensitive to the particular solution i.e.
  $a^{(c)}_{I}\left[\sum^{\infty}_{n=0}\frac{1}{{\cal N}_{p,(n)}}\left({\cal  M}_{(n)}\right)^{I}_{J}{\cal P}^{J}_{(n)}\right]= 0.$
  \item  \underline{\textcolor{red}{\bf Rule~II:}}\\
  The operator corresponding to the particular solution is insensitive to the complementary solution i.e.
  $a^{(p)}_{I(n)}\left[\frac{1}{{\cal N}_p}{\cal  M}^{I}_{J}{\cal P}^{J}\right]=0.$
  \end{itemize}
  For simplification we quantify the components of $a_{I} =(a_{\sigma},a^{\dagger}_{\sigma})$ for $\sigma=\pm 1$ as:
  \bea\label{xxq1} a_{\sigma}&=&\sum_{q={\bf R},{\bf L}}\left\{\left[\gamma_{q\sigma}b_{q}+\delta^{*}_{q\sigma}b^{\dagger}_{q}\right]+\sum^{\infty}_{n=0}\left[\bar{\gamma}_{q\sigma,n}\bar{b}_{q,n}+\bar{\delta}^{*}_{q\sigma,n}\bar{b}^{\dagger}_{q,n}\right]\right\},\\ \label{xxq2} a^{\dagger}_{\sigma}&=&\sum_{q={\bf R},{\bf L}}\left\{\left[\gamma^{*}_{q\sigma}b^{\dagger}_{q}+\delta_{q\sigma}b_{q}\right]+\sum^{\infty}_{n=0}\left[\bar{\gamma}^{*}_{q\sigma,n}\bar{b}^{\dagger}_{q,n}+\bar{\delta}_{q\sigma,n}\bar{b}_{q,n}\right]\right\}.\eea   
  Further the Bunch-Davies vacuum state can be expressed in terms of the direct product of ${\bf R}$ and ${\bf L}$ vacua by using Bogoliubov transformation as:
  \bea\label{qq1} |{\bf BD}\rangle &=&e^{\hat{\cal O}}~\left(|{\bf R}\rangle \otimes|{\bf L}\rangle\right).\eea 
  Here we assume that the Hilbert space corresponding to the Bunch-Davies vacuum state ${\cal H}_{\bf BD}$ can be decomposed into two separable portions as,  
  ${\cal H}_{\bf BD}:={\cal H}_{\bf R}\otimes {\cal H}_{\bf L}$, where ${\cal H}_{\bf R}$ and ${\cal H}_{\bf L}$ are the Hilbert space corresponding to ${\bf R}$ and ${\bf L}$ vacuum state.
  In the present context operator $\hat{\cal O}$ is given by:
  \bea \label{qq2} \hat{\cal O}&=&\frac{1}{2}\sum_{i,j={\bf R},{\bf L}}m_{ij}~b^{\dagger}_{i}~b^{\dagger}_{j}+\frac{1}{2}\sum_{i,j={\bf R},{\bf L}}\sum^{\infty}_{n=0}\bar{m}_{ij,n}~\bar{b}^{\dagger}_{i,n}~\bar{b}^{\dagger}_{j,n},\eea
  where the coefficients $m_{ij}$ and $\bar{m}_{ij,n}$ will be determined later. In Eqn~(\ref{qq1}) we will keep only linear terms in the exponential for simplicity.
  
  Note that ${\bf R}$ and ${\bf L}$ vacuum states can be expressed as:  
  \bea   |{\bf R}\rangle&=& |{\bf R}\rangle_{(c)}+|{\bf R}\rangle_{(p)},~~~  |{\bf L}\rangle= |{\bf L}\rangle_{(c)}+|{\bf L}\rangle_{(p)},\eea 
  with $(c)$ and $(p)$ representing the complementary and particular part respectively. Further in principle one can express the particular part as:
      \bea |{\bf R}\rangle_{(p)}&=&\sum^{\infty}_{n=0}|{\bf R}\rangle_{(p),n},~~~
      |{\bf L}\rangle_{(p)}=\sum^{\infty}_{n=0}|{\bf L}\rangle_{(p),n}.\eea
      Also the annihilation operator satisfy:
      \bea b_{q}|{q}\rangle_{(c)}&=& 0~~~~~~~~~~~\forall ~q=({\bf R},{\bf L}),\\
      \bar{b}_{q,n}|{q}\rangle_{(p)}&=& 0~~~~~~~~~~~\forall ~q=({\bf R},{\bf L}),~n=0,\cdots,\infty\eea
      as well as the following commutation relations:
      \bea 
      \left[ b_i,b^{\dagger}_j\right]&=&\delta_{ij},~~~~ \left[ b_i,b_j\right]=0=
      \left[ b^{\dagger}_i,b^{\dagger}_j\right].~~~~~~~~~~~~~~~\\ 
            \left[ \bar{b}_{i,n},\bar{b}^{\dagger}_{j,m}\right]&=&\delta_{ij}{\delta}_{nm},~~~~\left[ \bar{b}_{i,n},\bar{b}_{j,m}\right]= 0=
            \left[ \bar{b}^{\dagger}_{i,m},\bar{b}^{\dagger}_{j,m}\right].~~~~~~~~~~~
            \eea
            Further one can write the annihilation of Bunch-Davies vacuum in terms of the annihilations of ${\bf R}$ and ${\bf L}$ vacua as:
            \bea a_{\sigma}|{\bf BD}\rangle=\sum_{q={\bf R},{\bf L}}\sum^{4}_{s=1}{\cal A}^{(q)}_{s}=0,\eea
            where neglecting contribution from the higher powers of creation operators, ${\cal A}^{(q)}_{s}\forall s=1,2,3,4$ are defined as:
            \bea \sum_{q={\bf R},{\bf L}}{\cal A}^{(q)}_{1}&=&\sum_{q={\bf R},{\bf L}}\gamma_{q\sigma}b_q~e^{\hat{\cal O}}\left(|{\bf R}\rangle\otimes |{\bf L}\rangle\right)
            \approx\sum_{i,j={\bf R},{\bf L}}m_{ij}\gamma_{j\sigma}b^{\dagger}_{i}\left(|{\bf R}\rangle\otimes |{\bf L}\rangle\right),~~~~~~~~~~~~~~~~~~~\\
            \sum_{q={\bf R},{\bf L}}{\cal A}^{(q)}_{2}&=&\sum_{q={\bf R},{\bf L}}\delta^{*}_{q\sigma}b^{\dagger}_q~e^{\hat{\cal O}}\left(|{\bf R}\rangle\otimes |{\bf L}\rangle\right)
                        \approx\sum_{q={\bf R},{\bf L}}\delta^{*}_{q\sigma}b^{\dagger}_q\left(|{\bf R}\rangle\otimes |{\bf L}\rangle\right),~~~~~~~~~~~~~~~~~~~~~~~~~~\\
            \sum_{q={\bf R},{\bf L}}{\cal A}^{(q)}_{3}&=&\sum_{q={\bf R},{\bf L}}\sum^{\infty}_{n=0}\bar{\gamma}_{q\sigma,n}\bar{b}_{q,n}~e^{\hat{\cal O}}\left(|{\bf R}\rangle\otimes |{\bf L}\rangle\right)
                        \approx\sum_{i,j={\bf R},{\bf L}}\sum^{\infty}_{n=0}\bar{m}_{ij,n}\bar{\gamma}_{j\sigma,n}b^{\dagger}_{i,n}\left(|{\bf R}\rangle\otimes |{\bf L}\rangle\right),~~~~~~~~~~~\\
            \sum_{q={\bf R},{\bf L}}{\cal A}^{(q)}_{4}&=&\sum_{q={\bf R},{\bf L}}\sum^{\infty}_{n=0}\bar{\delta}^{*}_{q\sigma,n}\bar{b}^{\dagger}_{q,n}~e^{\hat{\cal O}}\left(|{\bf R}\rangle\otimes |{\bf L}\rangle\right)
                                               \approx\sum_{q={\bf R},{\bf L}}\sum^{\infty}_{n=0}\bar{\delta}^{*}_{q\sigma,n}\bar{b}^{\dagger}_{q,n}\left(|{\bf R}\rangle\otimes |{\bf L}\rangle\right).~~~~~~~~~~~~\eea
   This implies that the following condition holds good:
   \bea \left[\left(m_{ij}\gamma_{j\sigma}+\delta^{*}_{i\sigma}\right)b^{\dagger}_{i}+\sum^{\infty}_{n=0}\left(\bar{m}_{ij,n}\bar{\gamma}_{j\sigma,n}+\bar{\delta}^{*}_{i\sigma,n}\right)\bar{b}^{\dagger}_{i,n}\right]\left(|{\bf R}\rangle\otimes |{\bf L}\rangle\right)&=& 0.\eea
   Thus, the complementary part and particular part vanish independently as the solutions are independent of each other. Consequently, we get the following constraints:
         \bea \label{q1} \left(m_{ij}\gamma_{j\sigma}+\delta^{*}_{i\sigma}\right)&=& 0,\\
         \label{q22} \left(\bar{m}_{ij,n}\bar{\gamma}_{j\sigma,n}+\bar{\delta}^{*}_{i\sigma,n}\right)&=& 0~~~~\forall n.\eea
         Further using Eqn~(\ref{q1}) and Eqn~(\ref{q22}), the mass matrices corresponding to the complementary part and particular part can be expressed as: 
           \bea \label{q1a} m_{ij}&=& -\delta^{*}_{i\sigma}\left(\gamma^{-1}\right)_{\sigma j}\equiv\left(\begin{array}{ccc} m_{\bf RR} &~~~ m_{\bf RL} \\ m_{\bf LR} &~~~ m_{\bf LL}  \end{array}\right),~~ \label{q22a} \bar{m}_{ij,n}=-\bar{\delta}^{*}_{i\sigma,n}\left(\bar{\gamma}^{-1}\right)_{\sigma j,n}\equiv\left(\begin{array}{ccc} \bar{m}_{{\bf RR},n} &~~~ \bar{m}_{{\bf RL},n} \\ \bar{m}_{{\bf LR},n} &~~~ \bar{m}_{{\bf LL},n}  \end{array}\right).~~~~~~~~~~~\eea 
           Next substituting the right hand side of the above mentioned equations explicitly the entries of the mass matrices can be expressed for $i,j={\bf R},{\bf L}$ as:
   \be \label{ak1}
              m_{ij}= \footnotesize\displaystyle\left\{\begin{array}{ll}
                                                                \displaystyle \frac{\Gamma\left(2-ip\right)}{\Gamma\left(2+ip\right)}\frac{2i}{e^{2\pi p}-1}\left(\begin{array}{ccc} 0 &~~~ i\sinh p\pi \\ i\sinh p\pi &~~~ 0  \end{array}\right)
  \displaystyle \approx e^{i\theta}\frac{~e^{-p\pi}}{\sinh\pi p}\left(\begin{array}{ccc} 0 &~~~ i\sinh p\pi \\ i\sinh p\pi &~~~ 0  \end{array}\right)~~~~~~~~~~~~~~~~~  &
                                                                                                                          \mbox{\small {\textcolor{red}{\bf Case I}}}  
                                                                                                                         \\ 
                                                                         \displaystyle -\frac{\Gamma\left(\nu+\frac{1}{2}-ip\right)}{\Gamma\left(\nu+\frac{1}{2}+ip\right)}\frac{2~e^{i\pi\nu}}{e^{2\pi p}+e^{2i\pi\nu}}\left(\begin{array}{ccc} \cos\pi\nu &~~~ i\sinh p\pi \\ i\sinh p\pi &~~~ \cos\pi\nu  \end{array}\right)
\displaystyle\approx e^{i\theta}\frac{\sqrt{2}~e^{-p\pi}}{\sqrt{\cosh 2\pi p+\cosh 2\pi \nu}}\left(\begin{array}{ccc} \cos\pi\nu &~~~ i\sinh p\pi \\ i\sinh p\pi &~~~ \cos\pi\nu  \end{array}\right)                                                                          & \mbox{\small {\textcolor{red}{\bf Case II}}}.~~
                                                                                                                                   \end{array}
                                                                                                                         \right. \ee 
    Entries of the matrix $\bar{m}_{ij,n}$ can sismilarly written by replacing $p$ by $p_n$ in Eqn.~(\ref{ak1}).                                                                                                                                    
                           
                                                                                                 Before further discussion here we point out few important features:
 \begin{itemize}
 \item For the \textcolor{red}{\bf Case~I} we observe that for the complementary and particular part of the solution \bea m_{\bf RR}&=& 0=m_{\bf LL},~~~ \bar{m}_{{\bf RR},n}= 0=\bar{m}_{{\bf LL},n}.\eea But for \textcolor{red}{\bf Case~II} we find that \be m_{\bf RR}= m_{\bf LL}=e^{i\theta}\frac{\sqrt{2}~e^{-p\pi}\cos\pi\nu}{\sqrt{\cosh 2\pi p+\cosh 2\pi \nu}},~~~
 \bar{m}_{{\bf RR},n}= \bar{m}_{{\bf LL},n}=e^{i\theta}\frac{\sqrt{2}~e^{-p_n\pi}\cos\pi\nu}{\sqrt{\cosh 2\pi p_n+\cosh 2\pi \nu}}.\ee
 which is non vanishing for $0<\nu<3/2$ and $\nu>3/2$. A special case appear for $\nu=3/2$, where the result vanishes and one can get back the result of \textcolor{red}{\bf Case~I}. Another important observation is that simultaneously it is not possible to fix $p=0,\nu=3/2$ for complementary solution and $p_n=0,\nu=3/2$ for particular solution, as both of them give divergent contribution in the diagonal component of the mass matrix.  
 
 \item  For the \textcolor{red}{\bf Case~I} we see that for the complementary and particular part of the solution 
 \bea m_{\bf RL}&=& m_{\bf LR}=e^{i\left(\theta+\frac{\pi}{2}\right)}~e^{-p\pi},~~~
 \bar{m}_{{\bf RL},n}= \bar{m}_{{\bf LR},n}=e^{i\left(\theta+\frac{\pi}{2}\right)}~e^{-p_n\pi}.\eea But for \textcolor{red}{\bf Case~II} we find that \be m_{\bf RL}=m_{\bf LR}=\frac{e^{i\left(\theta+\frac{\pi}{2}\right)}\sqrt{2}~e^{-p\pi}\sinh p\pi}{\sqrt{\cosh 2\pi p+\cos 2\pi \nu}},~~~
  \bar{m}_{{\bf RL},n}= \bar{m}_{{\bf LR},n}=\frac{e^{i\left(\theta+\frac{\pi}{2}\right)}\sqrt{2}~e^{-p_n\pi}\sinh p_n\pi}{\sqrt{\cosh 2\pi p_n+\cos 2\pi \nu}}.\ee
  Here also, \textcolor{red}{\bf Case~II} coincides with \textcolor{red}{\bf Case~I} for $\nu=3/2$. Additionally, the non vanishing off diagonal components for both the cases indicate the signature of quantum entanglement, which finally gives rise to a non vanishing entanglement entropy. In what follows, we will explore this possibility in detail.
 
 \item Creation operators of $b$ oscillators can be redefined by absorbing the overall phase contribution $e^{i\theta}$.
                             \begin{figure*}[htb]
                             \centering
                             \subfigure[$\lambda_{\pm}$ vs $p$ plot for fixed $\nu$ for small axion mass limit.]{
                                 \includegraphics[width=7.2cm,height=5cm] {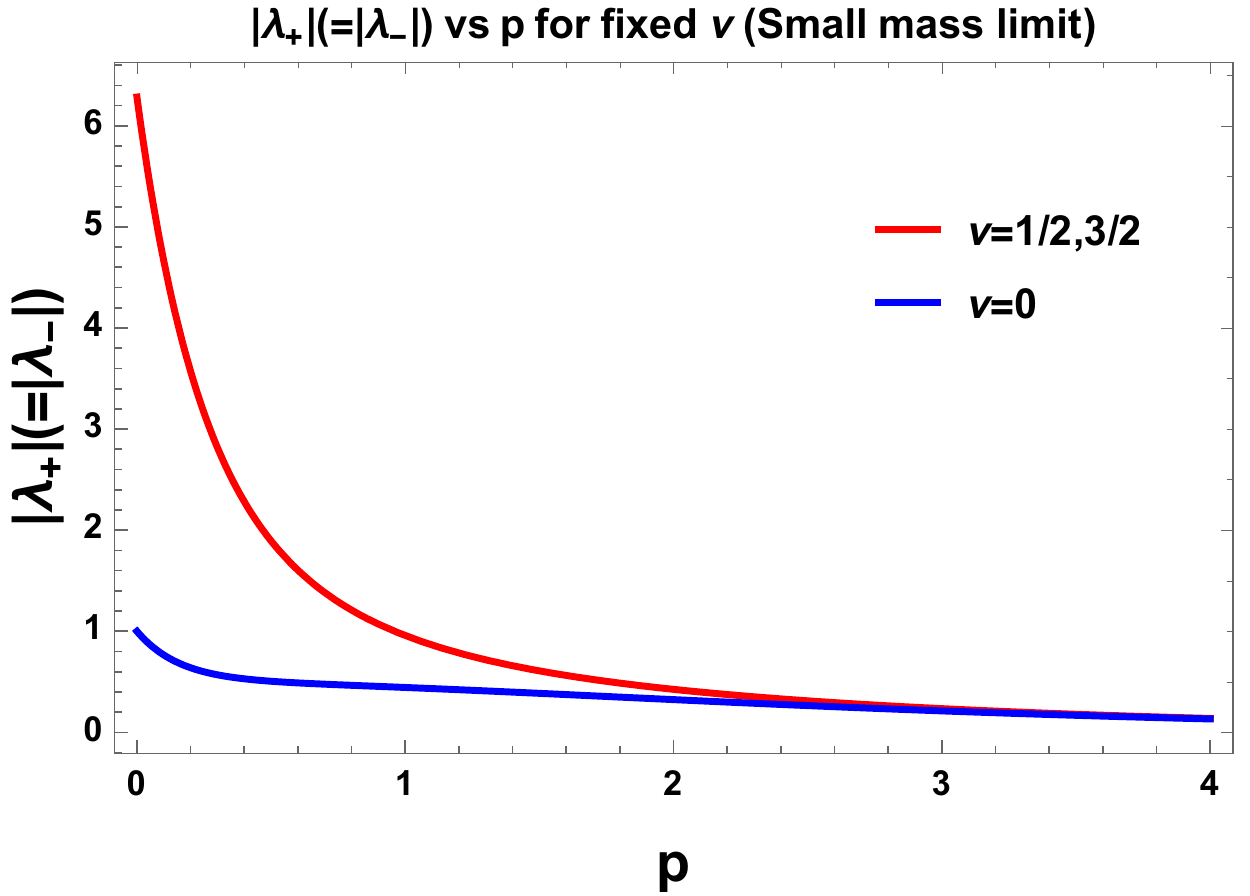}
                                 \label{saa1}
                             }
                             \subfigure[$\lambda_{\pm}$ vs $p$ plot for fixed $\nu$ for large axion mass limit.]{
                                 \includegraphics[width=7.2cm,height=5cm] {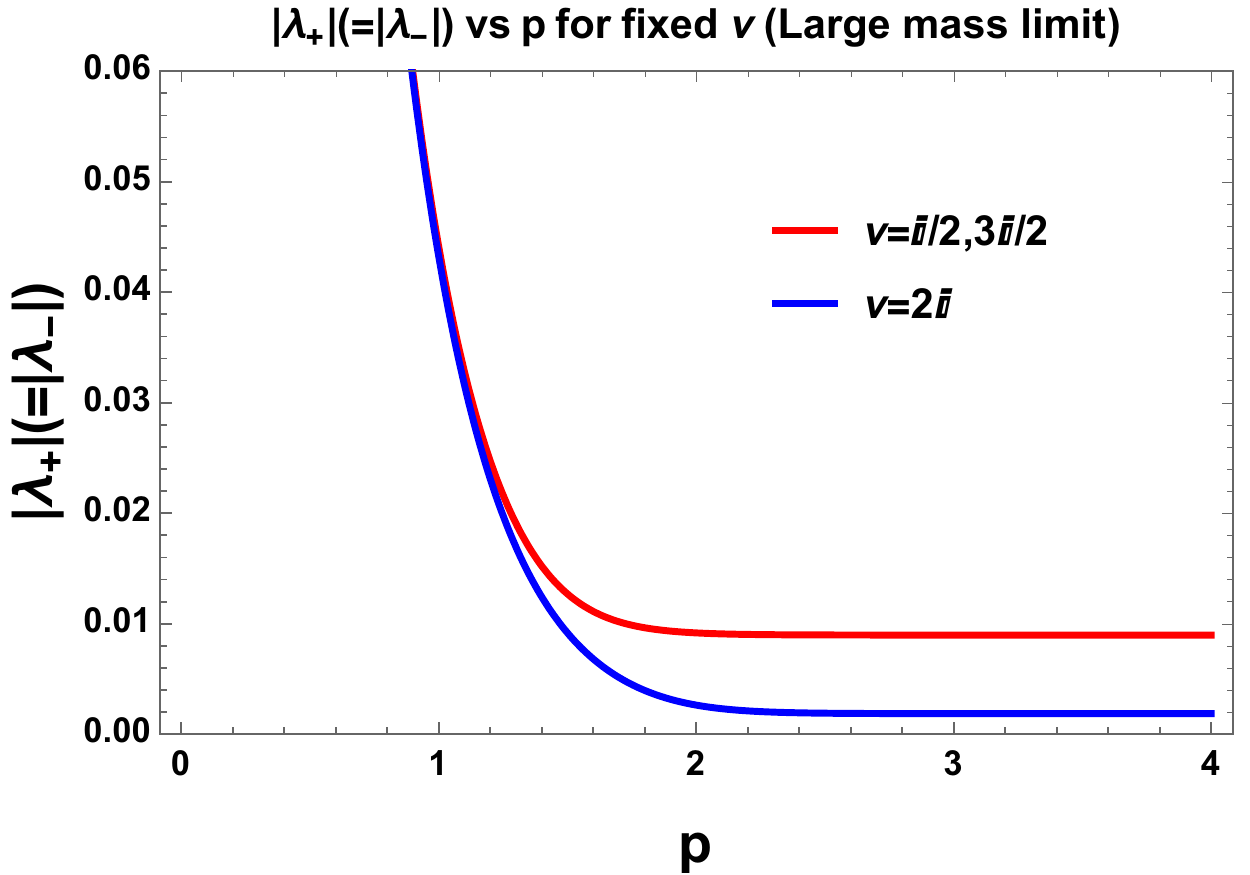}
                                 \label{saa2}
                                 }
                              \subfigure[$\lambda_{\pm}$ vs $|\nu|$ plot for fixed $p$ for large axion mass limit.]{
                                      \includegraphics[width=7.2cm,height=5cm] {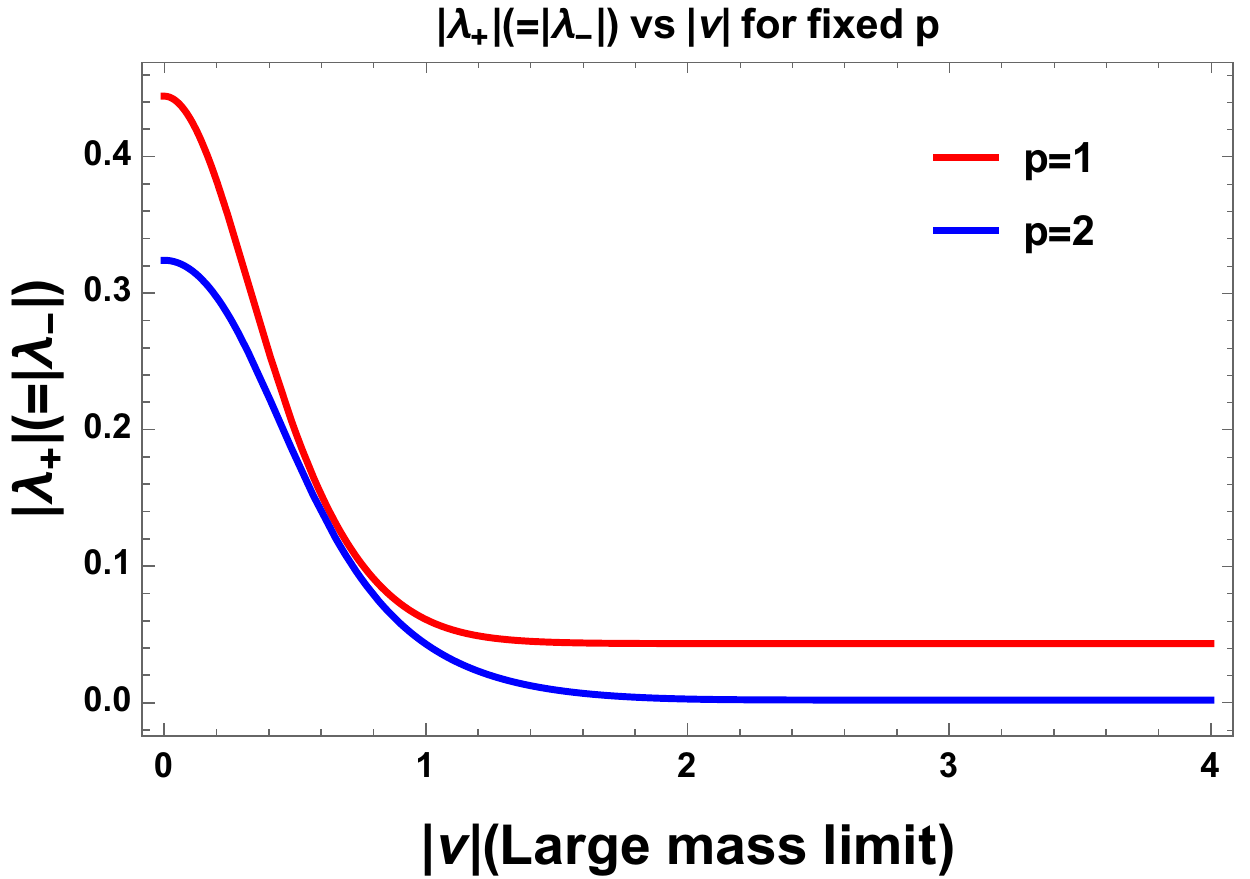}
                                      \label{saa3}   
                             }
                              \subfigure[$\lambda_{\pm}$ vs $|\nu|$ plot for fixed $p$ for small axion mass limit.]{
                                                                   \includegraphics[width=7.2cm,height=5cm] {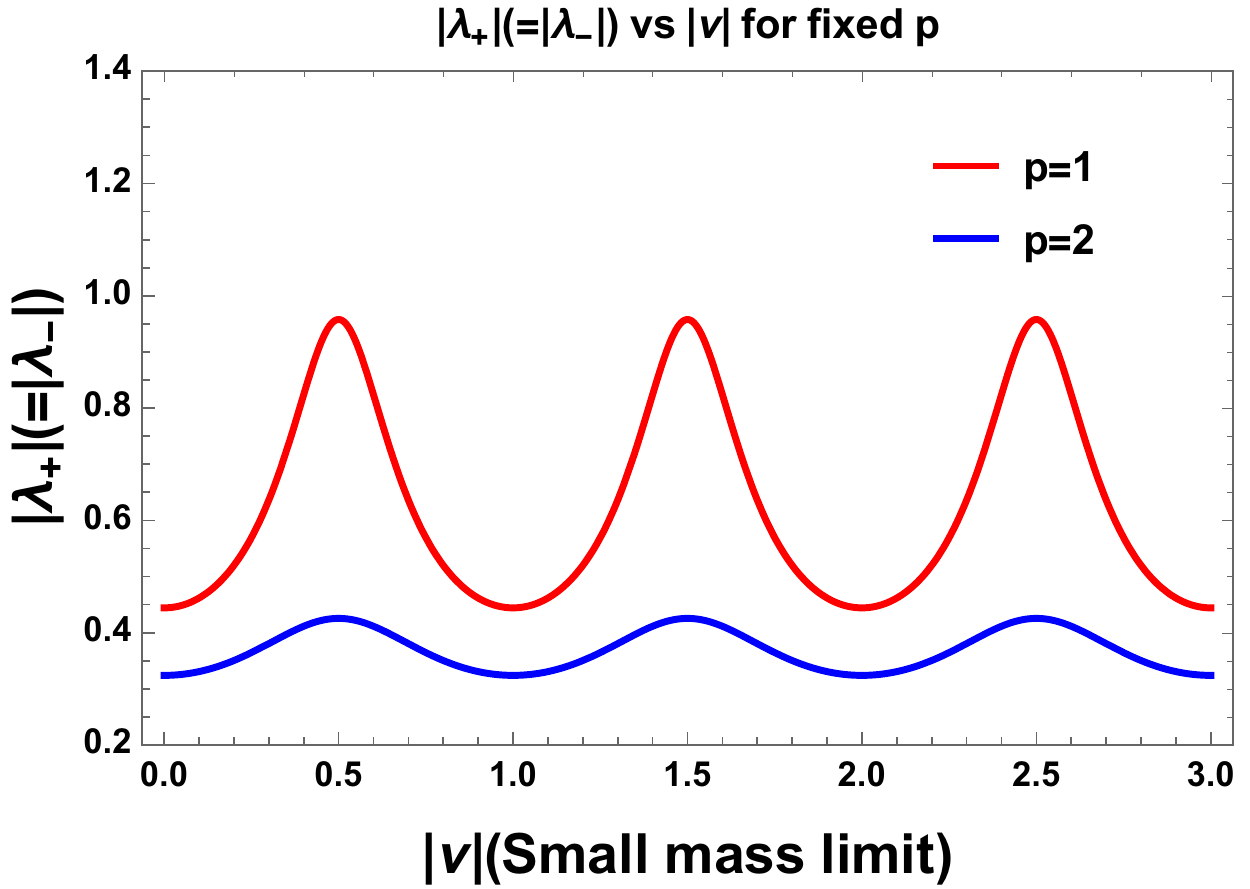}
                                                                   \label{saa4}   
                                                          }
 \subfigure[$\lambda_{\pm}$ vs $\nu^2$ plot for fixed $p$.]{
                                                                    \includegraphics[width=9.2cm,height=5cm] {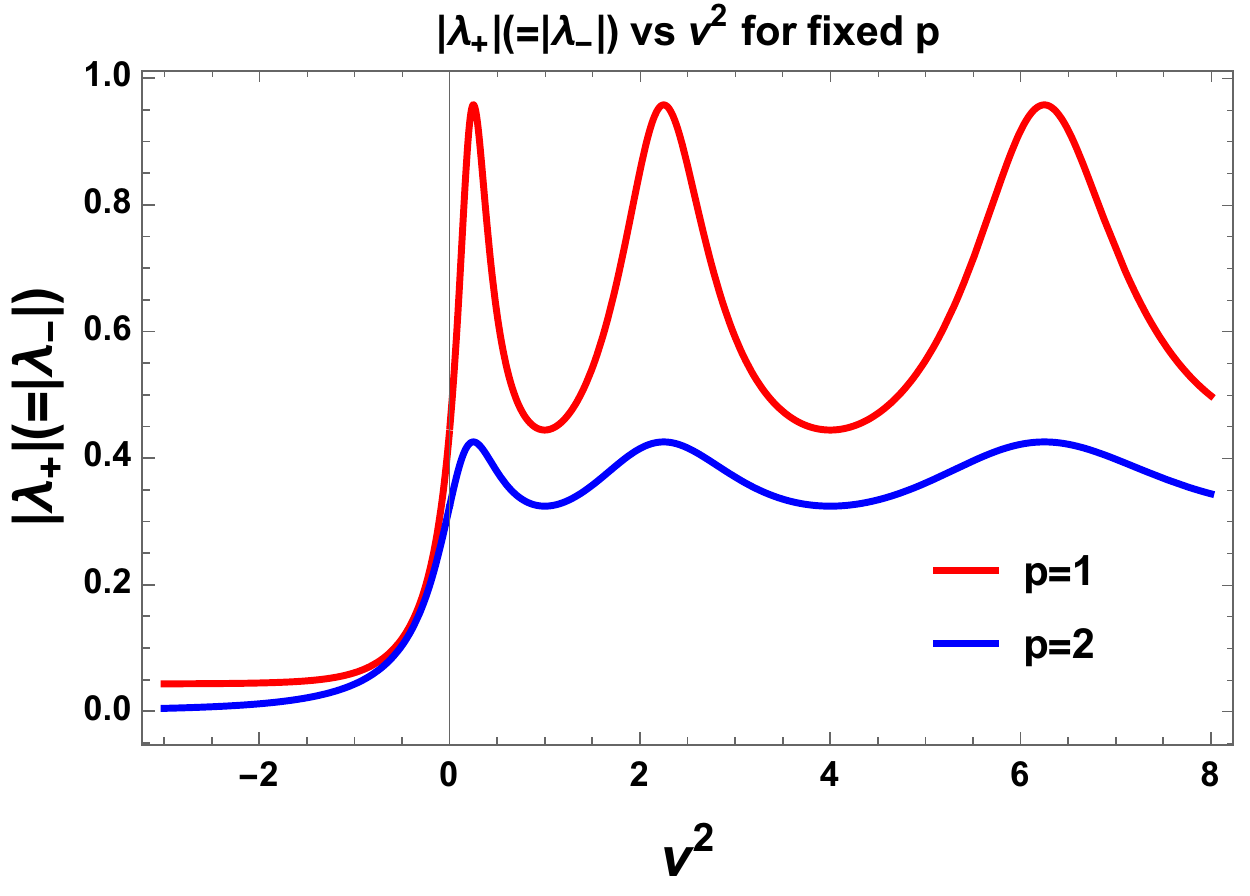}
                                                                    \label{saa5}   
                                                           }                            \caption[]{Behaviour of the eigenvalues  ($\lambda_{\pm}$) in de Sitter space  for $`+'$ branch of solution. This figure clearly shows that we cover both large and small axion mass limiting situations.} 
                             \label{sgg1}
                             \end{figure*}
      
 \item The eigenvalues for the complementary part and particular part of the solution are given by:
 \bea \lambda_{\pm}&=&\frac{1}{2}\left[(m_{\bf LL}+m_{\bf RR})\pm \sqrt{(m_{\bf LL}-m_{\bf RR})^2+4m_{\bf RL}m_{\bf LR}}\right],\\
\lambda_{\pm,n}&=&\frac{1}{2}\left[(\bar{m}_{{\bf LL},n}+\bar{m}_{{\bf RR},n})\pm \sqrt{(\bar{m}_{\bf LL}-\bar{m}_{{\bf RR},n})^2+4\bar{m}_{{\bf RL},n}\bar{m}_{{\bf LR},n}}\right].\eea
Their explicit expressions are:
 \bea \underline{\textcolor{red}{\bf Case~I:}}\nonumber\\
 \lambda_{\pm}&=&\pm m_{\bf RL}=\pm~ e^{i\left(\theta+\frac{\pi}{2}\right)}~e^{-p\pi},\\
 \lambda_{\pm,n}&=&\pm \bar{m}_{{\bf RL},n}=\pm~e^{i\left(\theta+\frac{\pi}{2}\right)}~e^{-p_n\pi},\eea\bea
   \underline{\textcolor{red}{\bf Case~II:}}\nonumber\\
  \lambda_{\pm}&=&m_{\bf RR}\pm m_{\bf RL}=e^{i\theta}\frac{\sqrt{2}~e^{-p\pi}\left(\cos\pi\nu \pm i\sinh p\pi\right)}{\sqrt{\cosh 2\pi p+\cos 2\pi \nu}},\\
  \lambda_{\pm,n}&=& \bar{m}_{{\bf RR},n}\pm \bar{m}_{{\bf RL},n}=e^{i\theta}\frac{\sqrt{2}~e^{-p_n\pi}\left(\cos\pi\nu \pm i\sinh p_n\pi\right)}{\sqrt{\cosh 2\pi p_n+\cos 2\pi \nu}}.\eea
  However, this is not a physical basis as it is extremely complicated to trace over the ${\bf R}$ and ${\bf L}$ contributions when the Bunch-Davies vacuum state is represented using Eqn~(\ref{qq1}) and Eqn~(\ref{qq2}). 
  
  In fig.~(\ref{saa1}) and fig.~(\ref{saa2}), we have shown the behaviour of the eigenvalue  ($\lambda_{\pm}$) vs the momentum $p$ for fixed values of the mass parameter $|\nu|$ in the small and large axion mass limits. Here we cover the mass parameter range $0<|\nu|<3/2$ for both the cases. In fig.~(\ref{saa1}) for the small mass limiting range for large values of the momentum $p$ the two branches of solution for $|\nu|=0$ and $|\nu|=1/2,3/2$ coincides with each other and in the small values of the momentum $p$ the two branches of solution can be separately visualized. Exactly opposite situation appears when we consider the large mass limit in fig.~(\ref{saa2}). Further in fig.~(\ref{saa3}) and fig.~(\ref{saa4}), we have explicitly shown the behaviour of the eigenvalue ($\lambda_{\pm}$) vs mass parameter $|\nu|$ for the fixed values of the the momentum $p$ in the small and large mass limits. For fixed value of $p$ ($p=1$ and $p=2$) in the large mass limit, we initially get decaying behaviour and then after $|\nu|=3/2$ they saturate. On the other hand, for same values of $p$, in the large mass limit, we get oscillating behaviour. Combined effects for large and small mass limits are plotted in fig.~(\ref{saa5}), where we have explicitly shown that for $\nu^2 <0$ (for $p=1$ and $p=2$) we get distingushable behaviour of the mass eigen values. Once $\nu^2\rightarrow 0$ magnitude of the eigevalues increse and finally when $\nu^2>0$, both of the plots show apeariodic oscillations.
 \end{itemize}                                                                                     To find a suitable basis where we can trace over all contributions from ${\bf R}$ and ${\bf L}$ region we need to perform another Bogoliubov transformation incorporating new sets of operators, which are given by:
   \bea 
   \label{k1} c_{\bf R}&=& u~b_{\bf R}+v~b^{\dagger}_{\bf R},~~~c_{\bf L}= \bar{u}~b_{\bf L}+\bar{v}~b^{\dagger}_{\bf L}.\\
      \label{k3} C_{{\bf R},n}&=& U_n~b_{{\bf R},n}+V_n~b^{\dagger}_{{\bf R},n},~~~C_{{\bf L},n}= \bar{U}_n~b_{{\bf L},n}+\bar{V}_n~b^{\dagger}_{{\bf L},n},\eea   
      with the following constraints:
        \bea 
           |u|^2-|v|^2&=& 1,~~~|\bar{u}|^2-|\bar{v}|^2= 1.\\  
                      |U_n|^2-|V_n|^2&=& 1,~~~|\bar{U}_n|^2-|\bar{V}_n|^2= 1.\eea  
                      Using the above set of operators one can write the Bunch-Davies vacuum state in terms of new Bogoliubov transformed basis represented by ${\bf R}^{'}$ and ${\bf L}^{'}$ vacuum state as:
 \bea |{\bf BD}\rangle &=&e^{\hat{\cal O}}\left(|{\bf R}\rangle\otimes |{\bf L}\rangle\right)=\frac{1}{{\cal N}_{p}}e^{\hat{\cal Q}}\left(|{\bf R}^{'}\rangle\otimes |{\bf L}^{'}\rangle\right),\eea  
 where we have introduced a new operator $\hat{\cal Q}$, defined in the basis as:
   \bea \label{qaq2} \hat{\cal Q}&=&\gamma_{p}~c^{\dagger}_{\bf R}~c^{\dagger}_{\bf L}+\sum^{\infty}_{n=0}\Gamma_{p,n}~C^{\dagger}_{{\bf R},n}~C^{\dagger}_{{\bf L},n},\eea
     where $\gamma_{p}$ and $\Gamma_{p,n}$ are defined as the coefficients of the complementary and particular part of the operator, which play significant role to construct the density matrix as well as the entanglement entropy. In this context the exponential of the operator $e^{\hat{\cal Q}}$ is approximated as:
     \bea  e^{\hat{\cal Q}}&=&\sum^{\infty}_{k=0}\frac{1}{k!}\hat{\cal Q}^{k}=\left[{\rm I}+\hat{\cal Q}+\frac{1}{2}\hat{\cal Q}^2+\cdots\right]\approx \left[{\rm I}+\hat{\cal Q}\right].\eea                                                                             The overall normalization factor ${\cal N}_{p}$ is defined as:
     \bea {\cal N}_{p}&=& \left|e^{\hat{\cal Q}}\left(|{\bf R}^{'}\rangle\otimes |{\bf L}^{'}\rangle\right)\right|\approx\left[1-\left(|\gamma_p|^2+\sum^{\infty}_{n=0}|\Gamma_{p,n}|^2\right)\right]^{-1/2}.\eea 
     Here due to the second Bogoliubov transformation the direct product of the ${\bf R}$ and ${\bf L}$ vacuum state is connected with the direct product of the new ${\bf R}^{'}$ and ${\bf L}^{'}$ vacuum state as:
     \bea \left(|{\bf R}\rangle\otimes |{\bf L}\rangle\right)\rightarrow
     \left(|{\bf R}^{'}\rangle\otimes |{\bf L}^{'}\rangle\right)={\cal N}_p~e^{-\hat{\cal Q}}~e^{\hat{\cal O}}\left(|{\bf R}\rangle\otimes |{\bf L}\rangle\right).\eea
     Before going to the further details let us first mention the following useful commutation relations of the creation and annihilation operators of the ${\bf R}^{'}$ and ${\bf L}^{'}$ vacua as given by:
           \bea 
           \left[ c_i,c^{\dagger}_j\right]&=&\delta_{ij},~~~~ \left[ c_i,c_j\right]=0=
           \left[ c^{\dagger}_i,c^{\dagger}_j\right].~~~~~~~~~~~~~~~\\
                 \left[ C_{i,n},C^{\dagger}_{j,m}\right]&=&\delta_{ij}{\delta}_{nm},~~~~\left[ C_{i,n},C_{j,m}\right]= 0=
                 \left[ C^{\dagger}_{i,m},C^{\dagger}_{j,m}\right].~~~~~~~~~~~~~~~
                 \eea
      In this context, the operations of creation and annihilation operators defined on the Bunch-Davies vacuum state are appended bellow:
      \bea 
      \label{g1} c_{\bf R}|{\bf BD}\rangle &=&\gamma_{p}~c^{\dagger}_{\bf L}|{\bf BD}\rangle,~~~
     c_{\bf L}|{\bf BD}\rangle =\gamma_{p}~c^{\dagger}_{\bf R}|{\bf BD}\rangle,\\ 
      \label{g3} C_{{\bf R},n}|{\bf BD}\rangle &=&\Gamma_{p,n}~C^{\dagger}_{{\bf L},n}|{\bf BD}\rangle,~~~
          C_{{\bf L},n}|{\bf BD}\rangle =\Gamma_{p,n}~C^{\dagger}_{{\bf R},n}|{\bf BD}\rangle.\eea

            Further, one can express the annihilation operators after second Bogoliubov transformation in terms of the annihilation operator obtained after first Bogoliubov transformation as: 
            \bea c_{J}&=& b_{I}{\cal G}^{I}_{J},~~~
            C_{J(n)}=\bar{b}_{J(n)}\left({\cal G}_{(n)}\right)^{I}_{J}.\eea
            Here the matrices ${\cal G}^{I}_{J}$ and $\left({\cal G}_{(n)}\right)^{I}_{J}$ are defined as:
            \bea {\cal G}^{I}_{J}&=&\left(\begin{array}{ccc} U_q &~~~ V^{*}_q \\ V_q &~~~ U^{*}_q  \end{array}\right)
                    ,~~~~~~
                    \left({\cal G}_{(n)}\right)^{I}_{J}=\left(\begin{array}{ccc} \bar{U}_{ q,n} &~~~ \bar{V}^{*}_{\sigma q,n} \\ \bar{V}_{ q,n} &~~~ \bar{U}^{*}_{ q,n}  \end{array}\right),\eea
                    where the entries of the matrices are given by:
                    \bea U_q &\equiv& {\rm \bf diag}\left(u,\bar{u}\right),~~V_q \equiv {\rm \bf diag}\left(v,\bar{v}\right),~~ \bar{U}_{q,n} \equiv {\rm \bf diag}\left(U_n,\bar{U}_n\right),~~\bar{V}_{q,n} \equiv {\rm \bf diag}\left(V_n,\bar{V}_n\right).~~\eea
            Further using Eqn~(\ref{k1}) and Eqn~(\ref{k3}), in Eqn~(\ref{g1}) and Eqn~(\ref{g3}), we get the following sets of homogeneous equations for the two cases: 
 \bea \underline{\textcolor{red}{\bf For~Case-I:}}~~~~~~~~~~~~~~~~~~~~~~~~~~~~~~~\nonumber\\ 
 v-\gamma_{p} m_{\bf RL}\bar{v}^{*}&=& 0,~~
 \bar{v}-\gamma_{p} m_{\bf RL}v^{*}= 0,\\ 
 m_{\bf RL}u-\gamma_{p} \bar{u}^{*}&=& 0,~~
 m_{\bf RL}\bar{u}-\gamma_{p} u^{*}= 0,~~\\
  V_n-\Gamma_{p,n} \bar{m}_{{\bf RL},n}\bar{V}^{*}_n&=& 0,~~
   \bar{V}_n-\Gamma_{p,n} \bar{m}_{{\bf RL},n}V^{*}_n= 0,\\ 
   \bar{m}_{{\bf RL},n}U_n-\Gamma_{p,n} \bar{U}^{*}_n&=& 0,~~
   \bar{m}_{{\bf RL},n}\bar{U}_n-\Gamma_{p,n} U^{*}_n= 0, \eea 
  \bea \underline{\textcolor{red}{\bf For~Case-II:}}~~~~~~~~~~~~~~~~~~~~~~~~~~~~~~~\nonumber\\
  m_{\bf RR}u+v-\gamma_{p} m_{\bf RL}\bar{v}^{*}&=& 0,~~
   m_{\bf RR}\bar{u}+\bar{v}-\gamma_{p} m_{\bf RL}v^{*}= 0,\\ 
   m_{\bf RL}u-\gamma_{p} \bar{u}^{*}-\gamma_{p}m_{\bf RR}\bar{v}^{*}&=& 0,~~
   m_{\bf RL}\bar{u}-\gamma_{p} u^{*}-\gamma_{p}m_{\bf RR}v^{*}= 0,\\
    \bar{m}_{{\bf RR},n}U_n+V_n-\Gamma_{p,n} \bar{m}_{{\bf RL},n}\bar{V}^{*}_n&=& 0,~~
       \bar{m}_{{\bf RR},n}\bar{U}_n+\bar{V}_n-\Gamma_{p,n} \bar{m}_{{\bf RL},n}V^{*}_n= 0,\\ 
       \bar{m}_{{\bf RL},n}U_n-\Gamma_{p,n} \bar{U}^{*}_n-\Gamma_{p,n} \bar{m}_{{\bf RR},n}\bar{V}^{*}_n&=& 0,~~
       \bar{m}_{{\bf RL},n}\bar{U}_n-\Gamma_{p,n}  U^{*}_n-\Gamma_{p,n} \bar{m}_{{\bf RR},n}V^{*}_n= 0,~~~~~~~~~~\eea
       From these equations it is important to note that:
       \begin{enumerate}
       \item For the complementary part in the \textcolor{red}{\bf Case~I} considering the diagonal and off diagonal terms we get~\footnote{Henceforth we drop the additional phase factor $e^{i\theta}$ as already mentioned earlier.}:
       \bea m_{\bf RR}&=& m_{\bf LL}=m^{*}_{\bf RR}=0,\\
       m_{\bf RL}&=& m_{\bf LR}=-m^{*}_{\bf RL}=e^{i\frac{\pi}{2}}~e^{-p\pi},\eea
       Similarly for \textcolor{red}{\bf Case~II} for the diagonal and off diagonal terms we get:
        \bea m_{\bf RR}&=& m_{\bf LL}=m^{*}_{\bf RR}=\frac{\sqrt{2}~e^{-p\pi}\cos\pi\nu}{\sqrt{\cosh 2\pi p+\cos 2\pi \nu}},\\
       m_{\bf RL}&=& m_{\bf LR}=-m^{*}_{\bf RL}=e^{i\frac{\pi}{2}}\frac{\sqrt{2}~e^{-p\pi}\sinh p\pi}{\sqrt{\cosh 2\pi p+\cos 2\pi \nu}},\eea
       
       \item For the particular solution part in the \textcolor{red}{\bf Case~I} considering the diagonal and off diagonal terms we get:
               \bea \bar{m}_{{\bf RR},n}&=& \bar{m}_{{\bf LL},n}=\bar{m}^{*}_{{\bf RR},n}=0,\\
                     \bar{m}_{{\bf RL},n}&=& \bar{m}_{{\bf LR},n}=-\bar{m}^{*}_{{\bf RL},n}=e^{i\frac{\pi}{2}}~e^{-p_n\pi},\eea
                     Similarly for \textcolor{red}{\bf Case~II} for the diagonal and off diagonal terms we get:
                      \bea \bar{m}_{{\bf RR},n}&=& \bar{m}_{{\bf LL},n}=\bar{m}^{*}_{{\bf RR},n}=\frac{\sqrt{2}~e^{-p_n\pi}\cos\pi\nu}{\sqrt{\cosh 2\pi p_n+\cos 2\pi \nu}},\\
                     \bar{m}_{{\bf RL},n}&=& \bar{m}_{{\bf LR},n}=-\bar{m}^{*}_{{\bf RL},n}=e^{i\frac{\pi}{2}}\frac{\sqrt{2}~e^{-p_n\pi}\sinh p_n\pi}{\sqrt{\cosh 2\pi p_n+\cos 2\pi \nu}}.\eea
 \item  In the \textcolor{red}{\bf Case~II}, if we consider $\gamma_p$ and $\Gamma_{p,n}$ are purely imaginary i.e. 
   $\gamma^{*}_{p}=-\gamma_{p}$,~~~
   $\Gamma^{*}_{p,n}=-\Gamma_{p,n}$,
   and if we set,
  $v^{*}=\bar{v},~~
   u^{*}=\bar{u},~~ V^{*}_n=\bar{V}_n,~~ 
      U^{*}_n=\bar{U}_n$.   
      Consequently four sets of equation reduces to two sets of homogeneous equations for the two cases the normalization $|u|^2-|v|^2=1$ and $|U_n|^2-|V_n|^2=1$ are explicitly imposed for the complementary and particular part of the solution.         
       \end{enumerate}  
         Finally, the non trivial solutions obtained from these system of equations for the two cases can be expressed as: 
         \bea \underline{\textcolor{red}{\bf For~Case-I:}}~~~~~~~~~~~~~~~~~~~~~~~~~~~~~~~~~~~~~ 
         \gamma_{p}&=&m_{\bf RL},~\frac{1}{m_{\bf RL}}.\\
         \Gamma_{p,n}&=&\bar{m}_{{\bf RL},n},~\frac{1}{\bar{m}_{{\bf RL},n}}. \eea
                             \begin{figure*}[htb]
                             \centering
                             \subfigure[$|\gamma_{p}|$ vs $p$ plot for fixed $\nu$ for small mass limit.]{
                                 \includegraphics[width=7.2cm,height=4.7cm] {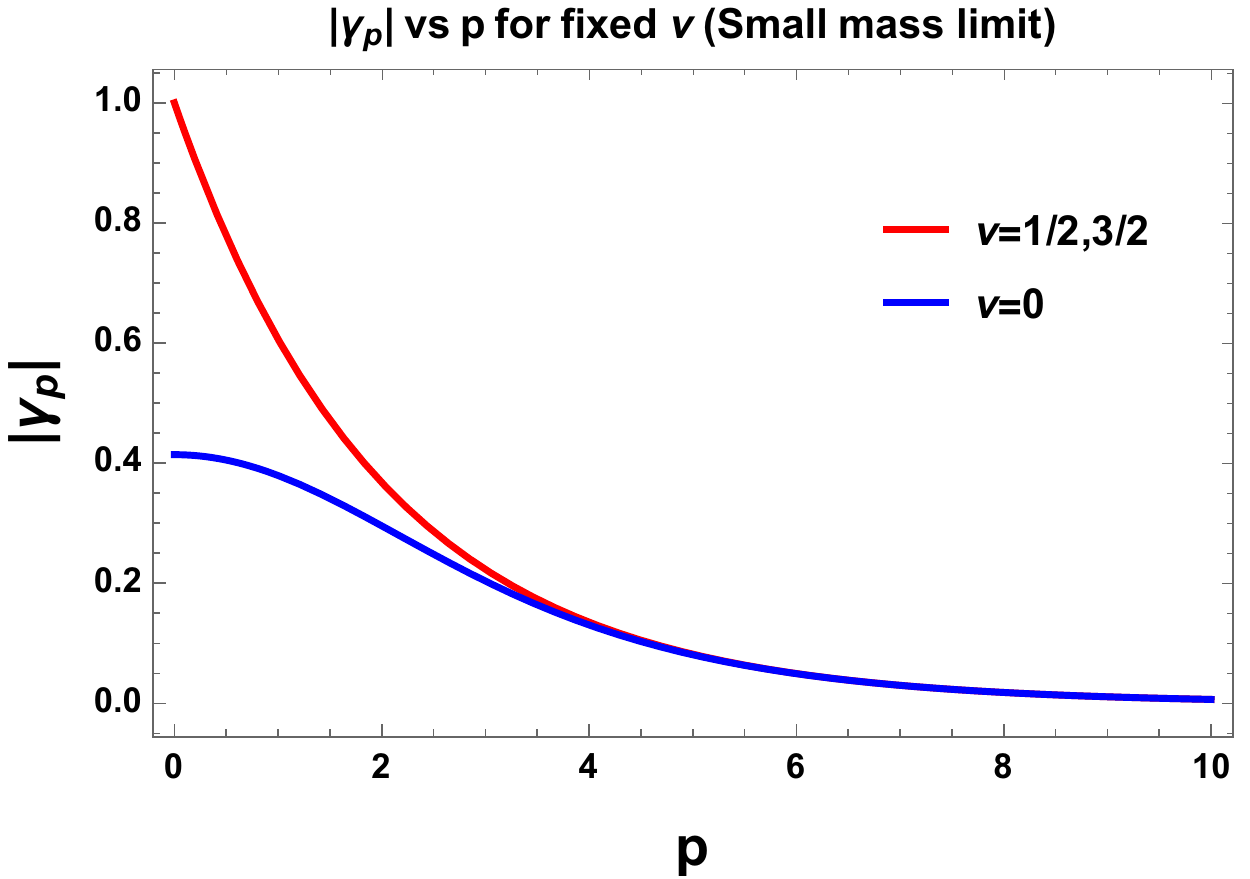}
                                 \label{gaa1}
                             }
                             \subfigure[$|\gamma_{p}|$ vs $p$ plot for fixed $\nu$ for small mass limit.]{
                                 \includegraphics[width=7.2cm,height=4.7cm] {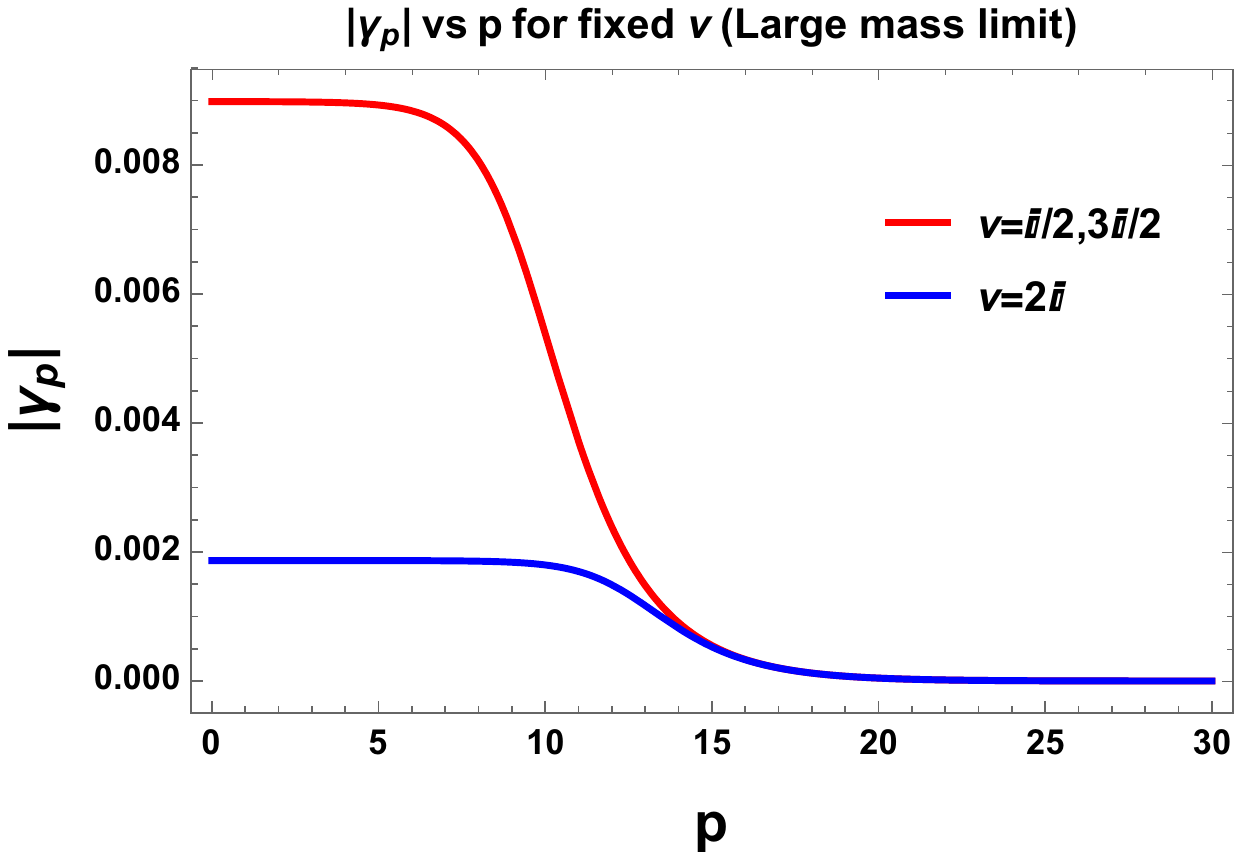}
                                 \label{gaa2}
                                 }
                              \subfigure[$|\gamma_{p}|$ vs $|\nu|$ plot for fixed $p$ for large mass limit.]{
                                      \includegraphics[width=7.2cm,height=4.7cm] {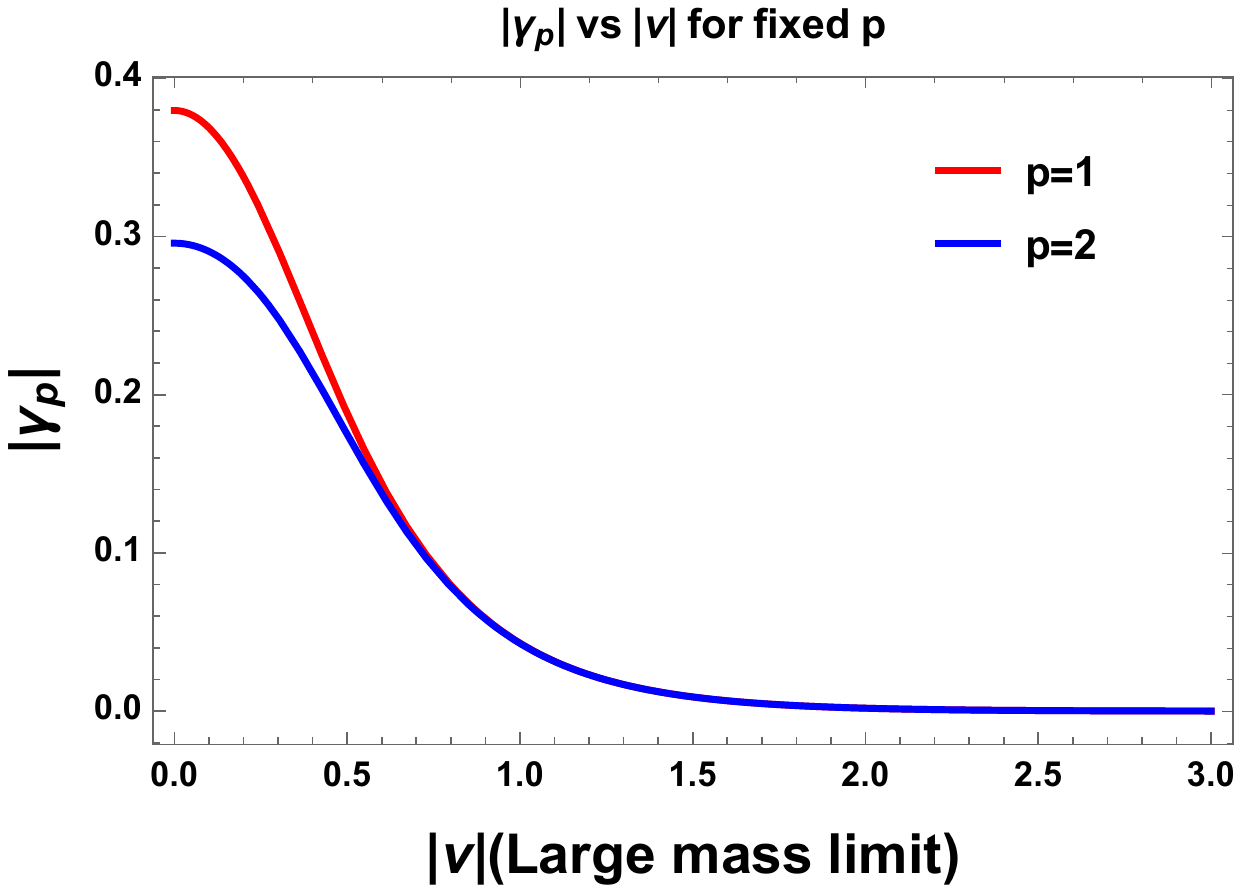}
                                      \label{gaa3}   
                             }
                              \subfigure[$|\gamma_{p}|$ vs $|\nu|$ plot for fixed $p$ for small mass limit.]{
                                                                   \includegraphics[width=7.2cm,height=4.7cm] {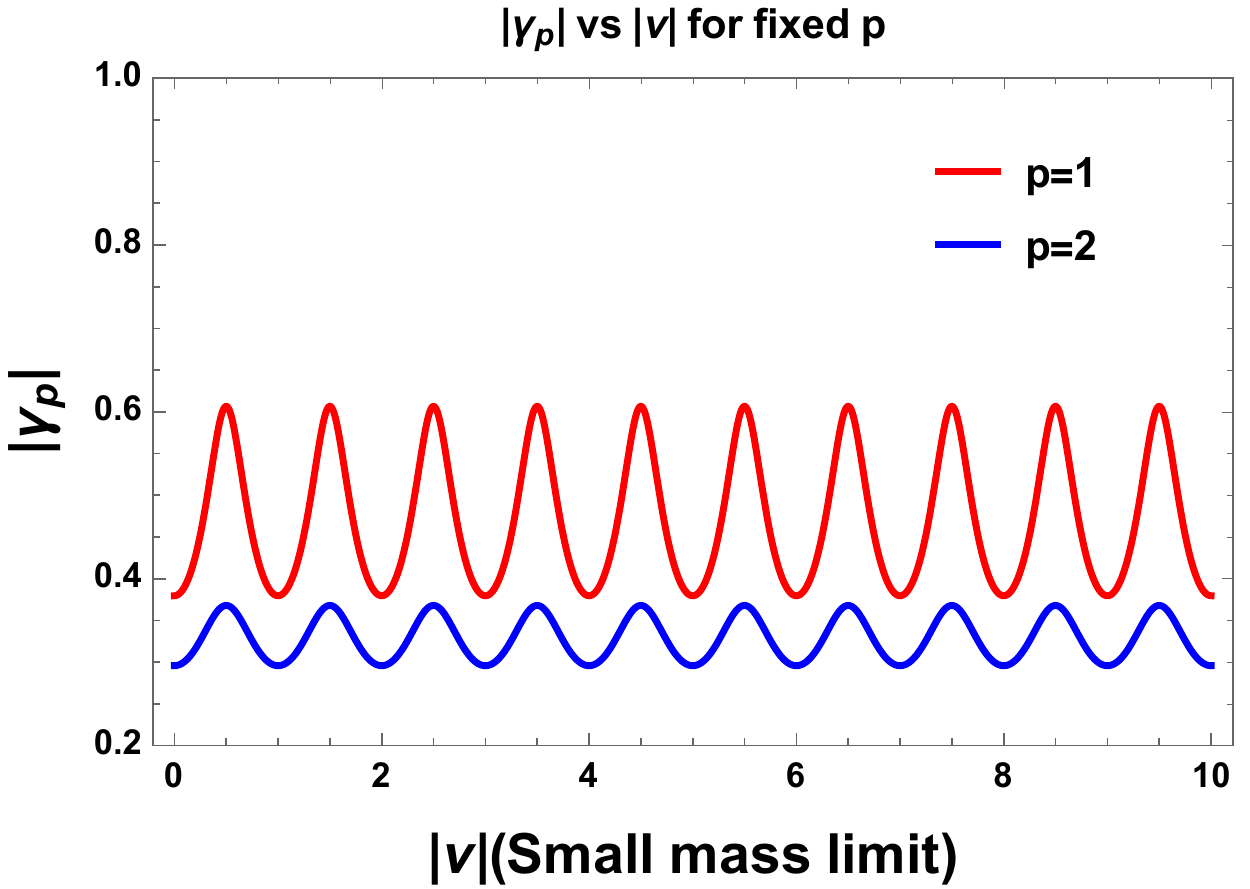}
                                                                   \label{gaa4}   
 }
                               \subfigure[$|\gamma_{p}|$ vs $\nu^2$ plot for fixed $p$.]{
                                                                    \includegraphics[width=9.2cm,height=4.7cm] {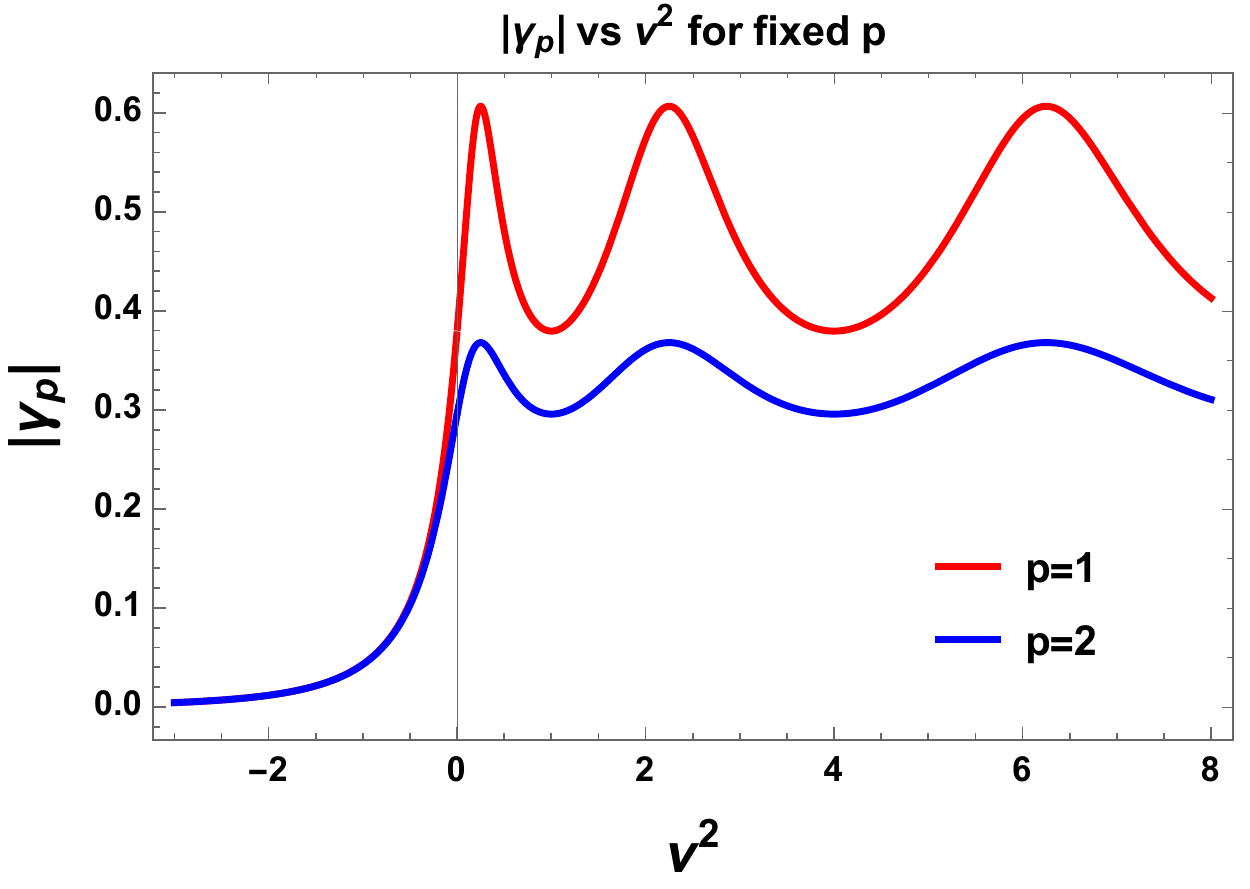}
                                                                    \label{gaa5}   
                                                           }                                                         
                                                          
                             \caption[Optional caption for list of figures]{Behaviour of $|\gamma_{p}|$ in de Sitter space  for $`+'$ branch of solution. This figure clearly shows that we cover both large and small mass limiting situations.} 
                             \label{ggg1}
                             \end{figure*}
     
        \bea
        \underline{\textcolor{red}{\bf For~Case-II:}}~~~
         \gamma_{p}&=&\frac{1}{2m_{\bf RL}}\left[\left(1+m^2_{\bf RL}-m^2_{\bf RR}\right) 
         \pm \sqrt{\left(1+m^2_{\bf RL}-m^2_{\bf RR}\right)^2-4m^2_{\bf RL}}\right].~~~~~~~~~~~\\
          \Gamma_{p,n}&=&\frac{1}{2\bar{m}_{{\bf RL},n}}\left[\left(1+\bar{m}^2_{{\bf RL},n}-\bar{m}^2_{{\bf RR},n}\right)
          \pm \sqrt{\left(1+\bar{m}^2_{{\bf RL},n}-\bar{m}^2_{{\bf RR},n}\right)^2-4\bar{m}^2_{{\bf RL},n}}\right].~~~~~~~~~~\eea          Further substituting the entries of the mass matrices one can further simplify the result as:   
          \bea \underline{\textcolor{red}{\bf For~Case-I:}}~~~~~~~~~~~~~~~~~~ 
                  \gamma_{p}&=& \pm i ~e^{\mp p\pi}.\\
                  \Gamma_{p,n}&=&\pm i ~e^{\mp p_n\pi}. \\
                  \underline{\textcolor{red}{\bf For~Case-II:}}~~~~~~~~~~~~~~~ 
                   \gamma_{p}&=&i\frac{\sqrt{2}}{\sqrt{\cosh 2\pi p +\cos 2\pi \nu}\pm\sqrt{\cosh 2\pi p +\cos 2\pi \nu+2}}.~~~~~~~~~~\\
                    \Gamma_{p,n}&=&i\frac{\sqrt{2}}{\sqrt{\cosh 2\pi p_n +\cos 2\pi \nu}\pm \sqrt{\cosh 2\pi p_n +\cos 2\pi \nu+2}}.~~~~~~~~~~\eea   
  In fig.~(\ref{gaa1}) and fig.~(\ref{gaa2}), we have explicitly shown the behaviour of $|\gamma_{p}|$ vs the momentum $p$ for the fixed values of the mass parameter $|\nu|$ in the small and large mass limits. Here we cover the mass parameter range $0<|\nu|<3/2$ for both of the cases. In fig.~(\ref{gaa1}) and fig.~(\ref{gaa2}), for the small mass limiting range for large values of the momentum $p$ the two branches of solution for $|\nu|=0$ and $|\nu|=1/2,3/2$ coincide with each other and in the small values of the momentum $p$ the two branches of solution can be separately visualized. Further in fig.~(\ref{gaa3}) and fig.~(\ref{gaa4}), we have shown the behaviour of $|\gamma_{p}|$ vs mass parameter $|\nu|$ for the fixed values of the the momentum $p$ in the small and large mass limits. For fixed value of $p$ at $p=1$ and $p=2$ in the large mass limiting situation, we initially get decaying behaviour and then after $|\nu|=3/2$ both of the results obtained for $p=1$ and $p=2$ saturates. On the other hand, for same values of $p$ in the large mass limit, we get oscillating behaviour. Combined effects for large and small mass limits are plotted in fig.~(\ref{gaa5}), where we have explicitly shown that for $\nu^2 <0$ for $p=1$ and $p=2$ we get distinguishable behaviour of the mass eigen values. Once $\nu^2\rightarrow 0$ magnitude of the eigevalues increase and finally when $\nu^2>0$ both of the plots show apeariodic behaviour.                                                              

    \subsection{Construction of density matrix}
    \label{ka33c}
         In this subsection our prime objective is construct the density matrix using the Bunch-Davies vacuum state which is expressed in terms of newly obtained sets of annihilation and creation operators in the Bogoliubov transformed frame. Most importantly, we already know that the full Bunch-Davies vaccum state can be expressed as a product of the vacuum state for each oscillator in the Bogoliubov transformed frame. Here each oscillators are labelled by the quantum numbers $p, l$ and $m$. After tracing over the right part of the Hilbert space we get the following expression for the density matrix for the left part of the Hilbert space as:
         \bea \label{ff1}(\rho_{\bf L})_{p,l,m}&=&{\bf \rm Tr}_{\bf R}|{\bf BD}\rangle \langle {\bf BD}|,\eea
         where the Bunch-Davies vacuum state takes the form:
         \bea \label{ddq1} |{\bf BD}\rangle &\approx&\left[1-\left(|\gamma_p|^2+\sum^{\infty}_{n=0}|\Gamma_{p,n}|^2\right)\right]^{1/2}\exp\left[\gamma_{p}~c^{\dagger}_{\bf R}~c^{\dagger}_{\bf L}+\sum^{\infty}_{n=0}\Gamma_{p,n}~C^{\dagger}_{{\bf R},n}~C^{\dagger}_{{\bf L},n}\right]\left(|{\bf R}^{'}\rangle\otimes |{\bf L}^{'}\rangle\right).~~~~~~~~~~\eea
         The above two eqns lead to the density matrix for the left part of the Hilbert space as:
      \bea \label{ff1x}(\rho_{\bf L})_{p,l,m}&=&\left(1-|\gamma_{p}|^2\right)\sum^{\infty}_{k=0}|\gamma_{p}|^{2k}|k;p,l,m\rangle\langle k;p,l,m|+f^{2}_{p}\sum^{\infty}_{n=0}\sum^{\infty}_{r=0}|\Gamma_{p,n}|^{2r}|n,r;p,l,m\rangle\langle n,r;p,l,m|,~~~~~~~~~\eea
      where $\gamma_{p}$ and $\Gamma_{p,n}$ are defined in the earlier section and have defined the source normalization factor ${\it f}_{p}$ as:
      \bea {\it f}_{p}&=&\left(\sum^{\infty}_{n=0}\frac{1}{1-|\Gamma_{p,n}|^2}\right)^{-1}.\eea 
      In Eqn.~(\ref{ff1x}), the states $|k;p,l,m\rangle$ and $|n,r;p,l,m\rangle$ are defined in terms of the quantum state in left Hilbert space as:
      \bea |k;p,l,m\rangle&=& \frac{1}{\sqrt{n!}}(c^{\dagger}_{\bf L})^{k}|{\bf L}^{'}\rangle,~~~
       |n,r;p,l,m\rangle= \frac{1}{\sqrt{r!}}(C^{\dagger}_{{\bf L},n})^{r}|{\bf L}^{'}\rangle.\eea
       Here we note the following crucial points:
       \begin{enumerate}
       \item The density matrix is diagonal for a given set of the ${\bf SO(1,3)}$ quantum numbers $p,l,m$. This leads to the total density matrix to take the form as:
       \bea \label{ff1xs}\rho_{\bf L}&=&\left(1-|\gamma_{p}|^2\right){\bf diag}\left(1,|\gamma_{p}|^2,|\gamma_{p}|^{4},|\gamma_{p}|^{6}\cdots\right) +f^{2}_{p}\sum^{\infty}_{n=0}{\bf diag}\left(1,|\Gamma_{p,n}|^2,|\Gamma_{p,n}|^{4},|\Gamma_{p,n}|^{6}\cdots\right),~~~~~~~~~\eea
       \item 
       To find out a suitable normalization of the total density matrix, we use the following two results:
              \bea \sum^{\infty}_{k=0}|\gamma_{p}|^{2k}&=&\lim_{k\rightarrow \infty}\frac{1-|\gamma_{p}|^{2k}}{1-|\gamma_{p}|^2}~~~~~\underrightarrow{|\gamma_{p}|<1}~~~~~ \frac{1}{1-|\gamma_{p}|^2},\\
              \sum^{\infty}_{n=0}\sum^{\infty}_{r=0}|\Gamma_{p,n}|^{2r}&=& \sum^{\infty}_{n=0}\lim_{r\rightarrow \infty}\frac{1-|\Gamma_{p,n}|^{2r}}{1-|\Gamma_{p,n}|^2}~~~~~\underrightarrow{|\Gamma_{p,n}|<1 \forall n}~~~~~ \sum^{\infty}_{n=0}\frac{1}{1-|\Gamma_{p,n}|^2}=f^{-1}_{p}.~~~~~~~~~~~~\eea
              Consequently using these results we get: 
              \bea {\bf Tr}\left[\left(1-|\gamma_{p}|^2\right){\bf diag}\left(1,|\gamma_{p}|^2,|\gamma_{p}|^{4},|\gamma_{p}|^{6}\cdots\right)\right]&=&\left(1-|\gamma_{p}|^2\right)\sum^{\infty}_{k=0}|\gamma_{p}|^{2k}=1,~~~~~~~\\  {\bf Tr}\left[f^2_p\sum^{\infty}_{n=0}{\bf diag}\left(1,|\Gamma_{p,n}|^2,|\Gamma_{p,n}|^{4},|\Gamma_{p,n}|^{6}\cdots\right)\right]&=&f^2_p\sum^{\infty}_{n=0}\sum^{\infty}_{r=0}|\Gamma_{p,n}|^{2r}=f_p,~~~~~\eea
              This implies that the normalization condition of this total density matrix is fixed by the expression, 
                     ${\bf Tr}\rho_{\bf L}
                     = 1+f_p$.
                    This is consistent with the ref.~\cite{Maldacena:2012xp} where
                     $f_p=0$. But to maintain always ${\bf Tr}\rho_{\bf L}= 1$ the total density matrix can be redefined by changing the normalization constant as:
                     \be\begin{array}{lll}\footnotesize\label{ff1xss}(\rho_{\bf L})_{p,l,m}=\underbrace{\frac{\left(1-|\gamma_{p}|^2\right)}{1+f_p}\sum^{\infty}_{k=0}|\gamma_{p}|^{2k}|k;p,l,m\rangle\langle k;p,l,m|}_{\textcolor{red}{\bf Complementary~part}}+\underbrace{\frac{f^{2}_{p}}{1+f_p}\sum^{\infty}_{n=0}\sum^{\infty}_{r=0}|\Gamma_{p,n}|^{2r}|n,r;p,l,m\rangle\langle n,r;p,l,m|}_{\textcolor{red}{\bf Particular~part}},\end{array}\ee
                   One may choose some other appropriate convention for normalization factors such that it maintains always ${\bf Tr}\rho_{\bf L}= 1$ even the presence of source~\footnote{ For an example here one can use the following equivalent ansatz for density matrix:
                                                             \be \label{ff1sx}(\rho_{\bf L})_{p,l,m}=\frac{1}{\left[\frac{1}{1-|\gamma_{p}|^2}+f_p\right]}\underbrace{\sum^{\infty}_{k=0}|\gamma_{p}|^{2k}|k;p,l,m\rangle\langle k;p,l,m|}_{\textcolor{red}{\bf Complementary~part}}+\frac{f^2_p}{\left[\frac{1}{1-|\gamma_{p}|^2}+f_p\right]}\underbrace{\sum^{\infty}_{n=0}\sum^{\infty}_{r=0}|\Gamma_{p,n}|^{2r}|n,r;p,l,m\rangle\langle n,r;p,l,m|}_{\textcolor{red}{\bf Particular~part}}.~~~~~~\ee
}.
                   
                     \item For each set of values of the ${\bf SO(1,3)}$ quantum numbers $p,l,m$, the density matrix yields $(\rho_{\bf L})_{p,l,m}$ and so that the total density matrix can be expressed as a product of all such possible contributions, 
                     \be \rho_{\bf L}= \prod^{\infty}_{p=0} \prod^{p-1}_{l=0} \prod^{+l}_{m=-l}(\rho_{\bf L})_{p,l,m}.\ee
                     This also indicates that in such a situation entanglement is absent among all states which carries non identical ${\bf SO(1,3)}$ quantum numbers $p,l,m$.
                     
                     \item Finally, the total density matrix can be written in terms of entanglement modular Hamiltonian of the axionic Bell pair as,  
                     $\rho_{\bf L}=e^{-\beta {\cal H}_{\bf ENT}},$
                     where at finite temperature $T_{\bf dS}$ of de Sitter space the parameter $\beta$ is defined as, $\beta=2\pi/T_{\bf dS}.$
                     This also implies that at finite temperature the modular Hamiltonian can be expressed as~\footnote{If we use the equivalent ansatz for density matrix in presence of axionic source, in that situation the modular Hamiltonian can be expressed as: \be {\cal H}_{\bf ENT}=-\frac{1}{\beta}\ln\left(\prod^{\infty}_{p=0} \prod^{p-1}_{l=0} \prod^{+l}_{m=-l}\left\{a_p\left[\sum^{\infty}_{k=0}|\gamma_{p}|^{2k}|k;p,l,m\rangle\langle k;p,l,m|
                     	+f^2_p\sum^{\infty}_{n=0}\sum^{\infty}_{r=0}|\Gamma_{p,n}|^{2r}|n,r;p,l,m\rangle\langle n,r;p,l,m|\right]\right\}\right).~~~~\ee}:
                     \bea {\cal H}_{\bf ENT}&=&-\frac{1}{\beta}\ln\left(\prod^{\infty}_{p=0} \prod^{p-1}_{l=0} \prod^{+l}_{m=-l}\left\{\frac{\left(1-|\gamma_{p}|^2\right)}{1+f_p}\sum^{\infty}_{k=0}|\gamma_{p}|^{2k}|k;p,l,m\rangle\langle k;p,l,m|\nonumber \right.\right.\\&&\left.\left.~~~~~~~~~~~~+\frac{f^{2}_{p}}{1+f_p}\sum^{\infty}_{n=0}\sum^{\infty}_{r=0}|\Gamma_{p,n}|^{2r}|n,r;p,l,m\rangle\langle n,r;p,l,m|\right\}\right).~~~~\eea
                       If we assume that the dynamical Hamiltonian in de Sitter space is represented by entangled Hamiltonian then for a given principal quantum number $p$ the Hamiltonian for axionic Bell pairs can be expressed as: 
                     \bea {\cal H}_{p}&=& \left[E_{p}c^{\dagger}_p c_{p}+\sum^{\infty}_{n=0}{\cal E}_{p,n}C^{\dagger}_{p,n}C_{p,n}\right].\eea
                     Acting this Hamiltonian on the Bunch-Davies vacuum state we find:
                     \bea {\cal H}_{p}|{\bf BD}\rangle &\approx&\left[1-\left(|\gamma_p|^2+\sum^{\infty}_{n=0}|\Gamma_{p,n}|^2\right)\right]^{1/2}\left[E_{p}c^{\dagger}_p c_{p}+\sum^{\infty}_{n=0}{\cal E}_{p,n}C^{\dagger}_{p,n}C_{p,n}\right]\nonumber\times\\
                     &&\exp\left[\gamma_{p}~c^{\dagger}_{\bf R}~c^{\dagger}_{\bf L}+\sum^{\infty}_{m=0}\Gamma_{p,m}~C^{\dagger}_{{\bf R},m}~C^{\dagger}_{{\bf L},m}\right]\left(|{\bf R}^{'}\rangle\otimes |{\bf L}^{'}\rangle\right)=E_{{\bf T},p}|{\bf BD}\rangle,~~~~~~~~ \eea
                     where 
                      the total energy spectrum of this sytem can be written as:
                                            \bea E_{{\bf T},p}&=& E_p+\sum^{\infty}_{n=0}{\cal E}_{p,n}, \eea
                     where the energy spectrum corresponding to the complementary and particular part of the wave function are given by:
                     \bea E_p &=& -\frac{1}{2\pi}\ln (|\gamma_p|^2),~~~~
                      {\cal E}_{p,n} = -\frac{1}{2\pi}\ln \left(B_{p,n}\right).\eea
                      One can also recast the total energy spectrum of this sytem as:
                       \bea E_{{\bf T},p}&=&-\frac{1}{2\pi}\ln\left(|\gamma_p|^2 O_p\right), \eea
                      where $O_{p}$ is defined as:
                      \bea O_{p} &=&\prod^{\infty}_{n=0}B_{p,n}~~ {\rm with}~~
                      B_{p,n}=\frac{1}{1-|\Gamma_{p,n}|^2}.\eea
                      Here in absence of the source term $B_{p,n}=1\forall n$ and consequently $\ln O_p =0$.
                                             \begin{figure*}[htb]
                                             \centering
                                             \subfigure[$E_{{\bf T},p}$ vs $p$ plot for fixed $\nu$ without source.]{
                                                 \includegraphics[width=7.2cm,height=5.2cm] {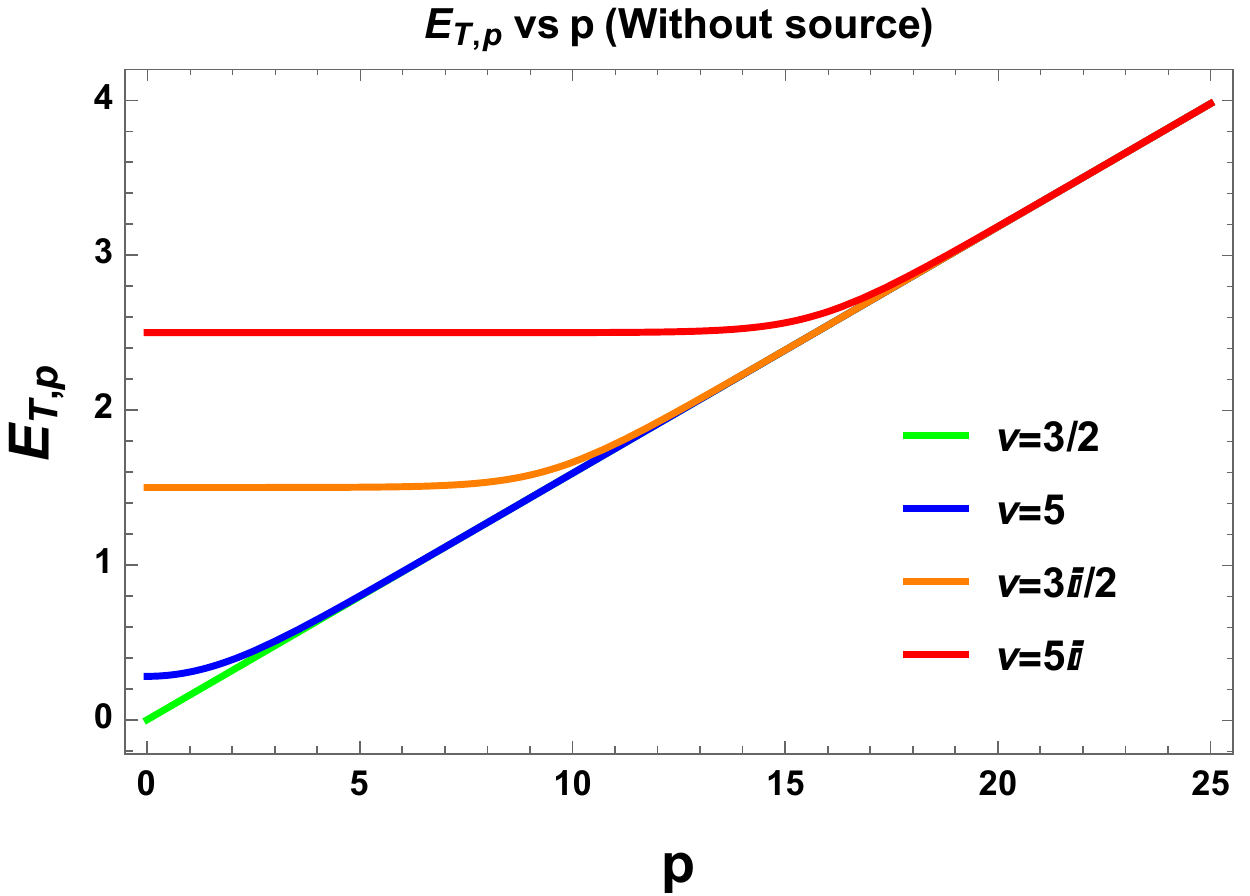}
                                                 \label{daa1}
                                             }
                                             \subfigure[$E_{{\bf T},p}$ vs $p$ plot for fixed $\nu$ with source.]{
                                                 \includegraphics[width=7.2cm,height=5.2cm] {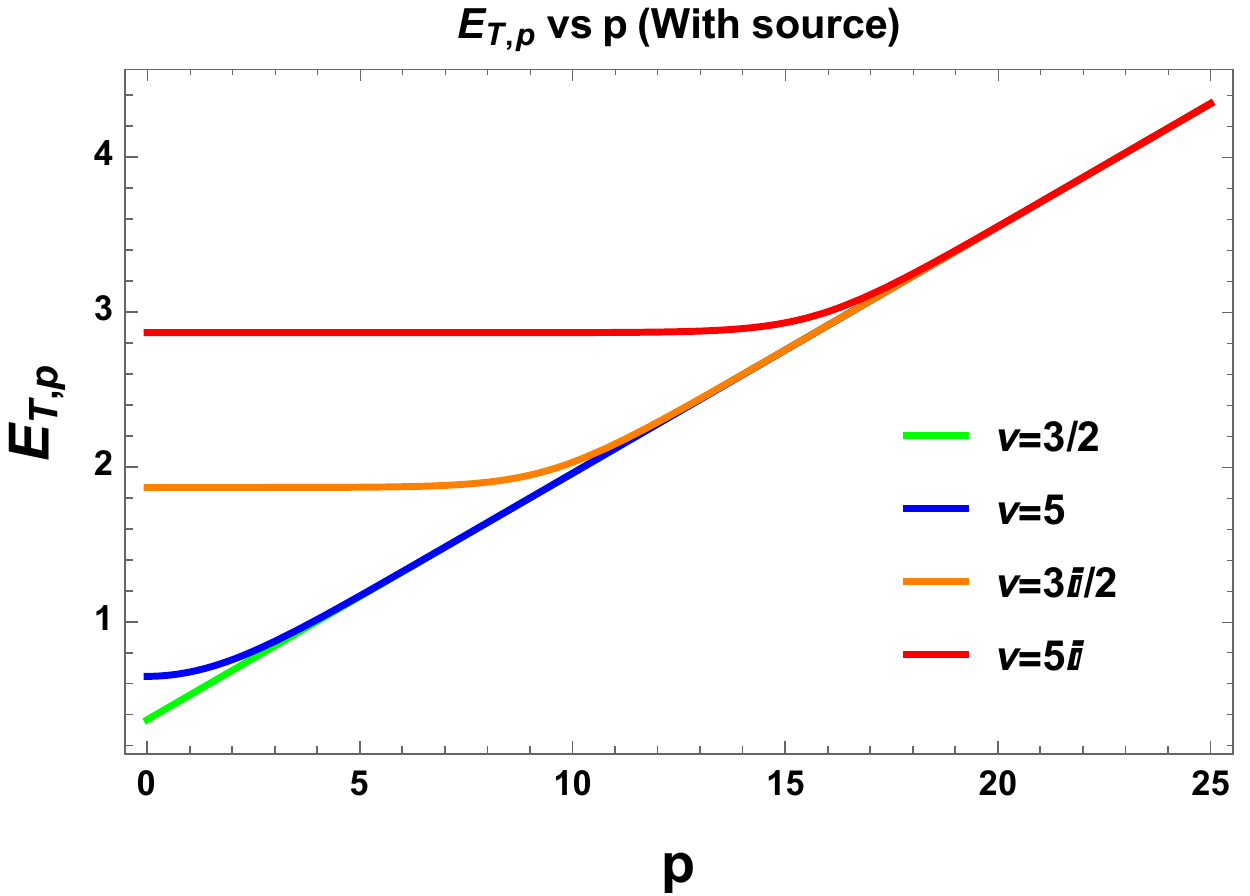}
                                                 \label{daa2}
                                                 }
                                              \subfigure[$E_{{\bf T},p}$ vs $\nu^2$ plot for fixed $p$ without source.]{
                                                      \includegraphics[width=7.2cm,height=5.2cm] {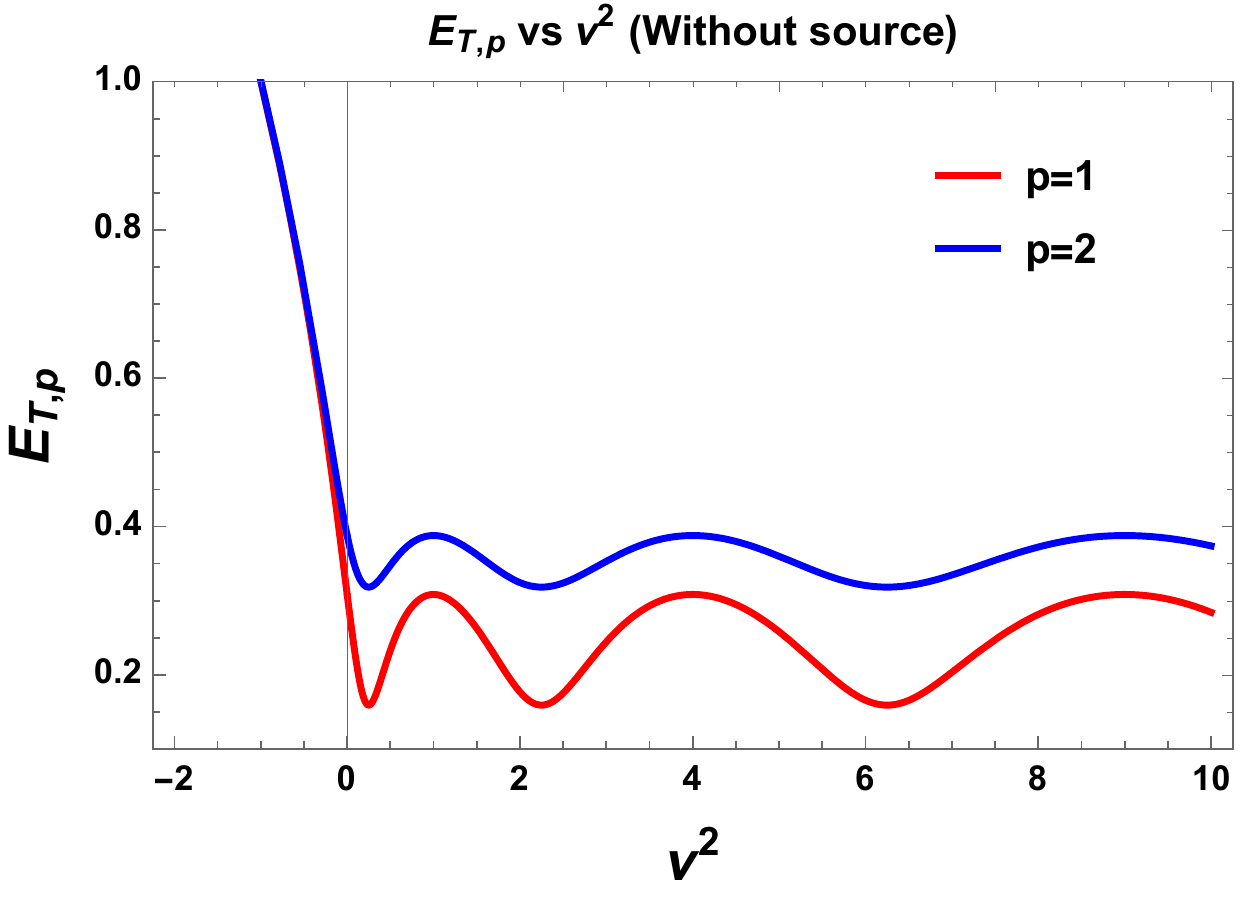}
                                                      \label{daa3}   
                                             }
                                              \subfigure[$E_{{\bf T},p}$ vs $\nu^2$ plot for fixed $p$ with source.]{
                                                                                   \includegraphics[width=7.2cm,height=5.2cm] {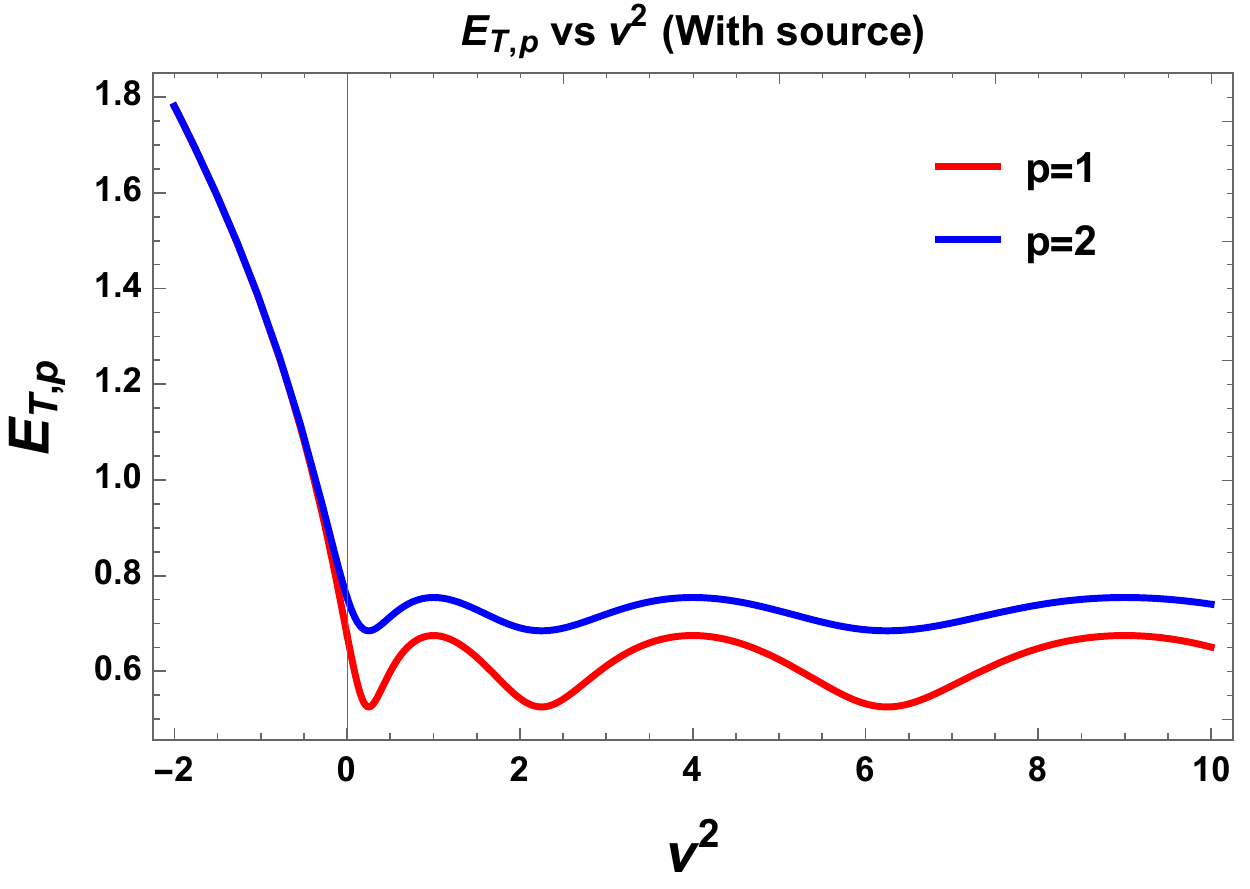}
                                                                                   \label{daa4}   
                 }
                     \caption[Optional caption for list of figures]{Behaviour of energy spectrum for axion in de Sitter space for $`+'$ branch of solution of $|\gamma_p|$ and $|\Gamma_{p,n}|$. } 
                                             \label{fggg1}
                                             \end{figure*}
                Note that, for conformally coupled axion ($\nu=1/2$) and for minimally coupled axion ($\nu=3/2$) the ${\bf SO(1,3)}$ principal quantum number $p$ dependent spectrum can be expressed as:
                 \bea E_p &=& \pm p, ~~~
                                      {\cal E}_{p,n} = -\frac{1}{2\pi}\ln \left(\frac{1}{1-e^{\mp 2p_n \pi}}\right).\eea
                                      In this case, the total energy spectrum is given by:
                \bea E_{{\bf T},p}&=&-\frac{1}{2\pi}\ln\left(\prod^{\infty}_{n=0}\frac{e^{\mp 2p\pi}}{1-e^{\mp 2p_n \pi}}\right)=\pm p-\frac{1}{2\pi}\ln \left(\prod^{\infty}_{n=0}\frac{1}{1-e^{\mp 2p_n \pi}}\right).\eea
                This implies that for conformally coupled axion ($\nu=1/2$) and for minimally coupled axion ($\nu=3/2$) the entangled Hamiltonian $({\cal H}_{\bf ENT})$ and the Hamiltonian for axion ${({\cal H}_p})_{{\bf R \times H^3}}$ are equivalent in absence of the linear source term in the effective action.  
                
                On the other hand, if we consider any arbitrary mass parameter $\nu$ in that case the ${\bf SO(1,3)}$ principal quantum number $p$ dependent spectrum can be expressed as:
                                 \bea E_p &=&-\frac{1}{2\pi}\ln \frac{2}{\left[\sqrt{\cosh 2\pi p +\cos 2\pi \nu}\pm\sqrt{\cosh 2\pi p +\cos 2\pi \nu+2}\right]^2}, \eea\bea
                                                      {\cal E}_{p,n} &=& -\frac{1}{2\pi}\ln 
                                                      \left(\frac{1}{1-\frac{2}{\left[\sqrt{\cosh 2\pi p_n +\cos 2\pi \nu}\pm \sqrt{\cosh 2\pi p_n +\cos 2\pi \nu+2}\right]^2}}\right).\eea
                                                      In this case, the total energy spectrum is given by:
                                \bea E_{{\bf T},p}&=&-\frac{1}{2\pi}\ln 2+\frac{1}{\pi}\ln\left[\sqrt{\cosh 2\pi p +\cos 2\pi \nu}\pm\sqrt{\cosh 2\pi p +\cos 2\pi \nu+2}\right]\nonumber\\
                                &&~~~~~~~-\frac{1}{2\pi}\ln 
          \left(\prod^{\infty}_{n=0}\frac{1}{1-\frac{2}{\left[\sqrt{\cosh 2\pi p_n +\cos 2\pi \nu}\pm \sqrt{\cosh 2\pi p_n +\cos 2\pi \nu+2}\right]^2}}\right) .\eea  
          This implies that for with arbitrary parameter $\nu$ the entangled Hamiltonian $({\cal H}_{\bf ENT})$ and the Hamiltonian for axion ${({\cal H}_p})_{{\bf R \times H^3}}$ are significantly differ as well in absence of the linear source term in the effective action.  
          
           In fig.~(\ref{daa1}) and fig.~(\ref{daa2}) we have shown the behaviour of the total energy spectrum for axion with respect to the ${\bf SO(1,3)}$ quantum number $p$ for a fixed value of the mass parameter $\nu$ without source contribution and with source contribution respectively. From both the plots it is clearly observed that for minimally coupled axion ($\nu=3/2$) (\textcolor{green}{\bf green}) the energy spectrum is linear. Also in this case with source the slope and intercept will change. Also it is important to mention that, for conformally coupled axion ($\nu=1/2$) the situation is exactly similar and the behaviour overlaps with result obtained for minimally coupled axion ($\nu=3/2$). Further if we increase the value of the mass parameter to $\nu=5$ (\textcolor{blue}{\bf blue}) then it shows small deviation for very small values of $p$. Next if we go the large mass range, where $\nu^2<0$ the energy spectrum shows significant deviation from the linearity for the large values of $p$. For example, we have considered $\nu=3i/2$ (\textcolor{orange}{\bf orange}) and $\nu=5i$ (\textcolor{red}{\bf red}) for our analysis. From both the plots it is clear that for $\nu=5i$ (\textcolor{red}{\bf red}) deviation from leniearity is more faster than the bevaiour obtained for $\nu=3i/2$ (\textcolor{orange}{\bf orange}).
           
           In fig.~(\ref{daa3}) and fig.~(\ref{daa4}) we have shown the behaviour of the total energy spectrum for axion with respect mass parameter $\nu^2$ for a fixed value of the ${\bf SO(1,3)}$ quantum number $p$ without source contribution and with source contribution respectively. From both the plots it is clearly observed that in presence of the source behaviour of the energy spectrum significantly changes compared to the result obtained for without source.
                                            
       \end{enumerate}

    \subsection{Computation of entanglement entropy}
    \label{ka33d}
          In this subsection we derive the expression for entanglement entropy in de Sitter space. In general the entanglement entropy can be written in terms of density matrix using Von Neumann measure as:
          \bea S(p,\nu)&=& -{\rm \bf Tr }\left[\rho_{\bf L}(p)\ln \rho_{\bf L}(p)\right],\eea
         where the parameter $\nu$ is defined earlier. For the \textcolor{red}{\bf Case~I} and \textcolor{red}{\bf Case~II}
                   the expression for the entanglement entropy in terms of the complementary and particular part of the obtained solution for a given ${\bf SO(1,3)}$ principal quantum number $p$ can be expressed as~\footnote{If we follow the equivalent ansatz of density matrix as mentioned in Eqn~(\ref{ff1sx}), the expression for the entanglement entropy in terms of the complementary and particular part of the obtained solution for a given ${\bf SO(1,3)}$ principal quantum number $p$ can be expressed as:\bea S(p,\nu)&=& -\ln a_p-a_p\frac{|\gamma_{p}|^2}{\left(1-|\gamma_{p}|^2\right)^2}\ln\left(|\gamma_{p}|^2\right)\left(1+f_p\left(1-|\gamma_{p}|^2\right)\right)\nonumber\\
                   &&-a_pf_p\ln\left(1+f_p\left(1-|\gamma_{p}|^2\right)\right)-a_pf^{2}_p\left(1-|\gamma_{p}|^2\right)\ln\left(1+f_p\right).~~~~~~~~~~~~~\eea
                    For our computation we will not further follow this ansatz of the density matrix.}:
\bea\label{oqw1} \boxed{S(p,\nu)=- \left(1+\frac{f_p}{1+f_p}\right)\left[\ln\left(1-|\gamma_{p}|^2\right)+\frac{|\gamma_{p}|^2}{\left(1-|\gamma_{p}|^2\right)}\ln\left(|\gamma_{p}|^2\right)\right]-\left(1-f_p\right)\ln\left(1+f_p\right)}~~~~~~~~~~~~~\eea             Then the quantifying formula for the entanglement entropy in de Sitter space in presence of axion can be expressed as a sum over all possible quantum states which carries ${\bf SO(1,3)}$ principal quantum number $p$. Technically, this sum can be written in terms of an integral as:
 \bea \sum_{\bf States}\sum^{\infty}_{p=0}\rightarrow V_{\bf H^3} \int^{\infty}_{p=0}~dp.\eea
       Consequently, the final expression for the entanglement entropy in $3+1$ D de Sitter space is given by the following expression:
       \bea\label{jjq} S(\nu)&=& \sum_{\bf States}\sum^{\infty}_{p=0}S(p,\nu)\rightarrow V_{\bf H^3} \int^{\infty}_{p=0}~dp~{\cal D}_{3}(p)~S(p,\nu),\eea 
       where ${\cal D}_3(p)$ characterize the density of quantum states corresponding to the radial functions on ${\bf H^3}$, 
       ${\cal D}_3(p)=\frac{p^2}{2\pi^2}.$
       The volume of the hyperboloid ${\bf H^3}$ is denoted by the overall factor $V_{\bf H^3}$. This volume is infinite and can be regularize with a large radial cutoff $r= L_c$ in the hyperboloid ${\bf H^3}$. Technically, here regularization in volume stands for embedding surface of entanglement from zero to finite conformal time. After performing the regularization the volume of the hyperboloid ${\bf H^3}$ for $r\leq  L_c$ can be written as~\footnote{In general for $D-1$-sphere, which is denoted as ${\bf S^{D-1}}$ volume can be expressed as $ V_{\bf S^{D-1}}=\frac{2\pi^{D/2}}{\Gamma\left(\frac{D}{2}\right)}.$}
       \bea V_{\bf H^3}&=& V_{\bf S^2}\int^{L_c}_{r=0}dr~\sinh^2r=\pi\left[\sinh 2L_c -2L_c\right]~~\underrightarrow{\rm {\bf large}~L_c~{\bf limit}}~~\frac{\pi}{2}\left[e^{2L_c} -4L_c\right],~~~~~~~~~\eea  
       where we use the fact that the volume of ${\bf S^2}$ is given by, $V_{\bf S^2}=4\pi$. Here one can interpret each of the term obtained in the regularized volume as:
       \begin{itemize}
       \item Here the entangling area ${\bf A_{ENT}}$ is given by:
       \bea {\bf A_{ENT}}&\sim&\pi\sinh 2L_c~~~~\underrightarrow{\rm {\bf large}~L_c~{\bf limit}}~~~~~\frac{\pi}{2}~e^{2L_c}.\eea
       \item  Consequently the second term is interpreted as logarithm of the entangling area ${\bf A_{ENT}}$, which is precisely the following factor: 
       \bea \ln{\bf A_{ENT}}&\sim&\ln(\pi\sinh 2L_c)~~~~\underrightarrow{\rm {\bf large}~L_c~{\bf limit}}~~~~~2L_c+\ln\left(\frac{\pi}{2}\right).\eea
       
       \end{itemize}
       Finally the volume of the hyperboloid ${\bf H^3}$ for $r\leq  L_c$ can be written in terms of the entangling area ${\bf A_{ENT}}$ as: 
       \bea V_{\bf H^3}&\sim &\left[{\bf A_{ENT}} -\pi\sinh^{-1}\left(\frac{{\bf A_{ENT}}}{\pi}\right)\right]~~~~\underrightarrow{\rm {\bf large}~L_c~{\bf limit}}~~\left[{\bf A_{ENT}} -\pi \ln {\bf A_{ENT}}+\pi\ln\left(\frac{\pi}{2}\right)\right],~~~~~~~~~~\eea 
       In this context the cutoff regulator $L_c$ is large and in this limit this is identified with the conformal time as, $L_c\sim -\ln \eta$. 
        We also define the regularized volume of the hyperboloid ${\bf H^3}$ as, 
        $V^{\bf REG}_{\bf H^3}=\frac{V_{\bf S^3}}{2}=2\pi.$
        Thus the volume of the hyperboloid ${\bf H^3}$ for $r\leq  L_c$ can be recast in terms of conformal time for large $L_c$ limit as:
         \bea V_{\bf H^3}&\sim &V^{\bf REG}_{\bf H^3}\left[\frac{1}{4\eta^2} +\ln\eta\right].\eea
               Finally, summing over all possible principal quantum number $p$ of ${\bf SO(1,3)}$ we get from Eqn.~(\ref{jjq}) the following expression for entanglement entropy in de Sitter space:     
               \bea S(\nu)&=&{\bf c_6}\left[\frac{1}{4\eta^2} +\ln\eta\right] ,\eea 
               where ${\bf c_6}$ is defined as:
               \bea {\bf c_6}\equiv S_{\bf intr}= \frac{1}{\pi}\int^{\infty}_{p=0}~dp~p^2~S(p,\nu),\eea 
               which represents the long range quantum entanglement in presence of axion. 
              
                     Further substituting the expression for entanglement entropy $S(p,\nu)$ computed in presence of axionic Bell pair and integrating over all possible ${\bf SO(1,3)}$ principal quantum number, lying within the window $0<p<\infty$, we get the following expression for the co-efficient ${\bf c_6}$:
               \bea \boxed{{\bf c_6}\equiv S_{\bf intr}= \left[\left(1+\frac{f_p}{1+f_p}\right){\cal I}_{1}+\left(1-f_p\right)\ln\left(1+f_p\right){\cal I}_2\right]}~,\eea
               where the integrals ${\cal I}_1$ and ${\cal I}_2$ can be written in $3+1$ dimensional space time  as:
               \bea {\cal I}_1&=&-\frac{1}{\pi}\int^{\infty}_{p=0}~dp~p^2~\left[\ln\left(1-|\gamma_{p}|^2\right)+\frac{|\gamma_{p}|^2}{\left(1-|\gamma_{p}|^2\right)}\ln\left(|\gamma_{p}|^2\right)\right],\\
               {\cal I}_2&=&-\frac{1}{\pi}\int^{\infty}_{p=0}~dp~p^2.\eea
               Here it is important to mention that:
               \begin{itemize}
               \item For both \textcolor{red}{\bf Case~I}  and \textcolor{red}{\bf Case~II} the integral ${\cal I}_2$ diverges in $3+1$ D within $0<p<\infty$. To make it finite we need to regularize this by introducing a cut-off $\Lambda_{\bf C}$ in $x=2\pi p$ as:
               \bea  {\cal I}_2&=&-\frac{1}{8\pi^4}\int^{\Lambda_{\bf C}}_{x=0}~dx~x^2=-\frac{\Lambda^3_{\bf C}}{24\pi^4}.\eea
               
               \item For \textcolor{red}{\bf Case~I} we have two possible solution for $\gamma_{p}$. For $\gamma_p=e^{-\pi p}$ the integral ${\cal I}_1$ is always finite for the \textcolor{red}{\bf Case~I}. But for the other solution of $\gamma_p=e^{\pi p}$ the integral ${\cal I}_1$ is divergent. To make the consistency throughout the analysis we put a cut-off $\Lambda_{\bf C}$ on the integral ${\cal I}_1$ obtained from both of the solutions of $\gamma_p$. Here we explicitly show that in the final expression for ${\cal I}_{1}$ after taking limit $\Lambda_{\bf C}\rightarrow \infty$ final result is finite for $\gamma_p=e^{-\pi p}$ and diverges for $\gamma_p=e^{\pi p}$. Here we introduce a change in variable by using $x=2\pi p$ and introducing the cut-off $\Lambda_{\bf C}$ for $\gamma_p=e^{-x/2}$ we get:
               \bea{\cal I}_1&=&-\frac{1}{8\pi^4}\int^{\Lambda_{\bf C}}_{x=0}~dx~x^2~\left[\ln\left(1-e^{-x}\right)-\frac{x~e^{-x}}{\left(1-e^{-x}\right)}\right]\nonumber\\
               &=&-\frac{1}{8\pi^4}\left[\frac{\Lambda^3_{\bf C}}{3} \left\{\Lambda_{\bf C}-4 \ln \left(1-e^{\Lambda_{\bf C}}\right)+\ln (\sinh \Lambda_{\bf C}-\cosh \Lambda_{\bf C}+1)\right\}\right.\nonumber\\&& \left.~~~~~~~~~~~~~~-4 \Lambda^2_{\bf C}\text{Li}_2\left(e^{\Lambda_{\bf C}}\right)+8 \Lambda_{\bf C} \text{Li}_3\left(e^{\Lambda_{\bf C}}\right)-8 \text{Li}_4\left(e^{\Lambda_{\bf C}}\right)+\frac{4 \pi ^4}{45}\right],~~~~~~\eea
               where $\text{Li}_n(x)\forall n=2,3,4$ is the polylogarithm function of $n$ th kind, which is defined as $\text{Li}_n(x)=\sum^{\infty}_{k=1}\frac{x^k}{k^n}.$
               After taking $\Lambda_{\bf C}\rightarrow \infty$ for $\gamma_p=e^{-x/2}$ we get,$\lim_{\Lambda_{\bf C}\rightarrow \infty}{\cal I}_{1}=\frac{1}{90}.$
              Further taking the source free limit in which $f_p\rightarrow 0$ we get:
                \bea \lim_{\Lambda_{\bf C}\rightarrow \infty, f_p\rightarrow 0}{\bf c_6}\equiv S_{\bf intr}=\frac{1}{90},\eea
               which is perfectly consistent with the results obtained from $\nu=1/2$ and $\nu=3/2$ cases without source in ref.~\cite{Maldacena:2012xp}. This also implies that in absence of the source with $\Lambda_{\bf C}\rightarrow \infty$ for this specific branch of solution the final result of the interesting part of the entanglement entropy is not very sensitive to the cut-off of ${\bf SO(1,3)}$ principal quantum number $p$.
            
                On the other hand, for $\gamma_p=e^{x/2}$ in $3+1$ D introducing the cut-off $\Lambda_{\bf C}$ in the rescaled principal quantum number $x$ we get:
                \bea{\cal I}_1&=&-\frac{1}{8\pi^4}\int^{\Lambda_{\bf C}}_{x=0}~dx~x^2~\left[\ln\left(1-e^{x}\right)+\frac{x~e^{x}}{\left(1-e^{x}\right)}\right]\nonumber\\
                 &=&-\frac{1}{8\pi^4}\left[-\Lambda^3_{\bf C} \ln \left(1-e^{\Lambda_{\bf C}}\right)-4 \Lambda^2_{\bf C} \text{Li}_2\left(e^{\Lambda_{\bf C}}\right)+8 \Lambda_{\bf C} \text{Li}_3\left(e^{\Lambda_{\bf C}}\right)-8 \text{Li}_4\left(e^{\Lambda_{\bf C}}\right)+\frac{4 \pi ^4}{45}\right],~~~~~~~~~~~~\eea
                 in which after taking the source free limit in which $f_p\rightarrow 0$ we get:
                    \bea \lim_{\Lambda_{\bf C}\rightarrow {\rm Large}, f_p\rightarrow 0}{\bf c_6}\equiv S_{\bf intr}&=& \frac{1}{8\pi^4}\left[\Lambda^3_{\bf C} \ln \left(1-e^{\Lambda_{\bf C}}\right)+4 \Lambda^2_{\bf C} \text{Li}_2\left(e^{\Lambda_{\bf C}}\right)\right.\nonumber\\&& \left.~~~~~~~-8 \Lambda_{\bf C} \text{Li}_3\left(e^{\Lambda_{\bf C}}\right)+8 \text{Li}_4\left(e^{\Lambda_{\bf C}}\right)-\frac{4 \pi ^4}{45}\right],\eea
                   which is perfectly
                   consistent with the results obtained from $\nu=1/2$ and $\nu=3/2$ cases without source for this branch of solution for $\gamma_p$.  In this context, this implies that in absence of the source with $\Lambda_{\bf C}\rightarrow {\rm Large}$ for this specific branch of solution the final result of the interesting part of the entanglement entropy is highly sensitive to the  cut-off of ${\bf SO(1,3)}$ principal quantum number $p$.

               \item For \textcolor{red}{\bf Case~II} we are dealing with the the situation where due to arbitrary $\nu$ dependence the expressions for the solutions of the $\gamma_p$ are very complicated, which are given by, 
                \be  \gamma_{p}=\frac{i\sqrt{2}}{\left[\sqrt{\cosh 2\pi p +\cos 2\pi \nu}\pm\sqrt{\cosh 2\pi p +\cos 2\pi \nu+2}\right]}.\ee Let us first analyze the representative integral ${\cal I}_{1}$ using both of the solutions obtained for arbitrary $\nu$. Following the previous logical argument here we also put a cut-off $\Lambda_{\bf C}$ to perform the integral on the rescaled ${\bf SO(1,3)}$ quantum number $x=2\pi p$ and after performing the integral we will check the behaviour of both of the results.
               First of all we start with the following integral with $``\pm"$ signature, as given by:
               \bea{\cal I}_1&=&-\frac{1}{8\pi^4}\int^{\Lambda_{\bf C}}_{x=0}~dx~x^2~ 
               \left[\ln\left(1-2G_{\pm}(x,\nu)\right)+\frac{2G_{\pm}(x,\nu)}{\left(1-2G_{\pm}(x,\nu)\right)}\ln\left(2G_{\pm}(x,\nu)\right)\right],~~~~~~~~~~~\eea
               where $G_{\pm}(x,\nu)$ is defined as:
               \bea G_{\pm}(x,\nu)&=&\frac{1}{\left[\sqrt{\cosh x +\cos 2\pi \nu}\pm\sqrt{\cosh x +\cos 2\pi \nu+2}\right]^{2}}.\eea
               However, it is not possible to solve this integral analytically using the exact mathematical structure. Small mass limiting situations are studied in $\nu=1/2$ as well in $\nu=3/2$ case which is appearing in \textcolor{red}{\bf Case~I}. Here we consider large mass limiting situation which is specifically important to study the physical imprints from \textcolor{red}{\bf Case~II}. In this large mass limiting situation we devide the total window of the ${\bf SO(1,3)}$ principal quantum number $p$ into two sub regions, as given by $0<p<|\nu|$ and $|\nu|<p<\Lambda_{\bf C}$ and in these consecutive region of interests the two solutions for $\gamma_p$ can be approximated as:
\bea
               \label{r2zzzzz}
      \displaystyle |\gamma_p| &=&\displaystyle\left\{\begin{array}{ll}
     \displaystyle e^{-\pi |\nu|}~~~~~~~~~~~~ &
                                                               \mbox{\small {\textcolor{red}{\bf for $0<p<|\nu|$}}}  
                                                              \\ 
              \displaystyle e^{-\pi p} & \mbox{\small { \textcolor{red}{\bf for $|\nu|<p<\Lambda_{\bf C}/2\pi$}}}.~~~~~~~~
                                                                        \end{array}
                                                              \right.\eea
   and
   \bea
                  \label{r2zzzzaz}
         \displaystyle |\gamma_p| &=&\displaystyle\left\{\begin{array}{ll}
        \displaystyle e^{\pi |\nu|}~~~~~~~~~~~~ &
                                                                  \mbox{\small {\textcolor{red}{\bf for $0<p<|\nu|$}}}  
                                                                 \\ 
                 \displaystyle e^{\pi p} & \mbox{\small { \textcolor{red}{\bf for $|\nu|<p<\Lambda_{\bf C}/2\pi$}}}.~~~~~~~~
                                                                           \end{array}
                                                                 \right.\eea
 Consequenty, for this large mass limiting situation the regularized specified integral ${\cal I}_1$ for the first solution for $|\gamma_p|$ can be written as:
                     \bea \label{Ar2zzzzz}
            \displaystyle {\cal I}_1 &=&\displaystyle\left\{\begin{array}{ll}
           \displaystyle -\frac{A(\nu)}{8\pi^4}\left[\ln\left(1-e^{-2\pi\nu}\right)-\frac{2\pi\nu~e^{-2\pi\nu}}{\left(1-e^{-2\pi\nu}\right)}\right]~~~~~~ &
                                                                     \mbox{\small {\textcolor{red}{\bf for $0<x<2\pi|\nu|$}}}  
                                                                    \\ 
                    \displaystyle -\frac{B(\nu,\Lambda_{\bf C})}{8\pi^4} & \mbox{\small { \textcolor{red}{\bf for $2\pi|\nu|<x<\Lambda_{\bf C}$}}}.~~
                                                                              \end{array}
                                                                    \right.\eea
         and for the second solution for $|\gamma_p|$ can be written as:
        \bea \label{Ar2zzzzzzz}
              \displaystyle {\cal I}_1 &=&\displaystyle\left\{\begin{array}{ll}
             \displaystyle -\frac{A(\nu)}{8\pi^4}\left[\ln\left(1-e^{2\pi\nu}\right)+\frac{2\pi\nu~e^{2\pi\nu}}{\left(1-e^{2\pi\nu}\right)}\right]~~~~~~ &
                                                                       \mbox{\small {\textcolor{red}{\bf for $0<x<2\pi|\nu|$}}}  
                                                                      \\ 
                      \displaystyle -\frac{C(\nu,\Lambda_{\bf C})}{8\pi^4} & \mbox{\small { \textcolor{red}{\bf for $2\pi|\nu|<x<\Lambda_{\bf C}$}}}.~~
                                                                                \end{array}
                                                                      \right.\eea
                                                                      Here $A(\nu)$, $B(\nu,\Lambda_{\bf C})$ abd $C(\nu,\Lambda_{\bf C})$ are defined as:
           \bea A(\nu)&=&\int^{2\pi\nu}_{x=0}~dx~x^2=\frac{8\pi^3}{3}\nu^3,\eea\bea
           B(\nu,\Lambda_{\bf C})&=&\int^{\Lambda_{\bf C}}_{x=2\pi\nu}~dx~x^2~\left[\ln\left(1-e^{-x}\right)-\frac{x~e^{-x}}{\left(1-e^{-x}\right)}\right]\nonumber\\
                               \displaystyle &=&\frac{1}{3} \left[{\Lambda^4_{\bf C}}-4 {\Lambda^3_{\bf C}} \ln \left(1-e^{\Lambda_{\bf C}}\right)+{\Lambda^3_{\bf C}} \ln (\sinh {\Lambda_{\bf C}}-\cosh {\Lambda_{\bf C}}+1)-12 {\Lambda^2_{\bf C}} \text{Li}_2\left(e^{\Lambda_{\bf C}}\right)\right.\nonumber\\ &&\left.\displaystyle ~~~+24 {\Lambda_{\bf C}} \text{Li}_3\left(e^{\Lambda_{\bf C}}\right)-24 \text{Li}_4\left(e^{\Lambda_{\bf C}}\right)-16\pi^4\nu^4 +32\pi^3 \nu^3 \ln \left(1-e^{2\pi\nu }\right)+24 \text{Li}_4\left(e^{2\pi\nu }\right)\right.\nonumber\\ &&\left.\displaystyle ~~~-8\pi^3\nu ^3 \ln (\sinh (2\pi\nu )-\cosh (2\pi\nu )+1)+48 \pi^2\nu^2 \text{Li}_2\left(e^{2\pi\nu }\right)-48 \pi\nu  \text{Li}_3\left(e^{2\pi\nu }\right)\right],~~~~~~~~~~~~\\
      C(\nu,\Lambda_{\bf C})&=&\int^{\Lambda_{\bf C}}_{x=2\pi\nu}~dx~x^2~\left[\ln\left(1-e^{x}\right)+\frac{x~e^{x}}{\left(1-e^{x}\right)}\right]\nonumber\\
                            \displaystyle &=&\left[-{\Lambda^3_{\bf C}} \ln \left(1-e^{\Lambda_{\bf C}}\right)-4 {\Lambda^2_{\bf C}} \text{Li}_2\left(e^{\Lambda_{\bf C}}\right)+8 {\Lambda_{\bf C}} \text{Li}_3\left(e^{\Lambda_{\bf C}}\right)-8 \text{Li}_4\left(e^{\Lambda_{\bf C}}\right)+8\pi^3\nu^3 \ln \left(1-e^{2\pi\nu}\right)\right.\nonumber\\&& \left.\displaystyle+16\pi^2\nu^2 \text{Li}_2\left(e^{2\pi\nu }\right)-16\pi\nu  \text{Li}_3\left(e^{2\pi\nu }\right)+8 \text{Li}_4\left(e^{2\pi\nu }\right)\right].                         \eea          
  Further to check the dependency of the cut-off and mass parameter $\nu$ in the final results obtained for both of the solution of $|\gamma_p|$ obtained for large mass limiting situation for the range  $0<x<2\pi |\nu|$ we take the limit $|\nu|>>1$, which gives the following results for the first solution for $|\gamma_p|$:
  \bea \label{Ar2zzzzx1vvz}
       \displaystyle \lim_{|\nu|>>1}{\cal I}_1 &\approx&\displaystyle
      \displaystyle  \frac{2\nu^4}{3}e^{-2\pi\nu}\left[1+{\cal O}\left(\nu^{-1}\right)\right].\eea
       Consequently in absence of source in the large mass limit the interesting part of the entanglement entropy can be written as: 
        \bea \lim_{|\nu|>>1, f_p\rightarrow 0}{\bf c_6}\equiv S_{\bf intr}\approx \frac{2\nu^4}{3}e^{-2\pi\nu}\left[1+{\cal O}\left(\nu^{-1}\right)\right],\eea
       For the second solution of $|\gamma_p|$, we can write:
   \bea \label{Ar2zzzzzzx2vvz}
         \displaystyle  \lim_{|\nu|>>1}{\cal I}_1 &=&\displaystyle
        \displaystyle -\frac{\nu^3}{3\pi}\left[\ln\left(1-e^{2\pi\nu}\right)+\frac{2\pi\nu~e^{2\pi\nu}}{\left(1-e^{2\pi\nu}\right)}\right], \eea  
         and in absence of source in the large mass limit the interesting part of the entanglement entropy can be written as: 
                 \bea \lim_{|\nu|>>1, f_p\rightarrow 0}{\bf c_6}\equiv S_{\bf intr}= -\frac{\nu^3}{3\pi}\left[\ln\left(1-e^{2\pi\nu}\right)+\frac{2\pi\nu~e^{2\pi\nu}}{\left(1-e^{2\pi\nu}\right)}\right].\eea
                 Both of the results obtained for large $|\nu|$ limiting situation in absence of the source are consistent with the results obtained  ref.~\cite{Maldacena:2012xp}.                                                                                                                                 
               \end{itemize}

 \begin{figure*}[htb]
 	\centering
 	\subfigure[Normalized entanglement entropy vs $\nu^2$ in $3+1$ D de Sitter space without axionic source ($f_p=0$).]{
 		\includegraphics[width=7.5cm,height=6.5cm] {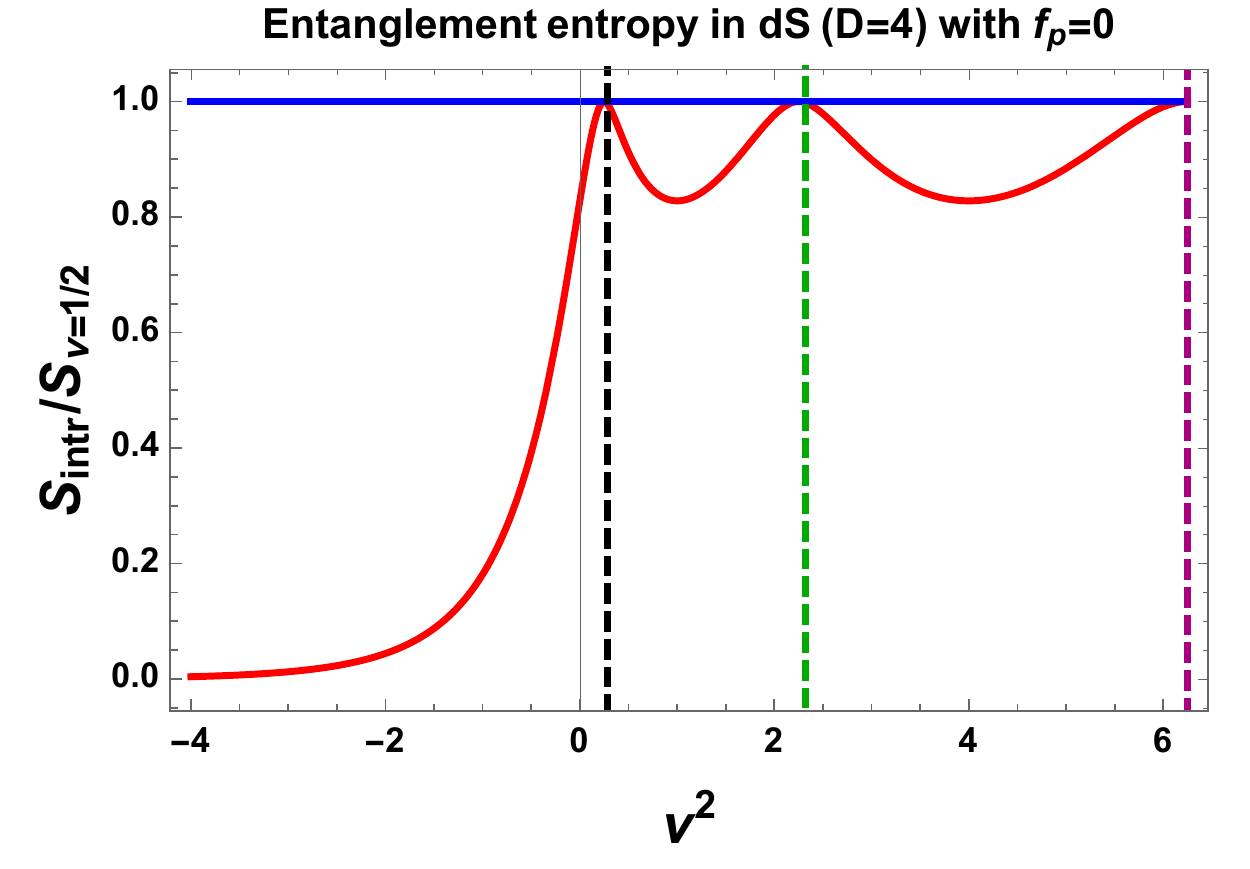}
 		\label{gax5}
 	}
 	\subfigure[Normalized entanglement entropy vs  $\nu^2$ in $3+1$ D de Sitter space with axionic source ($f_p=10^{-7}$).]{
 		\includegraphics[width=7.5cm,height=6.5cm] {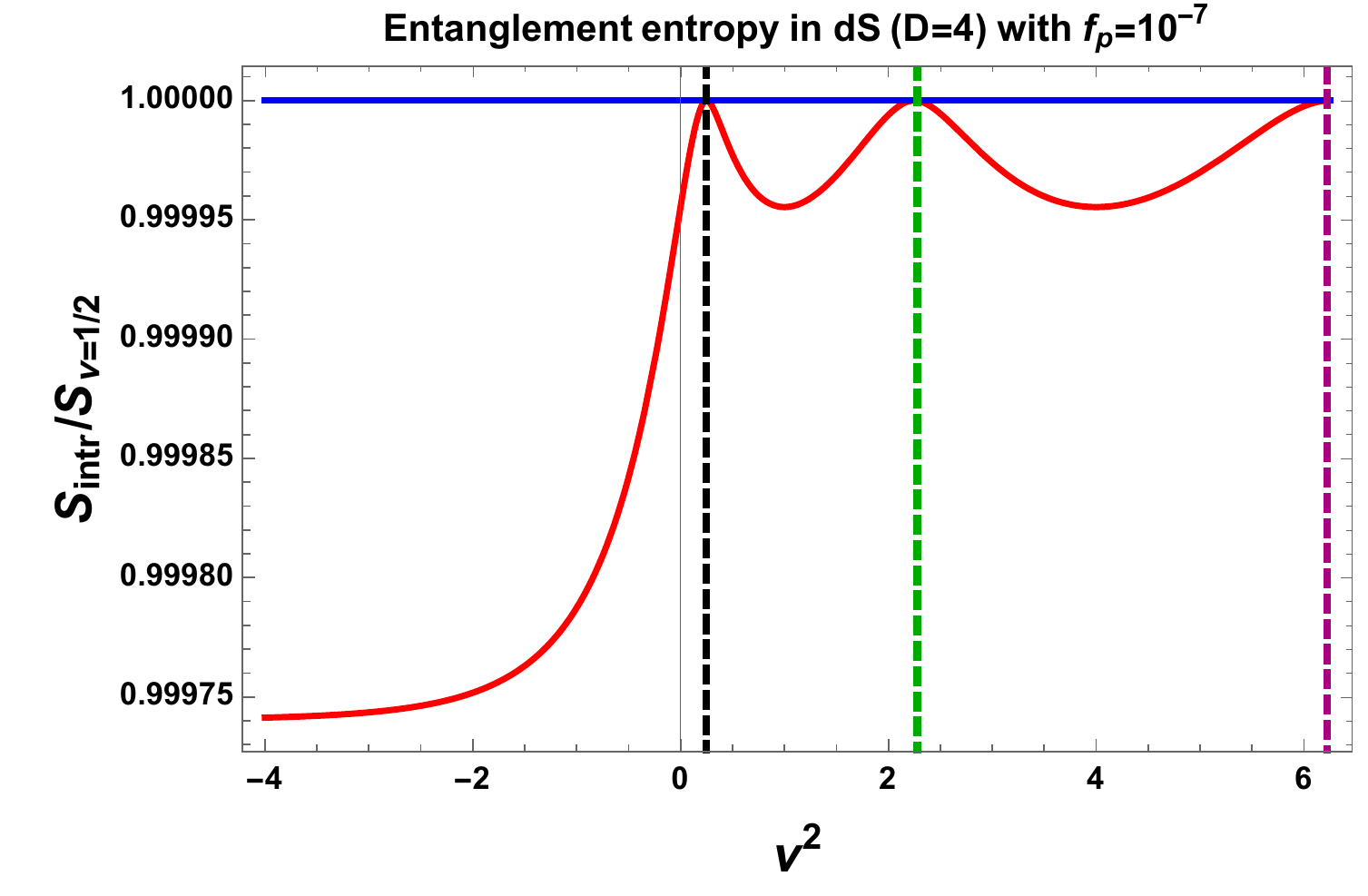}
 		\label{xgax5}
 	}
 	
 	\caption[Optional caption for list of figures]{Normalized entanglement entropy $S_{intr}/S_{\nu=1/2}$ vs mass parameter $\nu^2$ in $3+1$ D de Sitter space in absence of axionic source ($f_p=0$) and in presence of axionic source ($f_p=10^{-7}$)  for $`+'$ branch of solution of $|\gamma_p|$ and $|\Gamma_{p,n}|$. Here we set $\Lambda_{\bf C}=10000$. In both the cases we have normalized the entanglement entropy with the result obtained from $\nu=1/2$ which is the conformal limiting result.} 
 	\label{xxgg2}
 \end{figure*}                 
In fig.~(\ref{gax5}) and fig.~(\ref{xgax5}), we have explicitly shown the behaviour of entanglement entropy in $3+1$ D de Sitter space in absence ($f_p=0$) and in presence ($f_p=10^{-7}$) of axionic source with respect to the mass parameter $\nu^2$ where we consider both small and large mass limiting situation. In both the cases we have normalized the entanglement entropy with the result obtained from $\nu=1/2$ which is the conformal limiting result. In fig.~(\ref{gax5}), it is clearly observed that in absence of axionic source in the large mass limiting range where $\nu^2<0$, the normalized entanglement entropy $S_{intr}/S_{\nu=1/2}$ saturates to zero asymptotically for large values of $\nu^2$ in the negative axis. In the large mass limiting range actually the interesting part of the entanglement entropy for axion EPR Bell pair can be expressed in terms of its mass for the solution $\gamma_p=e^{-\pi|m_{axion}/H|}$ as:
\be \boxed{\lim_{m_{axion}>>H}S_{\bf intr}\approx \frac{2m^4_{axion}}{3H^4}e^{-\frac{2\pi m_{axion}}{H}}}~~,\ee
which implies less entanglement for large mass limiting situation. Further, if we increase the value of $\nu^2$ and move towards the $\nu^2>0$ region then the normalized entanglement entropy $S_{intr}/S_{\nu=1/2}$ will increase to its maximum value $\left(S_{intr}/S_{\nu=1/2}\right)_{max}=1$ at $\nu=1/2$. Further, between $3/2<\nu<5/2$ 
 normalized entanglement entropy $S_{intr}/S_{\nu=1/2}$ shows one oscillation and reaches its maximum value $\left(S_{intr}/S_{\nu=1/2}\right)_{max}=1$ again. Similar situation occurs  within the interval $5/2<\nu<7/2$ and the same oscillating feature will repeat itself. But all such oscillations are apeariodic in nature and for large positive values of the $\nu^2$ the period of oscillation increases. Similarly in fig.~(\ref{xgax5}), it is clearly depicted that in presence of axionic source in the large mass limiting range where $\nu^2<0$, the normalized entanglement entropy $S_{intr}/S_{\nu=1/2}$ saturates to zero asymptotically for large values of $\nu^2$ in the negative axis to a large value, say $\left(S_{intr}/S_{\nu=1/2}\right)_{sat}\sim 0.99975$, which is appearing in the representative figure. This is obviously higher than the result obtained from without source. On the other hand, in presence of axionic source if we increase the value of $\nu^2$ and move towards the $\nu^2>0$ region then the normalized entanglement entropy $S_{intr}/S_{\nu=1/2}$ will increase to its maximum value $\left(S_{intr}/S_{\nu=1/2}\right)_{max}=1$ at $\nu=1/2$, which is exactly similar to the result obtained without axionic source. However if we compare both the results then we see that in absence of the axionic source the rate of increase in the normalized value of the entanglement entropy is much higher compared to the result when axionic source is present. This feature clearly indicates that the normalized entanglement entropy for all values of the mass parameter $\nu^2$ is significantly higher in presence of the axionic source compared to the result obtained for without source. Further in between $3/2<\nu<5/2$ 
  normalized entanglement entropy $S_{intr}/S_{\nu=1/2}$ in presence of the axionic source shows one oscillation and reaches its maximum value $\left(S_{intr}/S_{\nu=1/2}\right)_{max}=1$ again very fast compared to the result obtained for without source. In presence of source, similar situation occurs in the subsequent intervals and the oscillating feature will repeat itself with increasing apeariodic nature for large positive values of the mass parameter $\nu^2$. But in presence of the source position of the minima in consecutive oscillations significantly increase compared to the case of without source. Additionally, it is important to note that for $\nu=1/2$ (\textcolor{red}{\bf red} dashed line), $\nu=3/2$ (\textcolor{green}{\bf green} dashed line) and $\nu=7/2$ (\textcolor{purple}{\bf purple} dashed line) we get same peak hights (maximum) of the oscillations and we get maximum normalized entanglement entropy at $\left(S_{intr}/S_{\nu=1/2}\right)_{max}=1$.  
  
  On the other hand, for the solution $\gamma_p=e^{\pi|m_{axion}/H|}$ the entanglement entropy for axion Bell pair in the large mass limiting range is very large:
   \be \lim_{m_{axion}>>H}S_{\bf intr}\approx -\frac{m^3_{axion}}{3\pi H^3}\left[\ln\left(1-e^{\frac{2\pi m_{axion}}{H}}\right)+\frac{2\pi m_{axion}}{H}\frac{e^{\frac{2\pi m_{axion}}{H}}}{\left(1-e^{\frac{2\pi m_{axion}}{H}}\right)}\right],\ee
   which implies very large entanglement for large mass limiting situation in a non trivial way. However, in this paper further we have not analyzed this case in detail due to its complicated structure. 
     \subsection{Computation of R$\acute{e}$nyi entropy}
     \label{ka33e}
      In this section we further use the expression for the density matrix to compute the R$\acute{e}$nyi entropy, which is defined by the following expression~\footnote{Here $q$ appearing below should not be confronted with $q=({\bf L},{\bf R})$ as mentioned in the previous section.}:
         \bea\label{oop1} S_{q}(p,\nu)&=& \frac{1}{1-q}\ln~{\rm \bf Tr}~\left[\rho_{\bf L}(p)\right]^q.~~~~~~~~~~{\rm with}~~q>0.\eea
         From this definition of R$\acute{e}$nyi entropy one can point out the following
         characteristics:
         \begin{itemize}
         \item Here by taking the limit $q\rightarrow 1$ one can produce the entanglement entropy for a given ${\bf SO(1,3)}$ principal quantum number $p$ and mass parameter $\nu$ i.e.
         $\lim_{q\rightarrow 1}S_{q}(p,\nu)= S(p,\nu).$
         Here $S_{q}(p,\nu)$ characterize the 
         R$\acute{e}$nyi entropy corresponding to single ${\bf SO(1,3)}$ principal quantum number $p$ and mass parameter $\nu$. However, here it is important to note that this statement is only true in absence of any source. In presence of a source one can generally write the following expression:
         \bea \lim_{q\rightarrow 1}S_{q}(p,\nu)&=& S(p,\nu)+\Delta S(p,\nu,f_p),\eea
         where $\Delta S(p,\nu,f_p)$ is the additional contribution which is appearing due to axion and vanishes in the limit, 
         $\lim_{f_p\rightarrow 0}\Delta S(p,\nu,f_p)=0.$
         \item  Further by taking the limit $q\rightarrow 0$ in the definition of R$\acute{e}$nyi entropy one can produce the Hartley entropy for a given ${\bf SO(1,3)}$ principal quantum number $p$ and mass parameter $\nu$ i.e.
                  $\lim_{q\rightarrow 0}S_{q}(p,\nu)= S_{\bf H}(p,\nu).$
                   This is actually a measure of the dimension of the density matrix and can be written as, 
                  $S_{\bf H}(p,\nu)=\ln \left({\bf dim}~ [{\cal H}_{\bf occupied}]\right),$
                  where ${\cal H}_{\bf occupied}$ represents the image of the density matrix.

         \item In the limit $q\rightarrow \infty$ R$\acute{e}$nyi entropy produces the largest eigenvalue of density matrix for a given ${\bf SO(1,3)}$ principal quantum number $p$ and mass parameter $\nu$ given by:
                           \bea \lim_{q\rightarrow \infty}S_{q}(p,\nu)&=& \lambda_{p,\nu}+\Delta \lambda_{p,\nu,f_p}=-\ln \left([\rho_{\bf L}]_{\bf max}\right),\eea where $[\rho_{\bf L}]_{\bf max}$ represents the largest eigenvalue of density matrix. Also $\Delta \lambda_{p,\nu,f_p}$ is the additional contribution arising from the source term and vanishes in the limit,
                                                      $\lim_{f_p\rightarrow 0}\Delta \lambda_{p,\nu,f_p}=0.$
         
         \item In absence of the axionic source, the quantifying formula for the R$\acute{e}$nyi entropy for a specified ${\bf SO(1,3)}$ principal quantum number $p$ and mass parameter $\nu$ can be expressed as:
         \bea S_{q}(p,\nu)&=&\frac{q}{1-q}\ln\left(1-|\gamma_p|^2\right)-\frac{1}{1-q}\ln\left(1-|\gamma_p|^{2q}\right).\eea
         Further the specific information about the long range quantum correlation in $3+1$ dimension is obtained by integrating over the ${\bf SO(1,3)}$ quantum number $p$ and a regularized volume integral over hyperboloid, as given by:
         \bea S_{q,{\bf intr}}&=&\frac{1}{\pi}\int^{\infty}_{p=0}dp~p^2~S_{q}(p,\nu),\eea
         which satisfy the constraint, 
         $\lim_{q\rightarrow 1}S_{q,{\bf intr}}=S_{{\bf intr}}\equiv{\bf c_{6}}.$
         But since our prime objective of this paper is to see the imprints of stringy axion, we have to compute the modified expression for the R$\acute{e}$nyi entropy for a given ${\bf SO(1,3)}$ principal quantum number $p$ and mass parameter $\nu$. In the next subsection we explicitly compute this result, in presence of the source.
         \end{itemize} 
    The general expression for the R$\acute{e}$nyi entropy is written in Eqn~(\ref{oop1}), where parameter $\nu$ is defined earlier. For the \textcolor{red}{\bf Case~I} and \textcolor{red}{\bf Case~II}
                      the expression for the R$\acute{e}$nyi entropy in terms of the complementary and particular part of the obtained solution for a given ${\bf SO(1,3)}$ principal quantum number $p$ can be expressed as:
   \bea S_{q}(p,\nu)&=&\left[\frac{q}{1-q}\ln\left(1-|\gamma_{p}|^2\right)-\frac{1}{1-q}\ln\left(1-|\gamma_{p}|^{2q}\right)\right]-\frac{q}{1-q}\ln\left(1+f_p\right)\nonumber\\&&~~~~~~~~~~~~~~~~~~~~~~~~~~~~~~+\frac{1}{1-q}\ln\left[1+\sum^{q}_{k=1}{}^{q}{\bf C}_{k} (f_p)^{k
      }\frac{\left(1-|\gamma_{p}|^2\right)^{-k}}{\left(1-|\gamma_{p}|^{-2k}\right)}\right].~~~~\eea
      Now to study the properties of the derived result we check the following physical limiting situations as given by:
      \begin{itemize}
      
      \item First of all we take the $q\rightarrow 1$ limit for which  we get the following simplified expression:
      \be \label{df1} \lim_{q\rightarrow 1}S_{q}(p,\nu)=-\ln\left(1-|\gamma_p|^2\right)-\frac{|\gamma_p|^2}{\left(1-|\gamma_p|^2\right)}\ln|\gamma_p|^2-\ln\left(1+f_p\right)+\ln\left(1-f_p\frac{|\gamma_p|^2}{\left(1-|\gamma_p|^2\right)^2}\right).\ee
      Further we take the difference between the results obtained from Eqn~(\ref{oqw1}) and Eqn~(\ref{df1}), which we define as:
      \bea \Delta S(p,\nu,f_p)&=&S(p,\nu)-\lim_{q\rightarrow 1}S_{q}(p,\nu)=-\frac{f_p}{1+f_p}\left[\frac{q}{1-q}\ln\left(1-|\gamma_{p}|^2\right)-\frac{1}{1-q}\ln\left(1-|\gamma_{p}|^{2q}\right)\right]\nonumber\\
      &&~~~~~~~~~~~~~~~~~~~~~~~~~~~~~~~+f_p\ln\left(1+f_p\right)-\ln\left(1-f_p\frac{|\gamma_p|^2}{\left(1-|\gamma_p|^2\right)^2}\right),\eea
      which shows that in presence of source, the entanglement entropy and R$\acute{e}$nyi entropy in the limit $q\rightarrow 1$ is not same. Now if we further take the limit $f_p\rightarrow 0$ then we get, 
      $\lim_{f_p\rightarrow 0}\Delta S(p,\nu,f_p)=0,$
      which is consistent with the result known for free theory. 
      
      \item Next we take the $q\rightarrow 0$ limit for which we get, 
            $\lim_{q\rightarrow 0}S_{q}(p,\nu)=\ln\left[\sum^{\infty}_{n=1}1\right]\rightarrow\infty.$
            This directly implies that the dimension of the image of the density matrix is given by for the present computation as, ${\bf dim}~ [{\cal H}_{\bf occupied}]\rightarrow \infty,$
      which is perfectly consistent with the fact that the density matrix in the present computation is infinite dimensional.  
      \item Finally, we take the $q\rightarrow \infty$ limit for which  we get the following simplified expression:
                  \bea \label{df1xz} \lim_{q\rightarrow \infty}S_{q}(p,\nu)&\approx&-\ln\left(1-|\gamma_{p}|^2\right)+\ln\left(1+f_p\right).~~~~~~~~~~~~\eea
                  This directly implies that the largest eigenvalue of density matrix given by for the present computation as, 
            $[\rho_{\bf L}]_{\bf max}= \left(1-|\gamma_{p}|^2\right)\frac{1}{1+f_p},$
            which is perfectly consistent with the fact that the density matrix in the present computation is infinite dimensional. Now if we further take the limit $f_p\rightarrow 0$ then we get, 
                  $\lim_{f_p\rightarrow 0}\Delta \lambda_{p,\nu,f_p}=0,$
                  which is consistent with the result known for free theory in absence of axion source contribution.
      \end{itemize}   
       Then the final expression for the interesting part of the R$\acute{e}$nyi entropy in $3+1$ D de Sitter space can be expressed as:
                     \bea  S_{q,\bf intr}= \frac{1}{\pi}\int^{\infty}_{p=0}~dp~p^2~S_{q}(p,\nu).\eea 
                     which represents the long range entanglement of the quantum state under consideration for the limit $q\rightarrow 1$. 
                     
                           Further substituting the explicit expression for entanglement entropy $S(p,\nu)$ computed in presence of axion and integrating over all possible ${\bf SO(1,3)}$ principal quantum number, lying within the window $0<p<\infty$, we get the following expression:
                     \bea \boxed{S_{q,\bf intr}= \left[{\cal J}_{1,q}+\ln\left(1+f_p\right){\cal J}_{2,q}+{\cal J}_{3,q}\right]}~,\eea
                     where the integrals ${\cal J}_{1,q}$, ${\cal J}_{2,q}$ and ${\cal J}_{3,q}$ are defined as:
                     \bea {\cal J}_{1,q}&=&\frac{1}{\pi}\int^{\infty}_{p=0}~dp~p^2~\left[\frac{q}{1-q}\ln\left(1-|\gamma_{p}|^2\right)-\frac{1}{1-q}\ln\left(1-|\gamma_{p}|^{2q}\right)\right],\\
                     {\cal J}_{2,q}&=&-\frac{1}{\pi}\frac{q}{1-q}\int^{\infty}_{p=0}~dp~p^2,\\
                                          {\cal J}_{3,q}&=&\frac{1}{\pi}\frac{1}{1-q}\int^{\infty}_{p=0}~dp~p^2~\ln\left[1+\sum^{q}_{k=1}{}^{q}{\bf C}_{k} f_p^{k
                                                }\frac{\left(1-|\gamma_{p}|^2\right)^{-k}}{\left(1-|\gamma_{p}|^{-2k}\right)}\right].\eea
                     Here it is important to note that:
                     \begin{itemize}
                     \item For both \textcolor{red}{\bf Case~I}  and \textcolor{red}{\bf Case~II} the integral ${\cal J}_{2,q}$ diverges in $3+1$ D. To make it finite we need to regularize this by introducing a change in variable by using $x=2\pi p$ and introducing a cut-off $\Lambda_{\bf C}$ in the rescaled principal quantum number $x$ as given by:
                     \bea  {\cal J}_{2,q}&=&-\frac{1}{8\pi^4}\frac{q}{1-q}\int^{\Lambda_{\bf C}}_{x=0}~dx~x^2=-\frac{\Lambda^3_{\bf C}}{24\pi^4}\frac{q}{1-q}.\eea

                     \item For \textcolor{red}{\bf Case~I} we have two possible solution for $\gamma_{p}$. For $\gamma_p=e^{-\pi p}$ the integrals ${\cal J}_{1,q}$ and ${\cal J}_{3,q}$ are always finite for the \textcolor{red}{\bf Case~I}. But for the other solution of $\gamma_p=e^{\pi p}$ the integrals ${\cal J}_{1,q}$ and ${\cal J}_{3,q}$ are divergent. To have the consistency throughout the analysis we put a cut-off $\Lambda_{\bf C}$ on the integrals ${\cal J}_{1,q}$ and ${\cal J}_{3,q}$ obtained from both the solutions of $\gamma_p$. Here after introducing the cut-off $\Lambda_{\bf C}$ in $x=2\pi p$ for $\gamma_p=e^{-x/2}$ we get:
                                    \bea {\cal J}_{1,q}&=&\frac{1}{8\pi^4}\int^{\Lambda_{\bf C}}_{x=0}~dx~x^2~~\left[\frac{q}{1-q}\ln\left(1-e^{-x}\right)-\frac{1}{1-q}\ln\left(1-e^{-xq}\right)\right]\nonumber\\
                     &=&\frac{1}{8\pi^4}\frac{1}{45 (q-1) q^3}\left[-15 \Lambda_{\bf C}^3 q^4 \ln \left(-e^{-\Lambda_{\bf C}}\right)+15 \Lambda_{\bf C}^3 q^3 \ln \left(-e^{-\Lambda_{\bf C} q}\right)+45 \Lambda_{\bf C}^2 q^4 \text{Li}_2\left(e^{\Lambda_{\bf C}}\right)\right.\nonumber\\&& \left.~~~~~~~~~~~~~~~~~~~~~~~~~~~~-45 \Lambda_{\bf C}^2 q^2 \text{Li}_2\left(e^{\Lambda_{\bf C} q}\right)-90 \Lambda_{\bf C} q^4 \text{Li}_3\left(e^{\Lambda_{\bf C}}\right)+90 q^4 \text{Li}_4\left(e^{\Lambda_{\bf C}}\right)\right.\nonumber\\&& \left.~~~~~~~~~~~~~~~~~~~~~~~~~~~~+90 \Lambda_{\bf C} q \text{Li}_3\left(e^{\Lambda_{\bf C} q}\right)-90 \text{Li}_4\left(e^{\Lambda_{\bf C} q}\right)-\pi ^4 q^4+\pi ^4\right],~~~~~~~~~~~~\\ {\cal J}_{3,q}&=&\frac{1}{8\pi^4}\frac{1}{1-q}\int^{\Lambda_{\bf C}}_{x=0}~dx~x^2~~\ln\left[1+\sum^{q}_{k=1}{}^{q}{\bf C}_{k} (f_p)^{k
                                                                     }\frac{\left(1-e^{-x}\right)^{-k}}{\left(1-e^{xk}\right)}\right].~~~~~~\eea
                     Taking ${\bf \Lambda}_{\bf C}\rightarrow \infty$ for $\gamma_p=e^{-x/2}$ we get:
                     \bea \lim_{{\bf \Lambda}_{\bf C}\rightarrow \infty}{\cal J}_{1,q}&=&\frac{ (q+1) \left(q^2+1\right)}{360 q^3},\eea
                     and further taking $q\rightarrow 1$, we get, 
                     $\lim_{{\bf \Lambda}_{\bf C}\rightarrow \infty,q\rightarrow 1}{\cal J}_{1,q}=\frac{1}{90}.$
                     Exact analytical form of the integral ${\cal J}_{3,q}$ for any $q$ is not computable. But in the limiting case $f_p\rightarrow 0$ we get, 
                     $\lim_{\Lambda_{\bf C}\rightarrow \infty,q\rightarrow 1,f_p\rightarrow 0}{\cal J}_{3,q}=0.$
                    This implies that in the source free limit where $f_p\rightarrow 0$ we get:
                      \bea \lim_{\Lambda_{\bf C}\rightarrow \infty, f_p\rightarrow 0,q\rightarrow 1}S_{q,\bf intr}= \frac{1}{90}=S_{\bf intr},\eea
                     which is consistent with the results obtained from $\nu=1/2$ and $\nu=3/2$ cases without source in $3+1$ D. 
                     
                      On the other hand, for $\gamma_p=e^{x/2}$ in $3+1$ D introducing the cut-off $\Lambda_{\bf C}$ in the rescaled principal quantum number $x$ we get:
                       \bea {\cal J}_{1,q}&=&\frac{1}{8\pi^4}\int^{\Lambda_{\bf C}}_{x=0}~dx~x^2~~\left[\frac{q}{1-q}\ln\left(1-e^{x}\right)-\frac{1}{1-q}\ln\left(1-e^{xq}\right)\right]\nonumber\\
                                           &=&\frac{1}{8\pi^4}\frac{1}{45 (q-1) q^3}\left[45 \Lambda_{\bf C}^2 q^4 \text{Li}_2\left(e^{\Lambda_{\bf C}}\right)-45 \Lambda_{\bf C}^2 q^2 \text{Li}_2\left(e^{\Lambda_{\bf C} q}\right)-90 \Lambda_{\bf C} q^4 \text{Li}_3\left(e^{\Lambda_{\bf C}}\right)\right.\nonumber\\&& \left.~~~~~~~~~~~~~~+90 q^4 \text{Li}_4\left(e^{\Lambda_{\bf C}}\right)+90 \Lambda_{\bf C} q \text{Li}_3\left(e^{\Lambda_{\bf C} q}\right)-90 \text{Li}_4\left(e^{\Lambda_{\bf C} q}\right)-\pi ^4 q^4+\pi ^4\right],~~~~~~~~~~~~\\ {\cal J}_{3,q}&=&\frac{1}{8\pi^4}\frac{1}{1-q}\int^{\Lambda_{\bf C}}_{x=0}~dx~x^2~~\ln\left[1+\sum^{q}_{k=1}{}^{q}{\bf C}_{k} (f_p)^{k
                                                                                           }\frac{\left(1-e^{x}\right)^{-k}}{\left(1-e^{-xk}\right)}\right].~~~~~~\eea
                       After taking $q\rightarrow 1$ we get:
                                            \bea \lim_{q\rightarrow 1}{\cal J}_{1,q}&=&\frac{1}{8\pi^4}\left[\Lambda_{\bf C}^3 \ln \left(1-e^{\Lambda_{\bf C}}\right)+4 \Lambda_{\bf C}^2 \text{Li}_2\left(e^{\Lambda_{\bf C}}\right)-8 \Lambda_{\bf C} \text{Li}_3\left(e^{\Lambda_{\bf C}}\right)+8 \text{Li}_4\left(e^{\Lambda_{\bf C}}\right)-\frac{4 \pi ^4}{45}\right].~~~~~~~~~~\eea
                                            Exact analytical form of the integral ${\cal J}_{3,q}$ for any $q$ is not computable. But in the limiting case $f_p\rightarrow 0$ we get,
                                            $\lim_{\Lambda_{\bf C}\rightarrow {\rm Large},q\rightarrow 1,f_p\rightarrow 0}{\cal J}_{3,q}=0.$
                                           This implies that in the source free limit in which $f_p\rightarrow 0$ we get:,
                                            \bea \lim_{\Lambda_{\bf C}\rightarrow {\rm Large},q\rightarrow 1, f_p\rightarrow 0}S_{q,\bf intr}&=& S_{\bf intr}=c_{\bf 6}=\frac{1}{8\pi^4}\left[\Lambda_{\bf C}^3 \ln \left(1-e^{\Lambda_{\bf C}}\right)+4 \Lambda_{\bf C}^2 \text{Li}_2\left(e^{\Lambda_{\bf C}}\right)\right.\nonumber\\
                                            &&\left.~~~~~~~~~~~~-8 \Lambda_{\bf C} \text{Li}_3\left(e^{\Lambda_{\bf C}}\right)+8 \text{Li}_4\left(e^{\Lambda_{\bf C}}\right)-\frac{4 \pi ^4}{45}\right],~~~~~~\eea
                                           which is perfectly
                         consistent with the results obtained from $\nu=1/2$ and $\nu=3/2$ cases without source for this branch of solution for $\gamma_p$ in $3+1$ D.

                     \item For \textcolor{red}{\bf Case~II} following the similar procedure we get:
                     \bea {\cal J}_{1,q}&=&\frac{1}{8\pi^4}\int^{\Lambda_{\bf C}}_{x=0}~dx~x^2~~\left[\frac{q}{1-q}\ln\left(1-2G_{\pm}(x,\nu)\right)-\frac{1}{1-q}\ln\left(1-(2G_{\pm}(x,\nu)\right)^q)\right],~~~~~~~~~~~~\\ {\cal J}_{3,q}&=&\frac{1}{8\pi^4}\frac{1}{1-q}\int^{\Lambda_{\bf C}}_{x=0}~dx~x^2~~\ln\left[1+\sum^{q}_{k=1}{}^{q}{\bf C}_{k} (f_p)^{k
                                                                                          }\frac{\left(1-2G_{\pm}(x,\nu)\right)^{-k}}{\left(1-(2G_{\pm}(x,\nu))^{-k}\right)}\right],~~~~~~\eea
                    where $G_{\pm}(x,\nu)$ is defined earlier. Here we consider large mass limiting situation which is specifically important to study the physical imprints from \textcolor{red}{\bf Case~II} as mentioned earlier. In this large mass limiting situation we devide the total window of the ${\bf SO(1,3)}$ principal quantum number $p$ into two sub regions, as given by $0<p<|\nu|$ and $|\nu|<p<\Lambda_{\bf C}$:
      
      Consequenty, for this large mass limiting situation the regularized specified integral ${\cal J}_{1,q}$ and ${\cal J}_{3,q}$ for the first solution for $|\gamma_p|$ in $3+1$ D can be written as:
                     \bea \label{Ar2zzzza}
            \displaystyle {\cal J}_{1,q} &=&\footnotesize\displaystyle\left\{\begin{array}{ll}
           \displaystyle \frac{A(\nu)}{8\pi^4}\left[\frac{q}{1-q}\ln\left(1-e^{-2\pi\nu}\right)-\frac{1}{1-q}\ln\left(1-e^{-2\pi\nu q}\right)\right]~~~~~~ &
                                                                     \mbox{\small {\textcolor{red}{\bf for $0<x<2\pi|\nu|$}}}  
                                                                    \\ 
                    \displaystyle \frac{D(\nu,\Lambda_{\bf C},q)}{8\pi^4} & \mbox{\small { \textcolor{red}{\bf for $2\pi|\nu|<x<\Lambda_{\bf C}$}}}.~~
                                                                              \end{array}
                                                                    \right.\\ \label{ssAr2zzzzvdxx}
                                                                                               \displaystyle {\cal J}_{3,q} &=&\footnotesize\displaystyle\left\{\begin{array}{ll}
                                                                                              \displaystyle \frac{A(\nu)}{8\pi^4}\frac{1}{1-q}~\ln\left[1+\sum^{q}_{k=1}{}^{q}{\bf C}_{k} (f_p)^{k                                                                                                             }\frac{\left(1-e^{-2\pi\nu}\right)^{-k}}{\left(1-e^{2\pi \nu k}\right)}\right]~~ &
                                                                                                                                                        \mbox{\small {\textcolor{red}{\bf for $0<x<2\pi|\nu|$}}}  
                                                                                                                                                       \\ 
                                                                                                       \displaystyle \frac{1}{8\pi^4}\frac{1}{1-q}\int^{\Lambda_{\bf C}}_{x=2\pi\nu}~dx~x^2~~\ln\left[1+\sum^{q}_{k=1}{}^{q}{\bf C}_{k} f_p^{k
                                                                                                                                                                                                                            }\frac{\left(1-e^{-x}\right)^{-k}}{\left(1-e^{xk}\right)}\right] & \mbox{\small { \textcolor{red}{\bf for $2\pi|\nu|<x<\Lambda_{\bf C}$}}}.~~~
                                                                                                                                                                 \end{array}
                                                                                                                                                       \right.\eea
         and for the second solution for $|\gamma_p|$ can be written as:
        \bea \label{Ar2zzzzzza}
              \displaystyle {\cal J}_{1,q} &=&\footnotesize\displaystyle\left\{\begin{array}{ll}
             \displaystyle \frac{A(\nu)}{8\pi^4}\left[\frac{q}{1-q}\ln\left(1-e^{2\pi\nu}\right)-\frac{1}{1-q}\ln\left(1-e^{2\pi\nu q}\right)\right]~~~~~~ &
                                                                       \mbox{\small {\textcolor{red}{\bf for $0<x<2\pi|\nu|$}}}  
                                                                      \\ 
                      \displaystyle \frac{E(\nu,\Lambda_{\bf C},q)}{8\pi^4} & \mbox{\small { \textcolor{red}{\bf for $2\pi|\nu|<x<\Lambda_{\bf C}$}}}.~~
                                                                                \end{array}
                                                                      \right.\\ \label{sssAr2zzzzvdvv}
                                                                                                 \displaystyle {\cal J}_{3,q} &=&\footnotesize\displaystyle\left\{\begin{array}{ll}
                                                                                                \displaystyle \frac{A(\nu)}{8\pi^4}\frac{1}{1-q}~\ln\left[1+\sum^{q}_{k=1}{}^{q}{\bf C}_{k} f_p^{k }\frac{\left(1-e^{2\pi\nu}\right)^{-k}}{\left(1-e^{-2\pi \nu k}\right)}\right]~~ &
                                                                                                                                                          \mbox{\small {\textcolor{red}{\bf for $0<x<2\pi|\nu|$}}}  
                                                                                                                                                         \\ 
                                                                                                         \displaystyle \frac{1}{8\pi^4}\frac{1}{1-q}\int^{\Lambda_{\bf C}}_{x=2\pi\nu}~dx~x^2~~\ln\left[1+\sum^{q}_{k=1}{}^{q}{\bf C}_{k} (f_p)^{k
                                                                                                                                                                                                                              }\frac{\left(1-e^{x}\right)^{-k}}{\left(1-e^{-xk}\right)}\right] & \mbox{\small { \textcolor{red}{\bf for $2\pi|\nu|<x<\Lambda_{\bf C}$}}}.~~~
                                                                                                                                                                   \end{array}
                                                                                                                                                         \right.\eea                                            
                                                                      Here $A(\nu)$ is defined earlier and $D(\nu,\Lambda_{\bf C})$ and $E(\nu,\Lambda_{\bf C})$ are defined as:
           \bea 
           D(\nu,\Lambda_{\bf C},q)&=&\int^{\Lambda_{\bf C}}_{x=2\pi\nu}~dx~x^2~\left[\frac{q}{1-q}\ln\left(1-e^{-x}\right)-\frac{1}{1-q}\ln\left(1-e^{-xq}\right)\right]\nonumber\\
                               \displaystyle &=&\frac{1}{3 (q-1) q^3}\left[8\pi^3\nu^3 q^4 \ln \left(-e^{-2\pi\nu}\right)-8\pi^3\nu^3 q^3 \ln \left(-e^{-2\pi\nu q}\right)-12\pi^2 \nu^2 q^4 \text{Li}_2\left(e^{2\pi\nu}\right)\right.\nonumber\\
                               &&\left.~~~~~~~~~~~~~~~~~~+12\pi^2 \nu^2 q^2 \text{Li}_2\left(e^{2\pi\nu q}\right)+12 \pi\nu q^4 \text{Li}_3\left(e^{2\pi\nu}\right)-6 q^4 \text{Li}_4\left(e^{2\pi\nu}\right)\right.\nonumber\\
                                                                                                                            &&\left.~~~~~~~~~~~~~~~~~~- 12\pi\nu q \text{Li}_3\left(e^{2\pi\nu q}\right)+6 \text{Li}_4\left(e^{2\pi\nu q}\right)-\Lambda_{\bf C}^3 q^4 \ln \left(-e^{-\Lambda_{\bf C}}\right)\right.\nonumber\\
                                                                                                                                                                                                                                                        &&\left.~~~~~~~~~~~~~~~~~~+\Lambda_{\bf C}^3 q^3 \ln \left(-e^{-\Lambda_{\bf C} q}\right)+3 \Lambda_{\bf C}^2 q^4 \text{Li}_2\left(e^{\Lambda_{\bf C}}\right)-3 \Lambda_{\bf C}^2 q^2 \text{Li}_2\left(e^{\Lambda_{\bf C} q}\right)\right.\nonumber\\
                                                                                                                                                                                                                                                                                                                                                                                                                                                                                                                &&\left.~~~~~~~~~~~-6 \Lambda_{\bf C} q^4 \text{Li}_3\left(e^{\Lambda_{\bf C}}\right)+6 q^4 \text{Li}_4\left(e^{\Lambda_{\bf C}}\right)+6 \Lambda_{\bf C} q \text{Li}_3\left(e^{\Lambda_{\bf C} q}\right)-6 \text{Li}_4\left(e^{\Lambda_{\bf C} q}\right)\right],~~~~~~~~~~~~\eea\bea
      E(\nu,\Lambda_{\bf C},q)&=&\int^{\Lambda_{\bf C}}_{x=2\pi\nu}~dx~x^2~\left[\frac{q}{1-q}\ln\left(1-e^{x}\right)-\frac{1}{1-q}\ln\left(1-e^{xq}\right)\right]\nonumber\\
                            \displaystyle &=&\frac{1}{(q-1) q^3}\left[-4\pi^2\nu^2 q^4 \text{Li}_2\left(e^{2\pi\nu}\right)+4\pi^2\nu^2 q^2 \text{Li}_2\left(e^{2\pi\nu q}\right)+4\pi\nu  q^4 \text{Li}_3\left(e^{2\pi\nu}\right)\right.\nonumber\\
                                                                                                                                                                                                                                                                                    &&\left.~~~~~~~~~~~~~~~~~~-2 q^4 \text{Li}_4\left(e^{2\pi\nu}\right)-4\pi\nu q \text{Li}_3\left(e^{2\pi\nu q}\right)+2 \text{Li}_4\left(e^{2\pi\nu q}\right)\right.\nonumber\\ &&\left.~~~~~~~~~~~~~~~~~~+\Lambda_{\bf C}^2 q^4 \text{Li}_2\left(e^{\Lambda_{\bf C}}\right)-\Lambda_{\bf C}^2 q^2 \text{Li}_2\left(e^{\Lambda_{\bf C} q}\right)-2 \Lambda_{\bf C} q^4 \text{Li}_3\left(e^{\Lambda_{\bf C}}\right)\right.\nonumber\\ &&\left.~~~~~~~~~~~~~~~~~~+2 q^4 \text{Li}_4\left(e^{\Lambda_{\bf C}}\right)+2 \Lambda_{\bf C} q \text{Li}_3\left(e^{\Lambda_{\bf C} q}\right)-2 \text{Li}_4\left(e^{\Lambda_{\bf C} q}\right)\right].                                                   \eea

  Further to check the dependency of the cut-off and mass parameter $\nu$ in the final results obtained for both of the solution of $|\gamma_p|$ obtained for large mass limiting situation for the range  $0<x<2\pi |\nu|$ we take the limit $\Lambda_{\bf C}\rightarrow \infty$ and $q\rightarrow 1$, which gives the following results for the first solution for $|\gamma_p|$ in $3+1$ D:
        \bea \label{Ar2zzzzx1a}
             \displaystyle \lim_{\Lambda_{\bf C}\rightarrow \infty,q\rightarrow 1}{\cal J}_{1,q} &=&\displaystyle
            \displaystyle \frac{\nu^3}{3}\left(\frac{2 \nu}{e^{2 \pi  \nu}-1}-\frac{\ln \left(1-e^{-2 \pi  \nu}\right)}{\pi }\right).\eea
            Further taking $|\nu|>>1$ limiting case we finally get:
\bea \label{Ar2zzzzx1b}
             \displaystyle \lim_{\Lambda_{\bf C}\rightarrow \infty,q\rightarrow 1,|\nu|>>1}{\cal J}_{1,q} &\approx&\displaystyle
            \displaystyle  \frac{2\nu^4}{3}e^{-2\pi\nu}\left[1+{\cal O}\left(\nu^{-1}\right)\right].\eea
            On the other hand if we take the source limit $f_p\rightarrow 0$ then for the integral ${\cal J}_{3,q}$ we get, 
             $\lim_{\Lambda_{\bf C}\rightarrow \infty,q\rightarrow 1,|\nu|>>1,f_p\rightarrow 0}{\cal J}_{3,q} = 0.$
                        Consequently in absence of source in the large mass limit with $q\rightarrow 1$ the interesting part of the R$\acute{e}$nyi entropy can be written as: 
              \bea \lim_{\Lambda_{\bf C}\rightarrow \infty,q\rightarrow 1,|\nu|>>1, f_p\rightarrow 0}S_{q,\bf intr}\approx \frac{2\nu^4}{3}e^{-2\pi\nu}\left[1+{\cal O}\left(\nu^{-1}\right)\right]=S_{\bf intr},\eea
             For the second solution for $|\gamma_p|$ taking $|\nu|>>1$ limiting case we get:
                          \bea  \lim_{\Lambda_{\bf C}\rightarrow {\rm Large},q\rightarrow 1,|\nu|>>1}{\cal J}_{1,q} &=& \frac{\nu^3}{3}\left(\frac{2 \nu e^{2 \pi  \nu}}{e^{2 \pi  \nu}-1}-\frac{\ln \left(1-e^{2 \pi  \nu}\right)}{\pi }\right),\eea
               and $\lim_{\Lambda_{\bf C}\rightarrow \infty,q\rightarrow 1,|\nu|>>1,f_p\rightarrow 0}{\cal J}_{3,q} = 0.$                        
               Consequently, in absence of source in the large mass limit the interesting part of the R$\acute{e}$nyi entropy can be written as: 
                       \bea \lim_{\Lambda_{\bf C}\rightarrow {\rm Large},q\rightarrow 1,|\nu|>>1, f_p\rightarrow 0}S_{q,\bf intr}=\frac{\nu^3}{3}\left(\frac{2 \nu e^{2 \pi  \nu}}{e^{2 \pi  \nu}-1}-\frac{\ln \left(1-e^{2 \pi  \nu}\right)}{\pi }\right)=S_{\bf intr}.\eea
                       Both of the results obtained for large $\nu$ limiting situation in absence of the source are consistent with the results obtained in this paper. 
                       
                                                                                                                                                             \begin{figure*}[htb]
                                                                                                                                                             \centering
                                                                                                                                                             \subfigure[Normalized R$\acute{e}$nyi entropy vs $\nu^2$ in $3+1$ D de Sitter space without axionic source ($f_p=0$).]{
                                                                                                                                                                 \includegraphics[width=7.5cm,height=6.5cm] {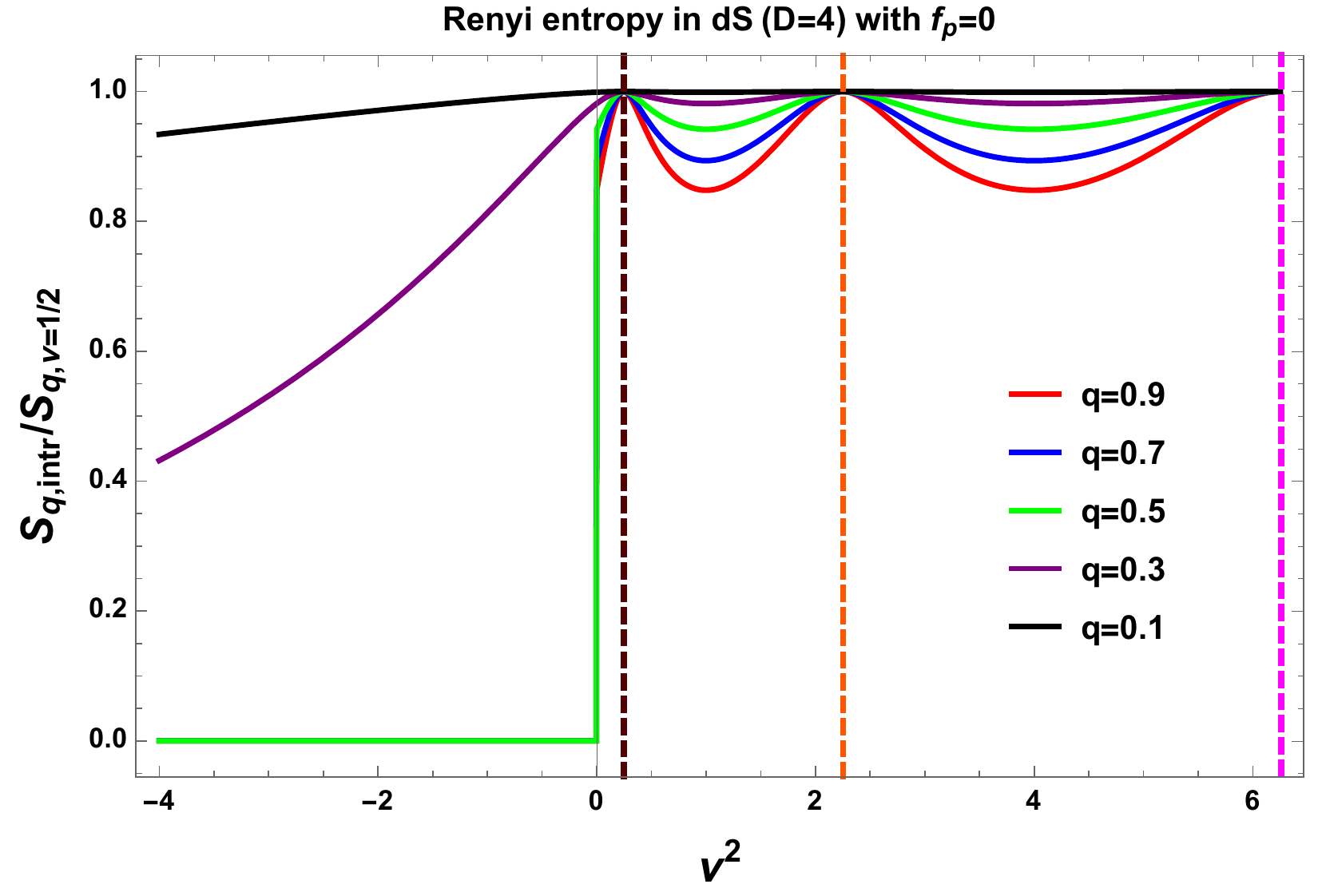}
                                                                                                                                                                 \label{zgax5}
                                                                                                                                                                 }
                                                                                                                                                             \subfigure[Normalized R$\acute{e}$nyi entropy vs $\nu^2$ in $3+1$ D de Sitter space with axionic source ($f_p=10^{-7}$).]{
                                                                                                                                                                 \includegraphics[width=7.5cm,height=6.5cm] {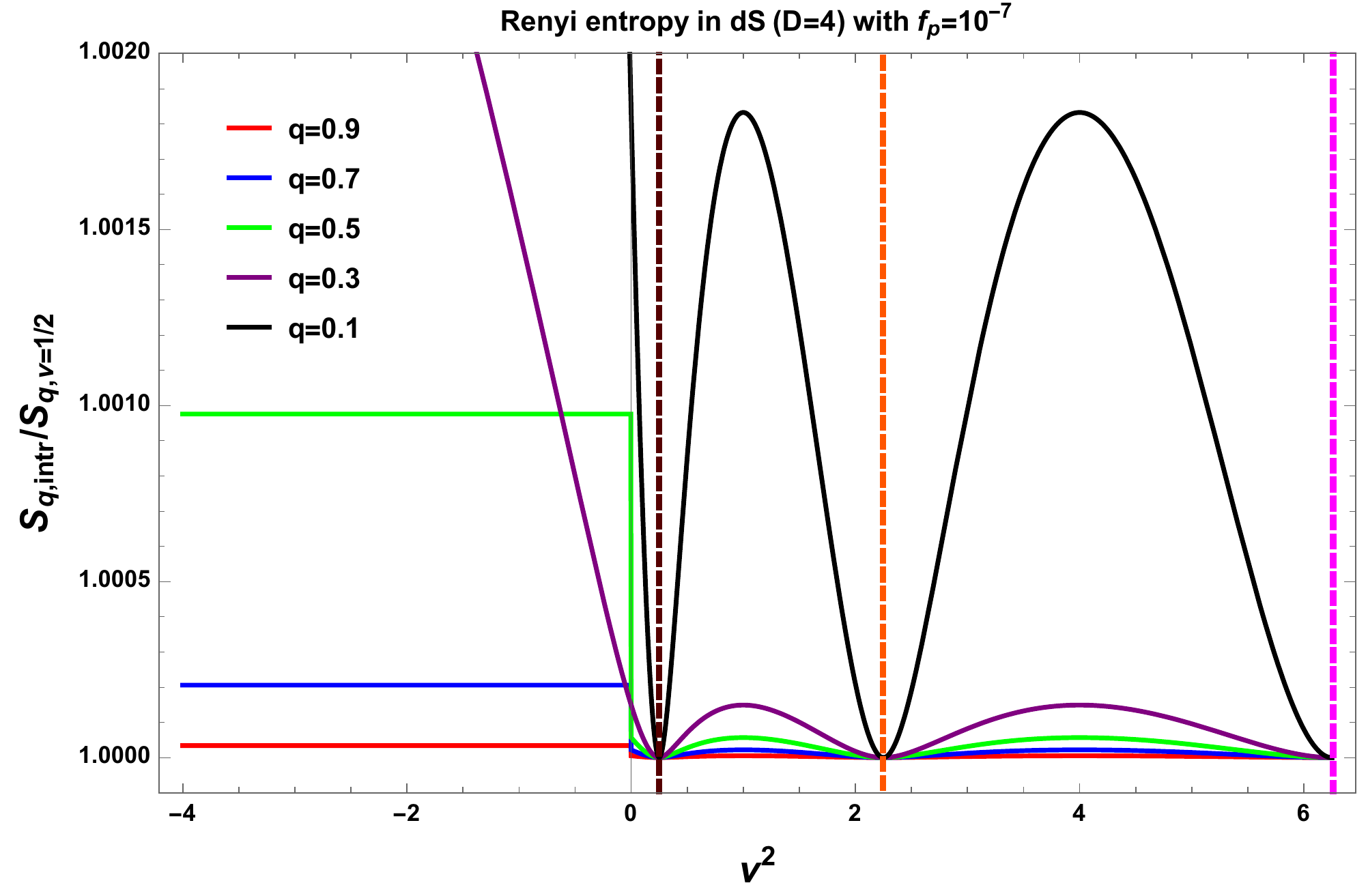}
                                                                                                                                                                 \label{zxgax5}
                                                                                                                                                                 }
                                                                                                                                                              
                                                                                                                                                             \caption[Optional caption for list of figures]{Normalized R$\acute{e}$nyi entropy $S_{intr}/S_{\nu=1/2}$ vs mass parameter $\nu^2$ in $3+1$ D de Sitter space in absence of axionic source ($f_p=0$) and in presence of axionic source ($f_p=10^{-7}$) for $q=0.1$ (\textcolor{black}{\bf black}),$q=0.3$ (\textcolor{purple}{\bf purple}),$q=0.5$ (\textcolor{green}{\bf green}),$q=0.7$ (\textcolor{blue}{\bf blue}),$q=0.9$ (\textcolor{red}{\bf red}) with $`+'$ branch of solution of $|\gamma_p|$ and $|\Gamma_{p,n}|$. Here we set $\Lambda_{\bf C}=10000$. In both the cases we have normalized the entanglement entropy with the result obtained from $\nu=1/2$ which is the conformal limiting result.} 
                                                                                                                                                             \label{zxxgg2}
                                                                                                                                                             \end{figure*}   
                                                                                                                                                                                                                                                                                                   
                                                                                                                                                               \begin{figure*}[htb]
                                                                                                                                                               \centering
                                                                                                                                                               \subfigure[Normalized R$\acute{e}$nyi entropy vs $\nu^2$ in $3+1$ D de Sitter space without axionic source ($f_p=0$).]{
                                                                                                                                                                   \includegraphics[width=7.5cm,height=6.5cm] {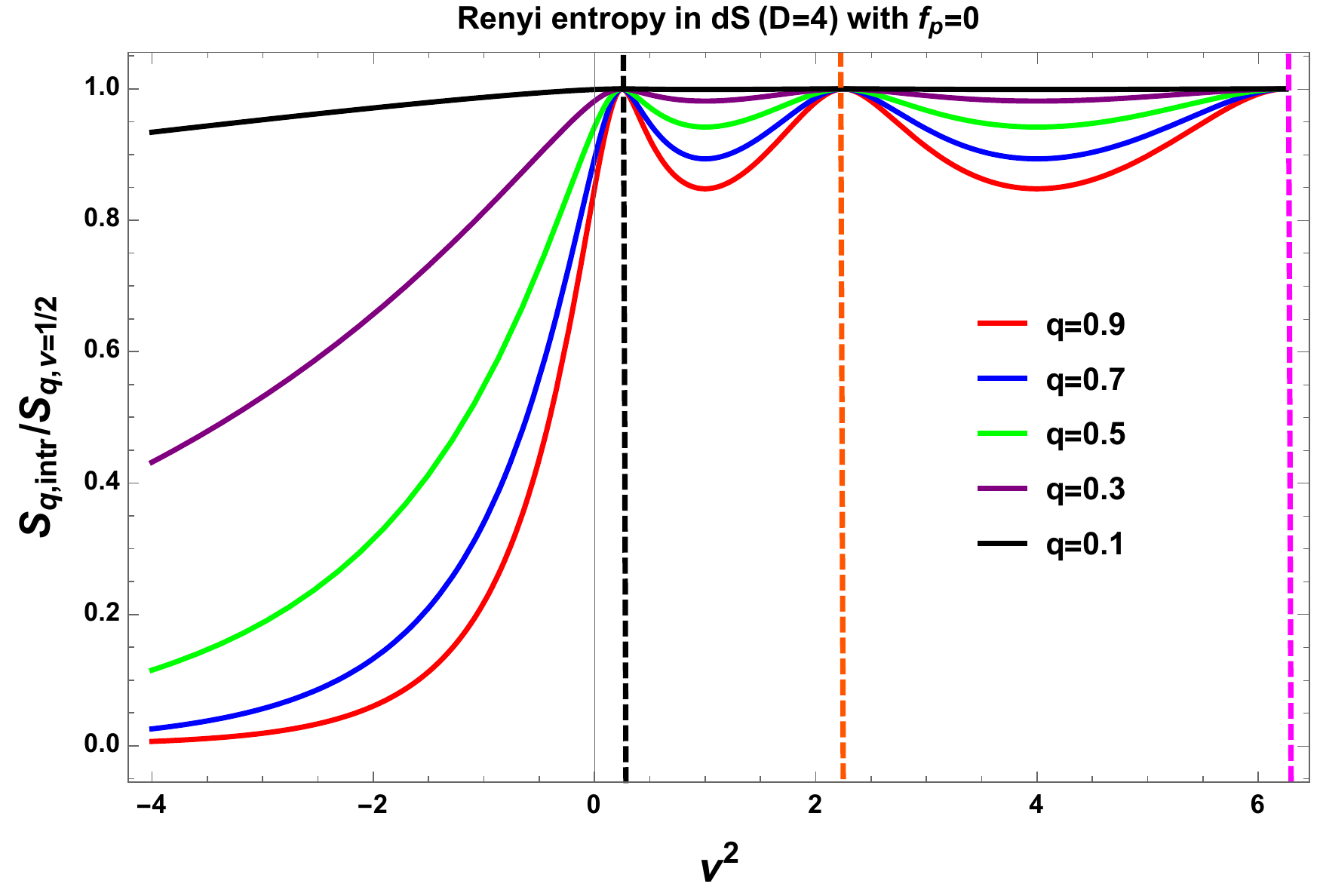}
                                                                                                                                                                   \label{vzgax5}
                                                                                                                                                                   }
                                                                                                                                                               \subfigure[Normalized R$\acute{e}$nyi entropy vs $\nu^2$ in $3+1$ D de Sitter space with axionic source ($f_p=10^{-7}$).]{
                                                                                                                                                                   \includegraphics[width=7.5cm,height=6.5cm] {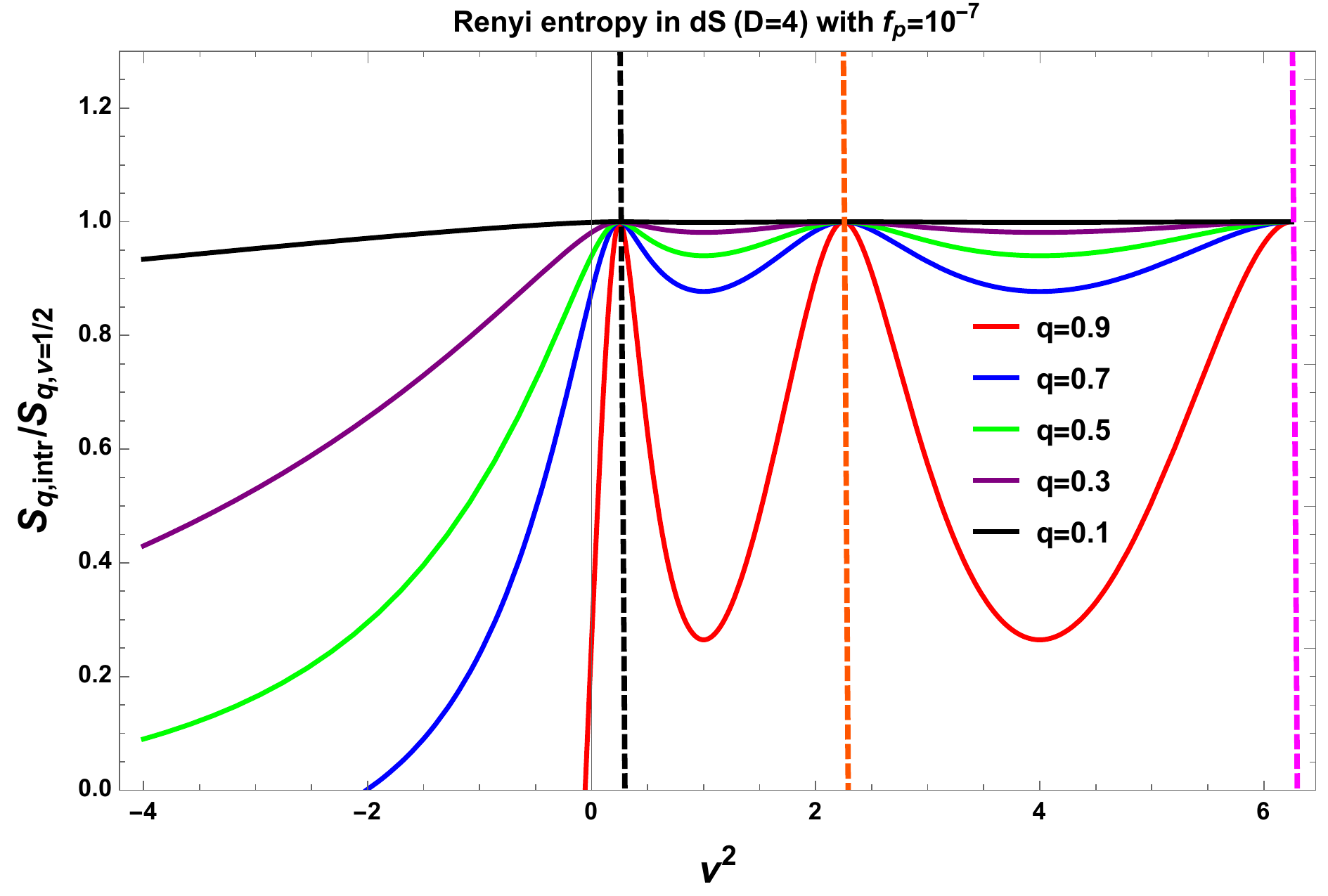}
                                                                                                                                                                   \label{vzxgax5}
                                                                                                                                                                   }
                                                                                                                                                                
                                                                                                                                                               \caption[Optional caption for list of figures]{Normalized R$\acute{e}$nyi entropy $S_{intr}/S_{\nu=1/2}$ vs mass parameter $\nu^2$ in $3+1$ D de Sitter space in absence of axionic source ($f_p=0$) and in presence of axionic source ($f_p=10^{-7}$) for $q=0.1$ (\textcolor{black}{\bf black}),$q=0.3$ (\textcolor{purple}{\bf purple}),$q=0.5$ (\textcolor{green}{\bf green}),$q=0.7$ (\textcolor{blue}{\bf blue}),$q=0.9$ (\textcolor{red}{\bf red}) with $`+'$ branch of solution of $|\gamma_p|$ and $|\Gamma_{p,n}|$. Here we set $\Lambda_{\bf C}=300$. In both the cases we have normalized the entanglement entropy with the result obtained from $\nu=1/2$ which is the conformal limiting result.} 
                                                                                                                                                               \label{vzxxgg2}
                                                                                                                                                               \end{figure*}   
                                                                                                                                                                                                                                                                                                     
                                                                                                             \begin{figure*}[htb]
                                                                                                                                              \centering
                                                                                                                                                             \subfigure[Normalized R$\acute{e}$nyi entropy vs $\nu^2$ in $3+1$ D de Sitter space without axionic source ($f_p=0$).]{
                                                                                                                                                                 \includegraphics[width=7.5cm,height=6.5cm] {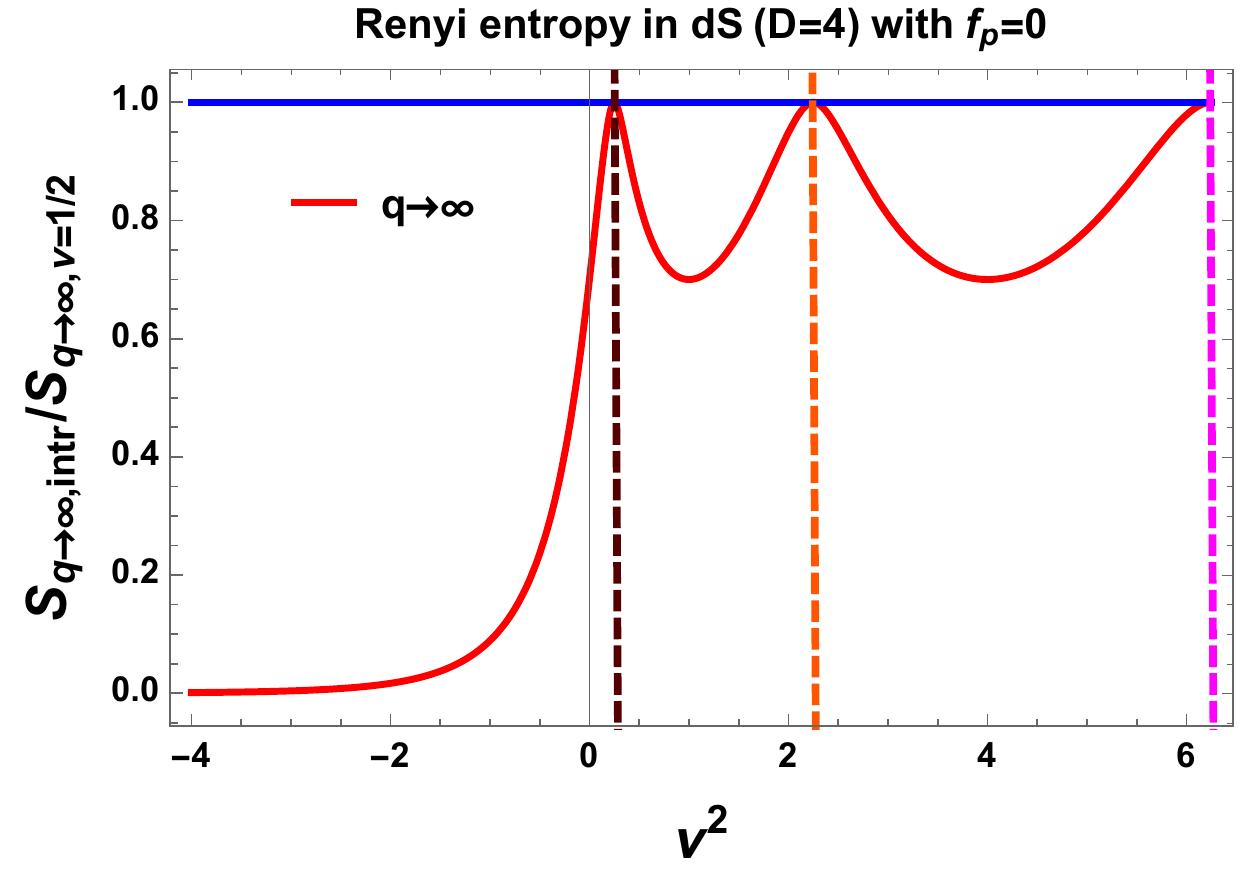}
                                                                                                                                                                 \label{azgax5}
                                                                                                                                                                 }
                                                                                                                                                             \subfigure[Normalized R$\acute{e}$nyi entropy vs $\nu^2$ in $3+1$ D de Sitter space with axionic source ($f_p=10^{-7}$).]{
                                                                                                                                                                 \includegraphics[width=7.5cm,height=6.5cm] {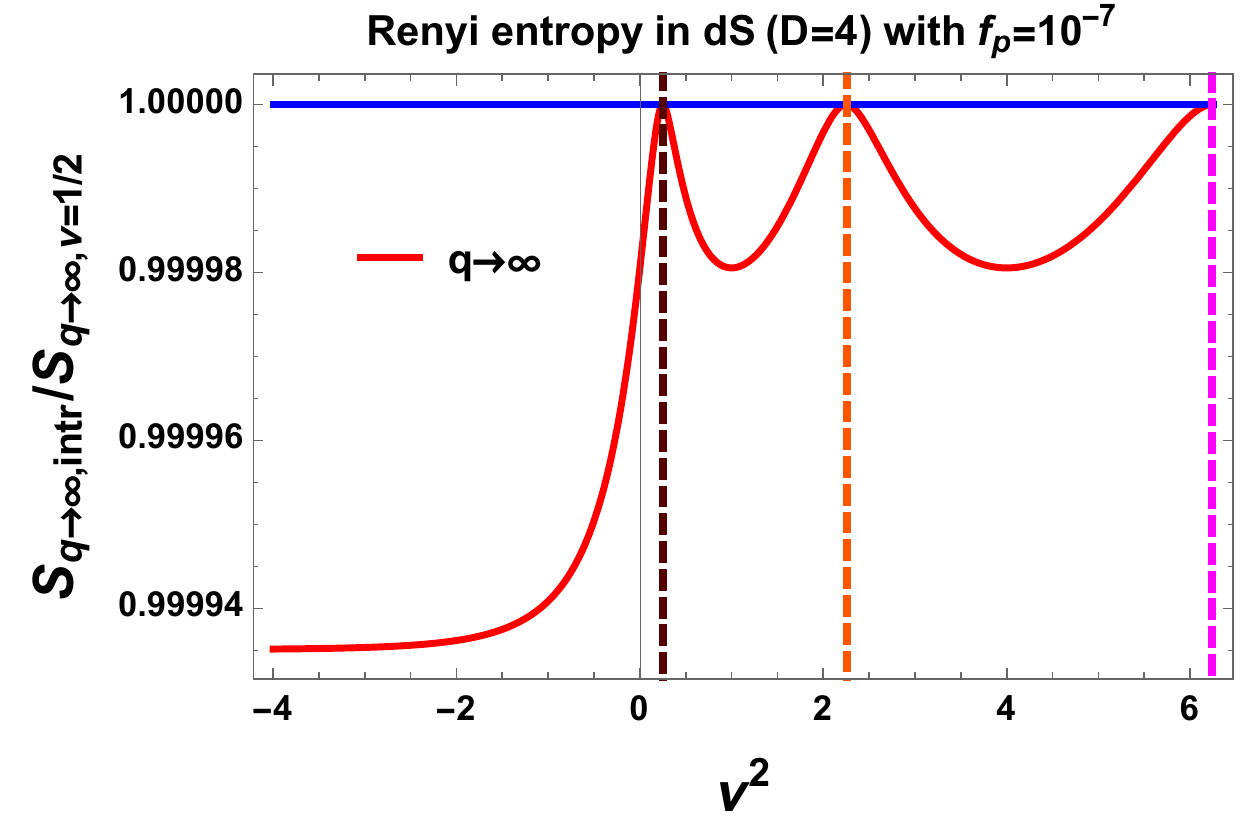}
                                                                                                                                                                 \label{azxgax5}
                                                                                                                                                                 }
                                                                                                                                                              
                                                                                                                                                             \caption[Optional caption for list of figures]{Normalized R$\acute{e}$nyi entropy $S_{intr}/S_{\nu=1/2}$ vs mass parameter $\nu^2$ in $3+1$ D de Sitter space in absence of axionic source ($f_p=0$) and in presence of axionic source ($f_p=10^{-7}$) for $q\rightarrow \infty$ with $`+'$ branch of solution of $|\gamma_p|$ and $|\Gamma_{p,n}|$, which signifies largest eigenvalue of density matrix. Here we set $\Lambda_{\bf C}=10000$. In both the cases we have normalized the entanglement entropy with the result obtained from $\nu=1/2$ which is the conformal limiting result.} 
                                                                                                                                                             \label{azxxgg2}
                                                                                                                                                             \end{figure*}

                     \end{itemize}    
 In fig.~(\ref{zgax5}) and fig.~(\ref{zxgax5}), we have explicitly shown the behaviour of R$\acute{e}$nyi entropy in $3+1$ D de Sitter space in absence ($f_p=0$) and presence ($f_p=10^{-7}$) of axionic source with respect to the mass parameter $\nu^2$ where we consider both small and large mass limiting situation with cutoff of rescaled principal quantum number fixed at the value $\Lambda_{\bf C}=10000$. In both the cases we have normalized the entanglement entropy with the result obtained from $\nu=1/2$ which is the conformal limiting result. In fig.~(\ref{zgax5}), it is clearly observed that in absence of axionic source in the large mass limiting range where $\nu^2<0$, the normalized R$\acute{e}$nyi entropy $S_{q,intr}/S_{q,\nu=1/2}$ saturates to zero suddenly through a jump at $\nu=0$ with $q=0.5$ (\textcolor{green}{\bf green}), $q=0.7$ (\textcolor{blue}{\bf blue}), $q=0.9$ (\textcolor{red}{\bf red}) for large values of $\nu^2$ in the negative axis. But for $q=0.1$ (\textcolor{black}{\bf black}) and $q=0.7$ (\textcolor{purple}{\bf purple}) $S_{q,intr}/S_{q,\nu=1/2}$ saturates to zero very slowly for the very large mass limiting range with $\nu^2<0$. On the other hand, if we increase the value of $\nu^2$ and move towards the $\nu^2>0$ region then the normalized R$\acute{e}$nyi entropy $S_{q,intr}/S_{q,\nu=1/2}$ will increase to its maximum value $\left(S_{q,intr}/S_{q,\nu=1/2}\right)_{max}=1$ at $\nu=1/2$ for all values of $q$, where $q>0$. Further in between $3/2<\nu<5/2$ 
  normalized R$\acute{e}$nyi entropy $S_{q,intr}/S_{q,\nu=1/2}$ shows one oscillation and reaches its maximum value $\left(S_{intr}/S_{\nu=1/2}\right)_{max}=1$ again exactly at $\nu=3/2$ for all values of $q$, where $q>0$. Here it is important to note that, as the value of $q$ increase the position (height) of the minima of the oscillation decrease and exactly reaches the result with normalized entanglement entropy once we go to the $q\rightarrow 1$ limiting situation. Here for $q=0.9$ (\textcolor{red}{\bf red}) we have represented this feature explicitly. Similar situation occurs  within the interval $5/2<\nu<7/2$ and the same oscillating feature will repeat itself for all values of $q$, where $q>0$. But all such oscillations are apeariodic in nature and for large positive values of the mass parameter $\nu^2$ the period of oscillation increases for all values of $q$, where $q>0$. Also in fig.~(\ref{zgax5}), for $\nu=1/2$ (\textcolor{black}{\bf black} dashed line), $\nu=3/2$ (\textcolor{orange}{\bf orange} dashed line) and $\nu=7/2$ (\textcolor{magenta}{\bf magenta} dashed line) we get same maxima of the oscillations at $\left(S_{q,intr}/S_{q,\nu=1/2}\right)_{max}=1$. In fig.~(\ref{zxgax5}), it is clearly depicted that in presence of axionic source in the large mass limiting range where $\nu^2<0$, the normalized R$\acute{e}$nyi entropy $S_{intr}/S_{\nu=1/2}$ saturates to unity for $q=0.9$ (\textcolor{red}{\bf red}) or other value for $q=0.5$ (\textcolor{green}{\bf green}) and $q=0.7$ (\textcolor{blue}{\bf blue}) through a jump at $\nu=0$ for large values of $\nu^2$ in the negative axis. Except from the jump at $\nu=0$ in normalized R$\acute{e}$nyi entropy for $q=0.9$ (\textcolor{red}{\bf red}) it replicates the limiting situation $q\rightarrow 1$ as required to produce normalized entanglement entropy. Such little deviation appears due to the presence of the axionic source term which is explicitly derived earlier. Most importantly for a given value of $q$ with $q>0$ the height of the maxima of oscillation gives the maximum value of the normalized R$\acute{e}$nyi entropy and its magnitude increases with decreasing value of $q$ which follows the restriction $q>0$. On the other hand, in presence of axionic source if we increase the value of $\nu^2$ and move towards the $\nu^2>0$ region then the normalized R$\acute{e}$nyi entropy $S_{q,intr}/S_{q,\nu=1/2}$ will decrease to its minimum value $\left(S_{q,intr}/S_{q,\nu=1/2}\right)_{min}=1$ at $\nu=1/2$, which is the maximum value of  normalized R$\acute{e}$nyi entropy without axionic source. Further in between $3/2<\nu<5/2$ 
   normalized R$\acute{e}$nyi entropy $S_{q,intr}/S_{q,\nu=1/2}$ in presence of the axionic source shows one oscillation and reaches its minimum value $\left(S_{q,intr}/S_{q,\nu=1/2}\right)_{max}=1$ at $\nu=3/2$. In presence of source contribution minimum value of normalized R$\acute{e}$nyi entropy is appearing in the subsequent intervals and the oscillating feature will repeat itself with increasing apeariodic nature for large positive values of the mass parameter $\nu^2$. But in presence of the source position of the minima in consecutive oscillations significantly increase compared to the result obtained without source. Additionally, it is important to note that in fig.~(\ref{zxgax5}), for $\nu=1/2$ (\textcolor{black}{\bf black} dashed line), $\nu=3/2$ (\textcolor{orange}{\bf orange} dashed line) and $\nu=7/2$ (\textcolor{magenta}{\bf magenta} dashed line) we get same minimum of the oscillations at $\left(S_{q,intr}/S_{q,\nu=1/2}\right)_{min}=1$.
   
    Further in fig.~(\ref{vzgax5}) and fig.~(\ref{vzxgax5}), we have explicitly shown the behaviour of R$\acute{e}$nyi entropy in $3+1$ D de Sitter space in absence ($f_p=0$) and presence ($f_p=10^{-7}$) of axionic source with respect to the mass parameter $\nu^2$ where we consider both small and large mass limiting situation with different cutoff of rescaled principal quantum number fixed at the small value $\Lambda_{\bf C}=300$. In both the cases we have normalized the entanglement entropy with the result obtained from $\nu=1/2$ which is the conformal limiting result. In fig.~(\ref{vzgax5}), it is clearly observed that in absence of axionic source in the large mass limiting range where $\nu^2<0$, the normalized R$\acute{e}$nyi entropy $S_{q,intr}/S_{q,\nu=1/2}$ saturates to different values depending on the different values of the parameter $q$ as given by,
    $q=0.1$ (\textcolor{green}{\bf black}), $q=0.3$ (\textcolor{blue}{\bf purple}), $q=0.5$ (\textcolor{green}{\bf green}), $q=0.7$ (\textcolor{blue}{\bf blue}), $q=0.9$ (\textcolor{red}{\bf red}) with large values of $\nu^2$ in the negative axis. On the other hand, if we increase the value of $\nu^2$ and move towards the $\nu^2>0$ region then we get the exactly similar features as appearing for fig.~(\ref{zgax5}). In fig.~(\ref{vzxgax5}), it is clearly depicted that in presence of axionic source in the large mass limiting range where $\nu^2<0$, the normalized R$\acute{e}$nyi entropy $S_{intr}/S_{\nu=1/2}$ saturates at different values for different values of the parameter $q$ for large values of $\nu^2$ in the negative axis. Most importantly for a given value of $q$ with $q>0$ the height of the maxima of oscillation increases with increasing value of $q$ with $\Lambda_{\bf C}=300 $. On the other hand, in presence of axionic source if we increase the value of $\nu^2$ and move towards the $\nu^2>0$ region then the normalized R$\acute{e}$nyi entropy $S_{q,intr}/S_{q,\nu=1/2}$ will increase to its maximum value $\left(S_{q,intr}/S_{q,\nu=1/2}\right)_{max}=1$ at $\nu=1/2$, which is also the maximum value of  normalized R$\acute{e}$nyi entropy without axionic source with $\Lambda_{\bf C}=300 $. Further in between $3/2<\nu<5/2$ 
       normalized R$\acute{e}$nyi entropy $S_{q,intr}/S_{q,\nu=1/2}$ in presence of the axionic source shows one oscillation and reaches its maximum value $\left(S_{q,intr}/S_{q,\nu=1/2}\right)_{max}=1$ at $\nu=3/2$. Additionally, it is important to note that in fig.~(\ref{vzxgax5}), for $\nu=1/2$ (\textcolor{black}{\bf black} dashed line), $\nu=3/2$ (\textcolor{orange}{\bf orange} dashed line) and $\nu=7/2$ (\textcolor{magenta}{\bf magenta} dashed line) we get same maximum of the oscillations at $\left(S_{q,intr}/S_{q,\nu=1/2}\right)_{max}=1$.  
    
    Next, in fig.~(\ref{azgax5}) and fig.~(\ref{azxgax5}), we have explicitly shown the behaviour of normalized R$\acute{e}$nyi entropy in $3+1$ D de Sitter space in absence ($f_p=0$) and presence ($f_p=10^{-7}$) of axionic source with respect to the mass parameter $\nu^2$ where we consider both small and large mass limiting situation alongwith the limiting case $q\rightarrow \infty$, which produces the largest eigenvalue of the density matrix. Here it is important to note that, the feature obatined for the $q\rightarrow\infty$ case is looks similar to the result obtained for normalized entanglement entropy in de Sitter space in $3+1$ D. However, the differences are appearing in the height of the minimum of the oscillations obtained from $\nu^2>0$ region. It also establishes that the results obtained for $q\rightarrow \infty$ and $q\rightarrow 1$ are not same. Here also in fig.~(\ref{azgax5}) and fig.~(\ref{azxgax5}), for $\nu=1/2$ (\textcolor{black}{\bf black} dashed line), $\nu=3/2$ (\textcolor{orange}{\bf orange} dashed line) and $\nu=7/2$ (\textcolor{magenta}{\bf magenta} dashed line) we get same maxima of the oscillations at $\left(S_{q,intr}/S_{q,\nu=1/2}\right)_{max}=1$.

\section{\textcolor{blue}{Conclusion}}
\label{ka44}
To summarize, in the present article, we have addressed the following points:
\begin{itemize}
	\item  Firstly we have started our discussion with the computational strategy for quantifying the expression for entanglement entropy in de Sitter space for a bipartite quantum field theory of axion with a linear contribution in the effective potential as appearing from string theory using a specific compactification technique.
	Here we consider two subclass of models for axion, in \textcolor{red}{\bf Case~I} only linear field contribution is appearing and for \textcolor{red}{\bf Case~II} additionally a quadratic field combination is appearing in the effective potential for axion in $\phi<f_a$ regime. 
	
	\item  Further, we have discussed about the basic setup for computing the all possible UV divergent and UV finite contribution of entanglement entropy in $3+1$ dimension. We have also discussed the importance of UV finite contribution  compared to the UV divergent part to quantify the interesting part of the entanglement entropy.  
	
	\item Next, we have discussed the geometrical construction and underlying symmetries of the system under consideration in $3+1$ D de Sitter space. Also we have discussed the string theory origin of axion effective potential and its role in the present context.
	
	\item  Further we have derived the expression for the wave function of axion which is the solution of the field equation in a hyperbolic slice, commonly known as open chart. We have explicitly shown that the total solution can be written as a sum of a complementary solution and a particular solution, which is a completely new solution in presence of axion source term in the equation of motion. 
	
	\item Next using the Bunch-Davies initial condition for the choice of the vacuum state we have written the total solution of the wave function in terms of creation and annihilation operators. Using this fact we construct a suitable basis by applying  Bogoliubov transformation for the Bunch-Davies vacuum state using which we construct the expression for density matrix by tracing over the exterior region of the prescribed axionic quantum field theory in de Sitter space.
	
	\item  Further, using the expression for the density matrix and also using the Von Neumann measure of entropy we have derived the new formula for entanglement entropy in $3+1$ dimensional de Sitter space
	in presence of axion source and we have also checked the consistency of our result by comparing with the result obtained in ref.~\cite{Maldacena:2012xp}, which was derived for a free massive scalar field without linear contribution in the effective potential. Here it is important to note that the large mass limiting range we have provided the exact analytical expression for the entanglement entropy. But to analyze the correctness of our derived result and to compare with ref.~\cite{Maldacena:2012xp} we have further used numerical techniques to study the behaviour of entanglement entropy with mass parameter $\nu^2$ for \textcolor{red}{\bf Case~I} and \textcolor{red}{\bf Case~II}.
	
	\item  Similarly we have computed the modified expression for the R$\acute{e}$nyi entropy in presence of axion source in $3+1$ dimensional de Sitter space using which after taking $q\rightarrow 1$ we have found that in presence of axion linear contribution it is not possible to get the exact analytical formula for entanglement entropy using Von Neumann entropy measure. Once we switch off the contribution from the linear term of the axion in the effective potential one can get back the exact formula for entanglement entropy as appearing in ref.~\cite{Maldacena:2012xp} by taking $q\rightarrow 1$ limit. Our analysis also clearly implies that the definition of R$\acute{e}$nyi entropy is not universal for any arbitrary structures of effective potential. Only for free theory where no such linear contribution appears, the definition is universal. Here the large mass limiting range we have provided the exact analytical expression for the R$\acute{e}$nyi entropy. But to analyze the correctness of our result and to compare with ref.~\cite{Maldacena:2012xp} we have further used numerical techniques to study the behaviour of entanglement entropy with mass parameter $\nu^2$ for \textcolor{red}{\bf Case~I} and \textcolor{red}{\bf Case~II}. Additionally, by using numerical technique and taking $q\rightarrow \infty$ we have studied the behaviour of largest eigenvalue of the density matrix with respect to the mass parameter $\nu^2$ for both the cases in presence of axion source.

	\item In this context it is important to mention here that, the appearance of non vanishing entanglement entropy in de Sitter space in $3+1$ D directly verifies the correctness of the existence of the one point function in primordial cosmology due to axionic pair. Moreover, it is one of the important facts of cosmological evolution of our universe and temperature fluctuations of the Cosmic Microwave Background (CMB) originated from quantum fluctuations during the initial inflationary era as it acts a theoretical tool using which one can able to break the degeneracy amongst the cosmological predictions of various models.  
	\end{itemize}

The future prospects of our work are appended below:
\begin{itemize}

\item In our analysis we choose Bunch-Davies vacuum state to specify the solution of wave function. Also it is used to determine rest of the quantities related to quantum entanglement. In future one can generalize this study of quantum entanglement in de Sitter space in presence of axion using $\alpha$ vacua \cite{Choudhury:2017iva}. 

\item Using this prescribed methodology one can further derive the expression for two point and any other higher $n$ point correlation function in superhorizon, subhorizon and horizon crossing scale to study the cosmological consequences of quantum fluctuation using Bunch-Davies vacuum as well as $\alpha$ vacua.

\item The condition for violation of Bell’s inequality in the context of primordial
cosmology are fixed by the further study of the quantum state in detail. In this connection computation of quantum discord, entanglement negativity and other possible strong measures in presence of axion will is also play important role. We have not
addressed all these issues in this paper, which one can address in future for completeness.
\end{itemize}
	\section*{\textcolor{blue}{Acknowledgments}}
	SC would like to thank Inter University Center for Atsronomy and Astrophysics, Pune for providing the Post-Doctoral
	Research Fellowship. SC take this opportunity to thank sincerely to  Shiraz Minwalla, Gautam Mandal and Varun Sahni for their constant support
	and inspiration. SC also thank the organizers of Indian String Meet 2016, Advanced String School 2017 and ST$^{4}$ 2017
	for providing the local hospitality during the work. SC also thank IOP, Bhubaneswar, CMI, Chennai, SINP, Kolkata and 
IACS, Kolkata for
	providing the academic visit during the work. SP acknowledges the J. C. Bose National Fellowship for support of his research. Last but not the least, We would all like to acknowledge our
	debt to the people of India for their generous and steady support for research in natural sciences, especially
	for string theory and cosmology.
\appendix

\vspace{0.2cm}
\section*{\textcolor{blue}{\Large Appendix}}
\label{ka55}

\section{ \textcolor{blue}{Derivation of entanglement entropy} }
\label{ka66a}
For the \textcolor{red}{\bf Case~I} and \textcolor{red}{\bf Case~II}
                  the expressions for the entanglement entropy can be expressed as a sum over four contributions, which can be expressed as:
                  \bea S(p,\nu)&=& \sum^{4}_{m=1}\Delta_{m}(p,\nu),\eea
                  where the four explicit contributions in terms of the functions $\Delta_{m}(p,\nu)\forall m=1,2,3,4$ as appearing in the entanglement entropy is given by:
                  \bea \Delta_{1}(p,\nu)&=&-{\bf \rm Tr}\left[\frac{\left(1-|\gamma_{p}|^2\right)}{1+f_p}\sum^{\infty}_{k=0}|\gamma_{p}|^{2k}|k;p,l,m\rangle\langle k;p,l,m|\right.\nonumber\\&& \left.~~~~~~~~~~~~~~~~~\ln\left(\frac{\left(1-|\gamma_{p}|^2\right)}{1+f_p}\sum^{\infty}_{k=0}|\gamma_{p}|^{2k}|k;p,l,m\rangle\langle k;p,l,m|\right)\right]\nonumber\\
                  &=&-\ln\left(1-|\gamma_{p}|^2\right)-\frac{|\gamma_{p}|^2}{\left(1-|\gamma_{p}|^2\right)}\ln\left(|\gamma_{p}|^2\right)+f_p\ln\left(1+f_p\right),\\
                   \Delta_{2}(p,\nu)&=&-{\bf \rm Tr}\left[\frac{f^2_p}{1+f_p}\sum^{\infty}_{n=0}\sum^{\infty}_{r=0}|\Gamma_{p,n}|^{2r}|n,r;p,l,m\rangle\langle n,r;p,l,m|\right.\nonumber\\&& \left.~~~~~~~~~~~~~~~~~\ln\left(\frac{\left(1-|\gamma_{p}|^2\right)}{1+f_p}\sum^{\infty}_{k=0}|\gamma_{p}|^{2k}|k;p,l,m\rangle\langle k;p,l,m|\right)\right]\nonumber\\
                                     &=&-\frac{f_p}{1+f_p}\left[\ln\left(1-|\gamma_{p}|^2\right)+\frac{|\gamma_{p}|^2}{\left(1-|\gamma_{p}|^2\right)}\ln\left(|\gamma_{p}|^2\right)\right],\\
                   \Delta_{3}(p,\nu)&=&-{\bf \rm Tr}\left[\frac{\left(1-|\gamma_{p}|^2\right)}{1+f_p}\sum^{\infty}_{k=0}|\gamma_{p}|^{2k}|k;p,l,m\rangle\langle k;p,l,m|\right.\nonumber\\&& \left.~~~~~~~~~~~~~~~~~\ln\left(1+\left(1-|\gamma_{p}|^2\right)^{-1}f^2_p 
                   \left(\sum^{\infty}_{k=0}|\gamma_{p}|^{2k}|k;p,l,m\rangle\langle k;p,l,m|\right)^{-1}\right.\right.\nonumber\\&& \left.\left.~~~~~~~~~~~~~~~~~~~~~~~~~\left(\sum^{\infty}_{n=0}\sum^{\infty}_{r=0}|\Gamma_{p,n}|^{2r}|n,r;p,l,m\rangle\langle n,r;p,l,m|\right)\right)\right]=-\frac{1}{1+f_p}\ln\left(1+f_p
                                                        \right),~~~~~~~\\
                 \Delta_{4}(p,\nu)&=&-{\bf \rm Tr}\left[\frac{f^2_p}{1+f_p}\sum^{\infty}_{n=0}\sum^{\infty}_{r=0}|\Gamma_{p,n}|^{2r}|n,r;p,l,m\rangle\langle n,r;p,l,m|\right.\nonumber\\&& \left.~~~~~~~~~~~~~~~~~\ln\left(1+\left(1-|\gamma_{p}|^2\right)^{-1}f^2_p 
                 \left(\sum^{\infty}_{k=0}|\gamma_{p}|^{2k}|k;p,l,m\rangle\langle k;p,l,m|\right)^{-1}\right.\right.\nonumber\\&& \left.\left.~~~~~~~~~~~~~~~~~~~~~~~~~\left(\sum^{\infty}_{n=0}\sum^{\infty}_{r=0}|\Gamma_{p,n}|^{2r}|n,r;p,l,m\rangle\langle n,r;p,l,m|\right)\right)\right]
                                                                         =-\frac{f_p}{1+f_p}                                             \ln\left(1+f_p
                                                                         \right).\eea
                                                                         Consequently the expression for the entanglement entropy in terms of the complementary and particular part of the obtaned solution can be expressed as:
\bea \boxed{S(p,\nu)=-\left(1+\frac{f_p}{1+f_p}\right)\left[\ln\left(1-|\gamma_{p}|^2\right)+\frac{|\gamma_{p}|^2}{\left(1-|\gamma_{p}|^2\right)}\ln\left(|\gamma_{p}|^2\right)\right]-\left(1-f_p\right)\ln\left(1+f_p\right)}.~~~~~~~\eea             
\section{ \textcolor{blue}{Derivation of R$\acute{e}$nyi entropy}}
\label{ka66b} 
For the \textcolor{red}{\bf Case~I} and \textcolor{red}{\bf Case~II}
                  the expressions for the R$\acute{e}$nyi entropy can be expressed as a sum over four contributions, which can be expressed as:
                  \bea S_{q}(p,\nu)&=& \sum^{3}_{m=1}V_{m,q}(p,\nu),\eea
                  where the four explicit contributions in terms of the functions $V_{m,q}(p,\nu)\forall m=1,2,3$ as appearing in the entanglement entropy is given by:
                  \bea V_{1,q}(p,\nu)&=&\frac{1}{1-q}\ln\left[{\bf \rm Tr}\left\{\left(\frac{\left(1-|\gamma_{p}|^2\right)}{1+f_p}\right)^q\left(\sum^{\infty}_{k=0}|\gamma_{p}|^{2k}|k;p,l,m\rangle\langle k;p,l,m|\right)^q\right\}\right]\nonumber\\
                  &=&\left[\frac{q}{1-q}\ln\left(1-|\gamma_{p}|^2\right)-\frac{1}{1-q}\ln\left(1-|\gamma_{p}|^{2q}\right)\right]-\frac{q}{1-q}\ln\left(1+f_p\right),\\
                   V_{2,q}(p,\nu)&=&\frac{1}{1-q}\ln\left[1+\sum^{q}_{k=1}{}^{q}C_{k}{\bf \rm Tr}\left\{\frac{(f_p)^{2k}}{(1+f_p)^k}\left(\sum^{\infty}_{n=0}\sum^{\infty}_{r=0}|\Gamma_{p,n}|^{2r}|n,r;p,l,m\rangle\langle n,r;p,l,m|\right)^{k}\right\}\right.\nonumber\\&& \left.~~~~~~~~~~~~~~~~~{\bf \rm Tr}\left\{\left(\frac{\left(1-|\gamma_{p}|^2\right)}{1+f_p}\right)^{-k}\left(\sum^{\infty}_{k=0}|\gamma_{p}|^{2k}|k;p,l,m\rangle\langle k;p,l,m|\right)^{-k}\right\}\right]\nonumber\\
                                     &=&\frac{1}{1-q}\ln\left[1+\sum^{q}_{k=1}{}^{q}{\bf C}_{k} (f_p)^{k
                                           }\frac{\left(1-|\gamma_{p}|^2\right)^{-k}}{\left(1-|\gamma_{p}|^{-2k}\right)}\right].\eea
                                                                         Consequently the expression for the entanglement entropy in terms of the complementary and particular part of the obtaned solution can be expressed as:
\bea S_{q}(p,\nu)&=&\left[\frac{q}{1-q}\ln\left(1-|\gamma_{p}|^2\right)-\frac{1}{1-q}\ln\left(1-|\gamma_{p}|^{2q}\right)\right]-\frac{q}{1-q}\ln\left(1+f_p\right)\nonumber\\&&~~~~~~~~~~~~~~~~~~~~~~~~~~~~~~+\frac{1}{1-q}\ln\left[1+\sum^{q}_{k=1}{}^{q}{\bf C}_{k} (f_p)^{k
      }\frac{\left(1-|\gamma_{p}|^2\right)^{-k}}{\left(1-|\gamma_{p}|^{-2k}\right)}\right].~~~~\eea           
    

\end{document}